\documentclass[showpacs,preprintnumbers,amsmath,amssymb]{revtex4}
\usepackage{amsmath,amssymb,graphics,epsfig,subfigure}
\usepackage{color}

\begin{document}

\renewcommand{\baselinestretch}{1.3}

\title{Photon sphere and reentrant phase transition of charged Born-Infeld-AdS black holes}

\author{Yu-Meng Xu$^{1}$,
Hui-Min Wang$^{1}$,
Yu-Xiao Liu$^{1}$,
    Shao-Wen Wei$^{1,2}$ \footnote{weishw@lzu.edu.cn}}

\affiliation{$^{1}$ Institute of Theoretical Physics $\&$ Research Center of Gravitation, Lanzhou University, Lanzhou 730000, People's Republic of China\\
    $^{2}$ Department of Physics and Astronomy, University of Waterloo, Waterloo, Ontario, Canada, N2L 3G1}
\date{\today}

\begin{abstract}
Comparing to the charged AdS black hole in GR, a new interesting phase transition, the reentrant phase transition, is observed in a charged Born-Infeld-AdS black hole system. It is worth to extend the study of the relationship between the photon sphere and the thermodynamic phase transition, especially the reentrant phase transition, to this black hole background. According to the number of the thermodynamic critical points, the black hole systems are divided into four cases with different values of Born-Infeld parameter $b$, where the black hole systems can have no phase transition, reentrant phase transition, or Van der Waals-like phase transition. For these different cases, we obtain the corresponding phase structures in pressure-temperature diagram and temperature-specific volume diagram. The tiny differences between these cases are clearly displayed. On the other hand, the radius $r_{\rm ps}$ and the minimal impact parameter $u_{\rm ps}$ of the photon sphere are calculated via the effective potential of the radial motion of photons. For different cases, $r_{\rm ps}$ and $u_{\rm ps}$ are found to have different behaviors. In particular, with the increase of $r_{\rm ps}$ or $u_{\rm ps}$, the temperature possesses a decrease-increase-decrease-increase behavior for fixed pressure if there exists the reentrant phase transition. While for fixed temperature, the pressure will show an increase-decrease-increase-decrease behavior instead. These behaviors are quite different from that of the Van der Waals-like phase transition. Near the critical point, the changes of $r_{\rm ps}$ and $u_{\rm ps}$ among the black hole phase transition confirm an universal critical exponent $\frac{1}{2}$. We also find that the temperature and pressure corresponding to the extremal points of $r_{\rm ps}$ and $u_{\rm ps}$ are highly consistent with the thermodynamic metastable curve for the black hole systems with different values of $b$. Furthermore, we also extend the corresponding study to the higher dimensional black holes cases. The results show that the photon sphere behaves quite different for the Van der Waals-like phase transition or the reentrant phase transition, and both the phase transitions can be reflected through the photon sphere.
\end{abstract}

\keywords{Black holes, thermodynamic, photon sphere}
\pacs{04.70.Dy, 04.50.Kd, 04.25.-g}

\maketitle

\section{Introduction}\label{sec:Introduction}

Since the establishment of the four thermodynamic laws of black holes \cite{Bekenstein,Bardeen}, phase transition continues to be an attractive and valuable subject in gravitational physics. Of particular interest is that in an anti-de Sitter (AdS) space, the negative cosmological constant $\Lambda$ was interpreted as the pressure $P$ \cite{Kastor,Dolan00,Cvetic}
\begin{equation}
  P=-\frac{\Lambda}{8\pi}.
\end{equation}
The corresponding conjugate quantity is the thermodynamic volume of the black hole system. After including this pressure and volume term, the black hole first law exactly coincides with that of an ordinary thermodynamic system. The inconsistency of the first law and Smarr formula for a rotating AdS black hole was also successfully solved. Moreover, the small-large black hole phase transition was identified with the liquid-gas phase transition of Van der Waals (VdW) fluid \cite{Kubiznak}. Besides it, more interesting phase transitions were found \cite{Gunasekaran,ZouZouZou,Altamirano,Mann,Frassino,Wei0,Kostouki,Wei1,Hennigar,ZouYue,
Hendi,Hendi2,Hendi3,Hendi4,Momeni,Chakraborty,Weisw}. In particular, near the critical point, black hole systems were also found to possess the same critical phenomena with the VdW fluid. This, to some extent, implies that these thermodynamic systems may share the similar microstructure, and the issue was further investigated in Ref. \cite{Weisw}.

Interestingly, the black hole phase transition information is also expected to be encoded or revealed by its dynamic and gravity properties. Therefore, studying the relationship between them can bridge the study of these two sides.

The quasinormal modes (QNMs) are dynamical perturbations, and they can provide us with the significant observational signatures to test the natural property of these black holes. The subject that probing the black hole phase transition by QNMs was numerically investigated in Ref. \cite{Pan}. For a charged Reissner-Nordstr\"om black hole, it was found that when the black hole charge exceeds the Davie point, the QNM frequency $\omega$ starts to display a spiral-like shape in the complex $\omega$ plane. The Davie point is related to the divergence point of the heat capacity. So this point actually measures a black hole phase transition between an unstable phase and a stable phase. Some further studies on this issue can be found in Refs. \cite{BertiBerti,He,He2,LinLin,Sup}.

For the VdW-like phase transition of a charged AdS black hole, the result given in Ref. \cite{Liu} suggests that the QNMs have different slopes along the small black hole phase or the large black hole phase. Thus this signature can be used to test the first-order small-large black hole phase transition. However, this method was argued to be accurate only for low pressure or temperature. While when the black hole system approaches to its critical case, the QNMs become more complicated, and thus one must be very careful with these cases. This method was also extended to other AdS black hole systems \cite{Liuzz,Mahapatra,Chabab,Zou2,Prasia0,Prasia2,Prasia,www,ZengZeng,YueYue}. All these results imply that the VdW-like phase transition can be reflected by the slope of the QNMs. Another interesting thing is whether the reentrant phase transition can be reflected by the behavior of QNMs, and how to distinguish it from the VdW-like phase transition. Since the isothermal and isobaric lines become more complicated. Thus we believe that beside the slope of the QNMs, more subtle information of the QNMs should be included in. Therefore, this issue deserves further study, and we will not discuss it here.

On the other hand, the geodesics of a test particle around a black hole has a close relation to the strong gravitational effects, such as the lensing and shadow. Studying the relationship between the geodesics and the black hole thermodynamics will give us a novel way to test the phase transition by using these observables. This will also open a new window to study black hole thermodynamics through the astronomical observation. Keeping this in mind, we studied the behavior of the photon sphere of a charged AdS black hole when the VdW-like phase transition occurs \cite{WeiLiuLiu}. Two key quantities, the radius and the minimal impact parameter of the photon sphere were found to behave quite differently whether there exists a VdW-like phase transition or not. These two quantities also get sudden changes when the phase transition takes place. Their changes decrease with the temperature and tend to vanish at the critical point. Further calculation also showed that these changes of the radius and the minimal impact parameter have a universal critical exponent $\frac{1}{2}$, which implies that these changes can serve as an order parameter to describe the small-large black hole phase transition. Subsequently, we extended our study to the rotating Kerr-AdS black holes \cite{WeiLiuLiu2}. Even when the black hole gets rotation, our result shows that the similar relationship between the photon ring and the phase transition also holds. Moreover, it was also found that the temperature and pressure corresponding to the extremal points of the radius and minimal impact parameter are highly consistent with the thermodynamic metastable curve for the Kerr-AdS black holes. This study was also generalized to other black hole backgrounds \cite{Bhamidipati,Han,Bhamidipatib,ZhangHan}. For example, in Ref. \cite{Bhamidipati} the authors studied the relation between the thermodynamic phase transition and the circular orbit for charged test particles. Photon orbits and thermodynamic phase transition were examined in Gauss-Bonnet AdS black holes \cite{Han}, massive black holes \cite{Bhamidipati}, and even in a general spherically symmetric spacetime \cite{ZhangHan}. In addition, the presence of photon orbit also reveals a York-Hawking-Page type phase transition of spacetimes \cite{Cvetic2,Cvetic3,Tang}.

Apart from the VdW-like phase transition, there are some other interesting phase transitions, such as the reentrant phase transition. Therefore, it is of great interest to examine the relationship between the photon sphere and the reentrant phase transition. As we know, the charged Born-Infeld (BI)-AdS black holes demonstrate a typical reentrant phase transition \cite{Gunasekaran,Hendi}. Therefore it is valuable to extend our previous study and examine the relationship for the BI-AdS black holes. These will reveal the novel behaviors of the photon sphere under the reentrant phase transition.

This work is organized as follows. In Sec. \ref{sec:phase_structure}, we give a brief review of the thermodynamics for the BI-AdS black holes. Then four different black hole cases with different values of the BI parameter $b$ are studied. The corresponding reentrant phase structure and VdW-like phase structure are displayed in the $P$-$T$ and $T$-$v$ diagrams, respectively. In each case, their thermodynamic properties are analyzed in detail. In Sec. \ref{sec:null_geodesic_and_critical}, we solve the radius and the minimal impact parameter of the photon sphere by using the effective potential of the radial motion. Then for different cases, we study the behaviors of the pressure and temperature as function of the $r_{\rm ps}$ or $u_{\rm ps}$, respectively. When the reentrant phase transition takes place, the temperature and pressure behave quite differently from that of the VdW-like phase transition. Furthermore, near the critical point, we confirm that the changes $\Delta r_{\rm ps}$ and $\Delta u_{\rm ps}$ have an universal critical exponent $\frac{1}{2}$, which is the same as that of the VdW-like phase transition for the charged-AdS black holes or rotating Kerr-AdS black holes. The temperature and pressure corresponding to the extremal points of $r_{\rm ps}$ and $u_{\rm ps}$ are also found to be highly consistent with the thermodynamic metastable curve for the BI-AdS black hole systems even there is no the reentrant phase transitions or VdW-like phase transition. In Sec. \ref{sec:higher-dim}, we study the relation between the photon sphere and the small-large black hole phase transition. The critical exponents of $\Delta r_{\rm ps}$ and $\Delta u_{\rm ps}$ are also calculated. Finally, the conclusions and discussions are presented in Sec. \ref{sec:conclusion}.

\section{Thermodynamics and phase structure}
\label{sec:phase_structure}

For the four dimensional charged BI-AdS black hole, it is found that due to the values of BI parameter, there may exist the VdW-like phase transition, reentrant phase transition \cite{Gunasekaran}. However, there only exists the VdW-like phase transition for the higher dimensional black hole cases \cite{ZouZouZou}. Among the study of the black hole phase transition, it is interesting to examine the phase diagram, by using which one could easily find the type of the phase transition in different parameter ranges. In this section, we would like to briefly review the phase transition. In Refs. \cite{Gunasekaran,ZouZouZou}, the phase diagram is given in $P$-$T$ diagram. However the coexistence region is only one curve in that diagram. In order to display the coexistence region, we would like to investigate the phase diagram in the $T$-$v$ diagram, and study the difference between the VdW-like phase transition and reentrant phase transition.

The action describing the BI-AdS black hole is
\begin{equation} \label{IEM}
I_{\rm EM} = -\frac{1}{16\pi}\int_M \sqrt{-g}\left(R+{\cal L}_{\rm BI}+\frac{6}{l^2}\right),
\end{equation}
where the BI term is
\begin{equation}
{\cal L}_{\rm BI}=4{b^2}\left(1-\sqrt{1+\frac{1}{2b^2}F^{\mu\nu}F_{\mu\nu}}\right).
\end{equation}
Here the AdS radius $l$ is related to the pressure as $P=\frac{3}{8\pi l^{2}}$.
The BI parameter $b$ denotes the maximal electromagnetic field strength, and has a relation with the string tension. Solving the corresponding Einstein equations, one can obtain the following black hole solution
\begin{eqnarray}
ds^2&=&-f(r)dt^2+\frac{dr^2}{f(r)}+r^2 (d\theta^{2}+\sin^{2}\theta d\phi^{2}),\label{BImetric}\\
F&=&\frac{Q}{\sqrt{r^4+Q^2/b^2}}dt\wedge dr\label{BIE}.
\end{eqnarray}
The metric function is given by
\begin{eqnarray}
f(r)&=&1-\frac{2M}{r}+\frac{r^2}{l^2}+\frac{2b^2}{r}\int_r^\infty \left(\sqrt{r^4+\frac{Q^2}{b^2}}-r^2\right)dr\nonumber\\
&=&1-\frac{2M}{r}+\frac{r^2}{l^2}+\frac{2b^2 r^2}{3}\left(1-\sqrt{1+\frac{Q^2}{b^2 r^4}}\right)+\frac{4Q^2}{3r^2}\,{}_2 F_1\left(\frac{1}{4},\frac{1}{2}; \frac{5}{4};-\frac{Q^2}{b^2 r^4}\right).\label{metricf}
\end{eqnarray}
Here $_2 F_1$ is the hypergeometry function. The parameters $M$ and $Q$ are the black hole mass and charge, respectively. Solving $f(r_{+})=0$, we can express the black hole mass as
\begin{equation}
M=\frac{r_{+}}{2}+\frac{r_{+}^3}{2l^2}+\frac{b^2 r_{+}^3}{3}\left(1-\sqrt{1+\frac{Q^2}{b^2 r_{+}^4}}\right)+\frac{2Q^2}{3r_{+}}\,{}_2 F_1\left(\frac{1}{4},\frac{1}{2}; \frac{5}{4};-\frac{Q^2}{b^2 r_{+}^4}\right),
\end{equation}
where $r_{+}$ is the radius of the black hole horizon. Using the `Euclidean trick', one can obtain the black hole temperature
\begin{equation}
\label{eq:T}
T=\frac{1}{4\pi r_+}\!\left(1\!+\!\frac{3r_+^2}{l^2}\!+\!2b^2 r_+^2\left(1\!-\!\sqrt{1\!+\!\frac{Q^2}{b^2 r_+^4}}\right)\right).
\end{equation}
According to the Bekenstein-Hawking area-entropy relation, the black hole entropy can be calculated as
\begin{equation}
S = \frac{A}{4}=\pi r_+^2,
\end{equation}
where $A=4\pi r_+^2$ is the area of the black hole horizon. The electric potential and the electric polarization measured at infinity with respect to the horizon are
\begin{eqnarray}
\Phi&=&\frac{Q}{r_+}\,{}_2 F_1\!\left(\frac{1}{4},\frac{1}{2};\frac{5}{4};-\frac{Q^2}{b^2 r_+^4}\right),\\
B&=&\frac{2}{3}br_{+}^{3}\left(1-\sqrt{1+\frac{Q^{2}}{b^{2}r_{+}^{4}}}+\frac{Q^{2}}{3br_{+}}\right){}_2 F_1\left(\frac{1}{4},\frac{1}{2}; \frac{5}{4};-\frac{Q^2}{b^2 r_{+}^4}\right).
\end{eqnarray}
The Gibbs free energy is
\begin{eqnarray}
G(T,P)&=&\frac{1}{4}\bigg(r_+-\frac{8\pi}{3} P r_+^3-\frac{2b^2r_+^3}{3}\left(1-\sqrt{1+\frac{Q^2}{b^2r_+^4}}\right)\nonumber\\
&&+\frac{8Q^2}{3r_+}\,{}_2F_1\!\left(\frac{1}{4},\frac{1}{2};\frac{5}{4};- \frac{Q^2}{b^2r_+^4}\right)\bigg).
\end{eqnarray}
Employing these thermodynamic quantities, one can check that the following first law and Smarr formula hold
\begin{eqnarray}
dM&=&TdS+VdP+\Phi dQ+Bdb,\\
M&=&2(TS-VP)+\Phi Q-Bb,
\end{eqnarray}
with the thermodynamic volume $V=\frac{4\pi r_{+}^3}{3}$. Identifying the specific volume $v = 2 r_+$, we can obtain the state equation from (\ref{eq:T})
\begin{equation}
\label{eq:state}
P =\frac{T}{v}-\frac {1}{2\pi\,{v}^{2}}-\frac{{b}^{2}}{4{\pi}} \Bigl( 1- \sqrt{1+\frac{16 {Q}^{2}}{{b}^{2}{v}^{4}}} \,\Bigr)\,.
\end{equation}
This state equation is found to have a VdW-like phase transition or a reentrant phase transition due to different values of the BI parameter $b$. The corresponding critical points can be obtained by solving
\begin{equation}
(\partial_v P)_{T}= 0,\quad (\partial^2_v P)_{T}= 0,
\end{equation}
and are given by \cite{Gunasekaran}
\begin{eqnarray}
T_{\rm c}&=&\frac{1-8xQ^2}{\pi v_{\rm c}},\nonumber\\
P_{\rm c}&=&\frac{1-16xQ^2}{2\pi v_{\rm c}^2}-\frac{b^2}{4\pi}\left(1-\frac{1}{v_{\rm c}^2x}\right),\nonumber\\
v_{\rm c}&=&\left( \frac{1}{x_{k}^2}-\frac{16 Q^2}{b^2} \right)^{\frac{1}{4}},
\end{eqnarray}
with
\begin{eqnarray}
x_k&=&2\sqrt{-\frac{p}{3}}\cos\left(\frac{1}{3}\arccos\left(\frac{3q}{2p}\sqrt{\frac{-3}{p}}\right)-\frac{2\pi k}{3}\right)\,, ~~~ k=0, 1, 2\,,\\
p&=&-\frac{3b^{2}}{32Q^{2}},\quad q=\frac{b^{2}}{256Q^{4}}.
\end{eqnarray}
For $x_{2}$, the value of the corresponding critical point is always a complex number. Therefore we have two critical points at most. Following Ref. \cite{Gunasekaran}, we can divide the parameter range of $b$ into four cases:\\

\par
\noindent \textbf{Case I:} $b<b_{0}$. This case is similar to the Schwarzschild-AdS black hole. A stable large black hole phase and an unstable small black hole phase are presented. However, there do not exist the VdW-like phase transition and the critical point.\\

\par
\noindent \textbf{Case II:} $b_{0}<b<b_{1}$. Two positive critical points $c_0$ and $c_1$ are presented in this case. However, $c_0$ has a higher Gibbs free energy, and thus it is global unstable. For this case, there exist a first-order phase transition and a `zeroth'-order phase transition, which is a typical reentrant phase transition.\\

\par
\noindent \textbf{Case III:} $b_{1}<b<b_{2}$. This case is similar to the case II, where a reentrant phase transition exists. The only difference is that the critical point $c_0$ has a negative pressure.\\

\par
\noindent \textbf{Case IV:} $b_{2}<b$. Only one critical point $c_1$ exists for this case. The phase transition for this case is the VdW type.

\begin{figure}
\centering
\subfigure[Isothermal curves]{\includegraphics[width=0.38\textwidth]{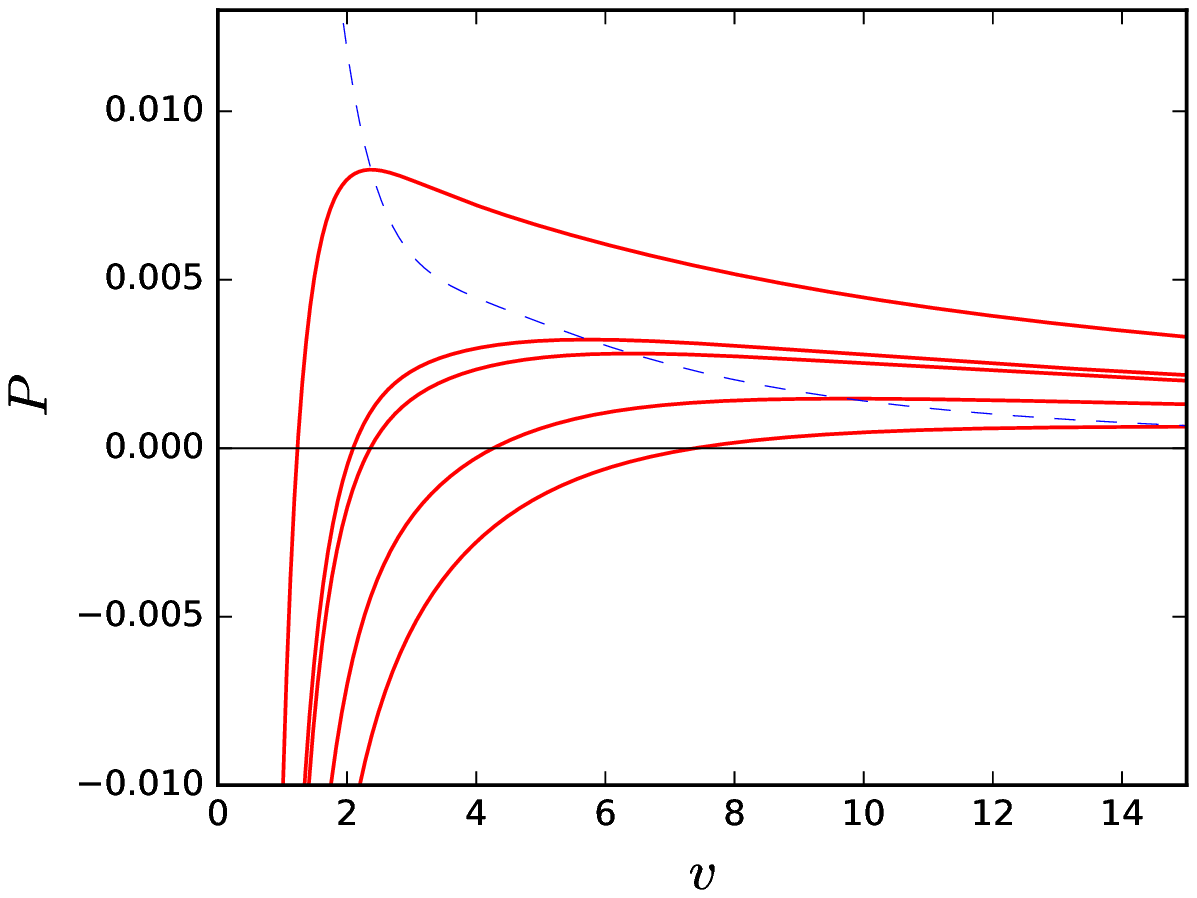}}
\subfigure[Isobaric curves]{\includegraphics[width=0.36\textwidth]{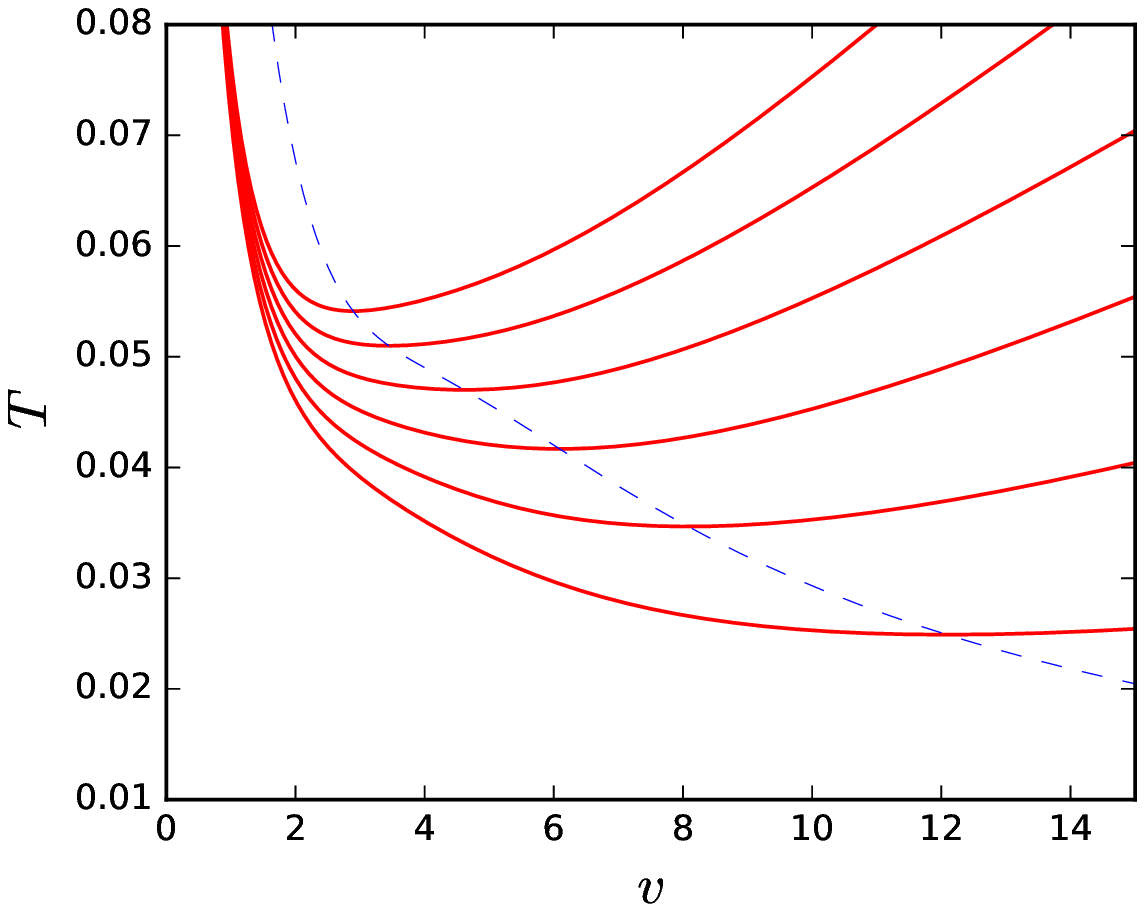}}
\caption{Case I: Isothermal and isobaric curves of the BI-AdS black holes for $b=0.3 < b_0$. (a) Isothermal curves with the temperature $T$=0.02, 0.03, 0.04, 0.043, and 0.06 from bottom to top. (b) Isobaric curves with the pressure $P$=0.001, 0.002,  0.003, 0.004, 0.005, and 0.006 from bottom to top. The dashed curves describe the extremal points for the isothermal and isobaric curves.}\label{figPvTv_b0}
\end{figure}

\begin{figure}
\centering
\subfigure[Isothermal curves] {\includegraphics[width=0.38\textwidth]{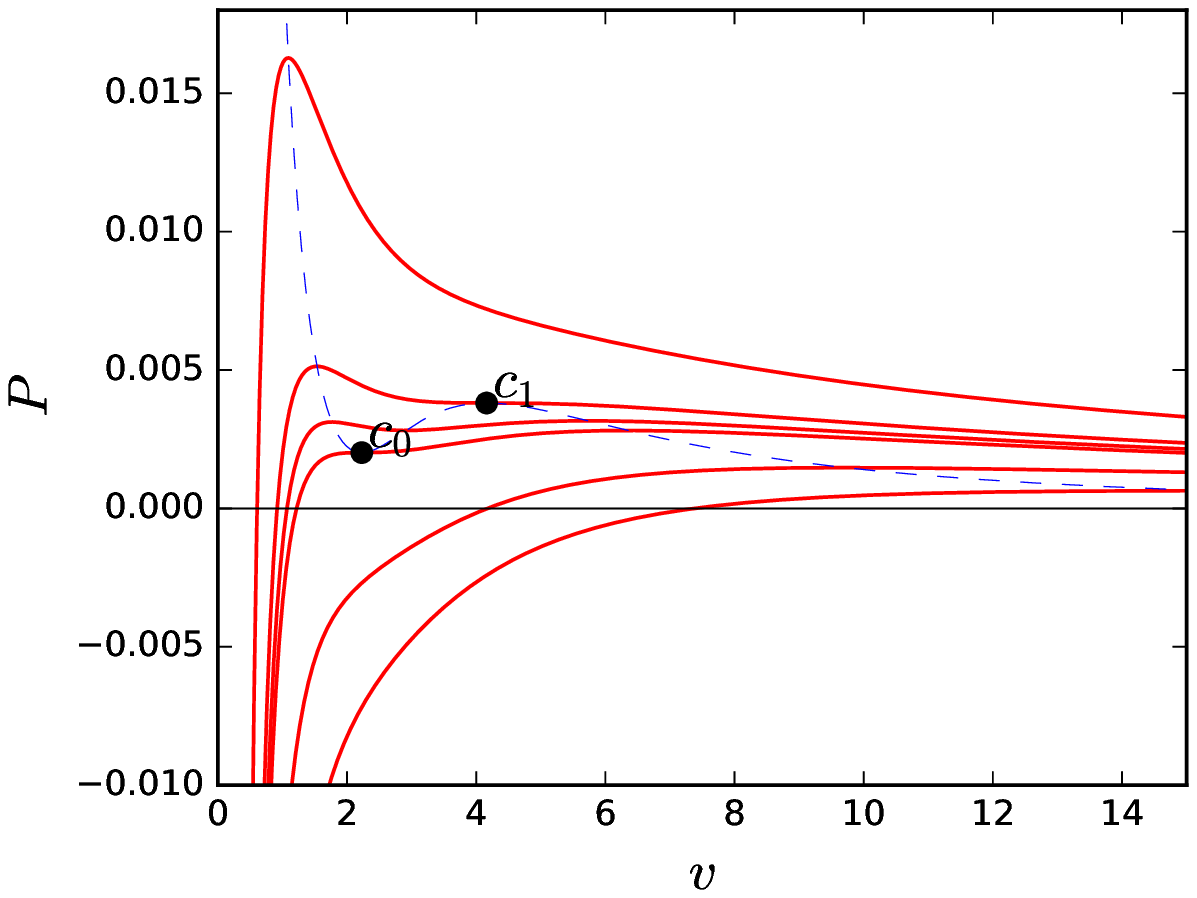}}
\subfigure[Isobaric curves] {\includegraphics[width=0.38\textwidth]{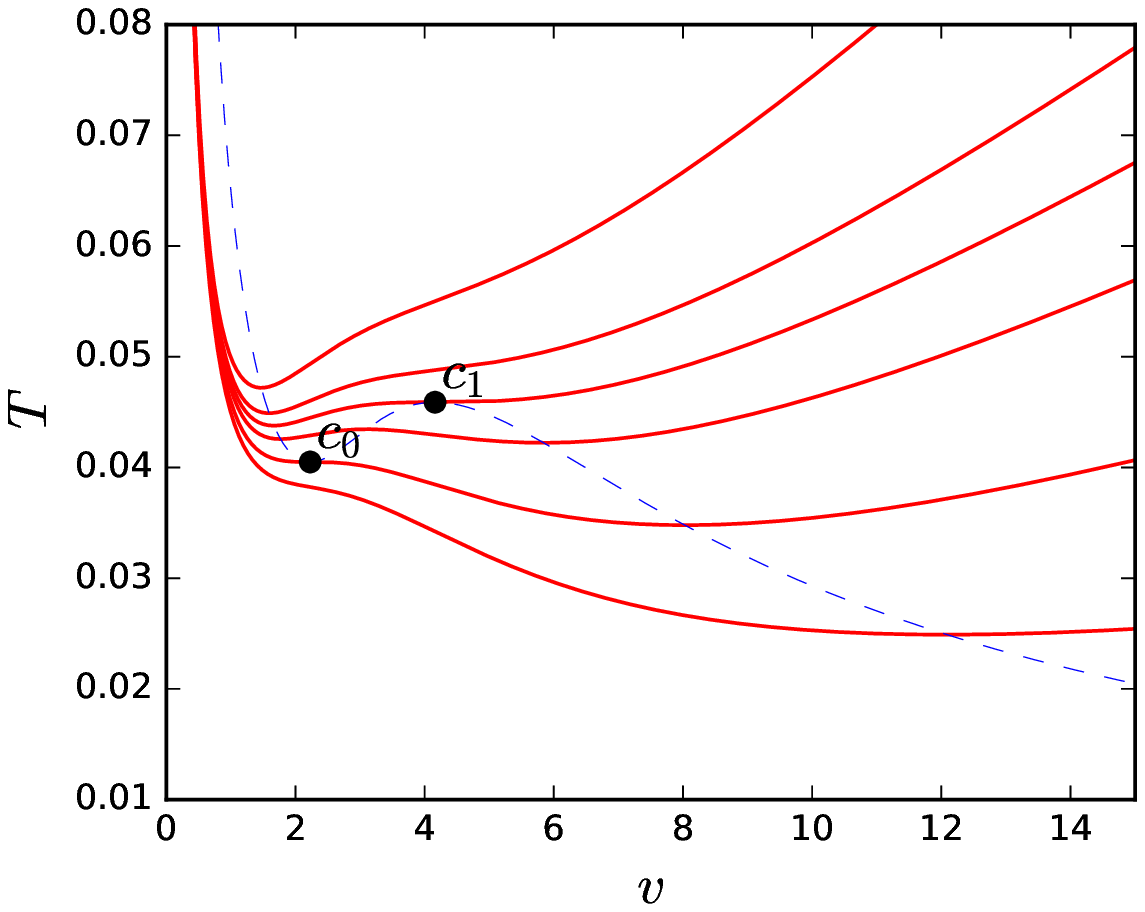}}
\caption{Case II: Isothermal and isobaric curves of the BI-AdS black holes for $b$=0.4 $\in$ ($b_0$, $b_1$). (a) Isothermal curves with the temperature $T$=0.02, 0.03, 0.040492 ($T_{\rm c0}$), 0.043, 0.045890 ($T_{\rm c1}$) and 0.06 from bottom to top. (b) Isobaric curves with the pressure $P$=0.001, 0.002016 ($P_{\rm c0}$),  0.003, 0.003807 ($P_{\rm c1}$), 0.0045 and 0.006 from bottom to top. Different from Fig. \ref{figPvTv_b0}, for some certain values of the pressure and the temperature, the number of the extremal points can be more than one. These two critical points are marked with the black dots, which both have positive pressure and temperature. However, the critical point locating at $(T_{\rm c0}, P_{\rm c0})$ has a higher Gibbs free energy, so it is not global stable one.}\label{figPvTv_b01}
\end{figure}

\begin{figure}
\centering
\subfigure[Isothermal curves] {\includegraphics[width=0.38\textwidth]{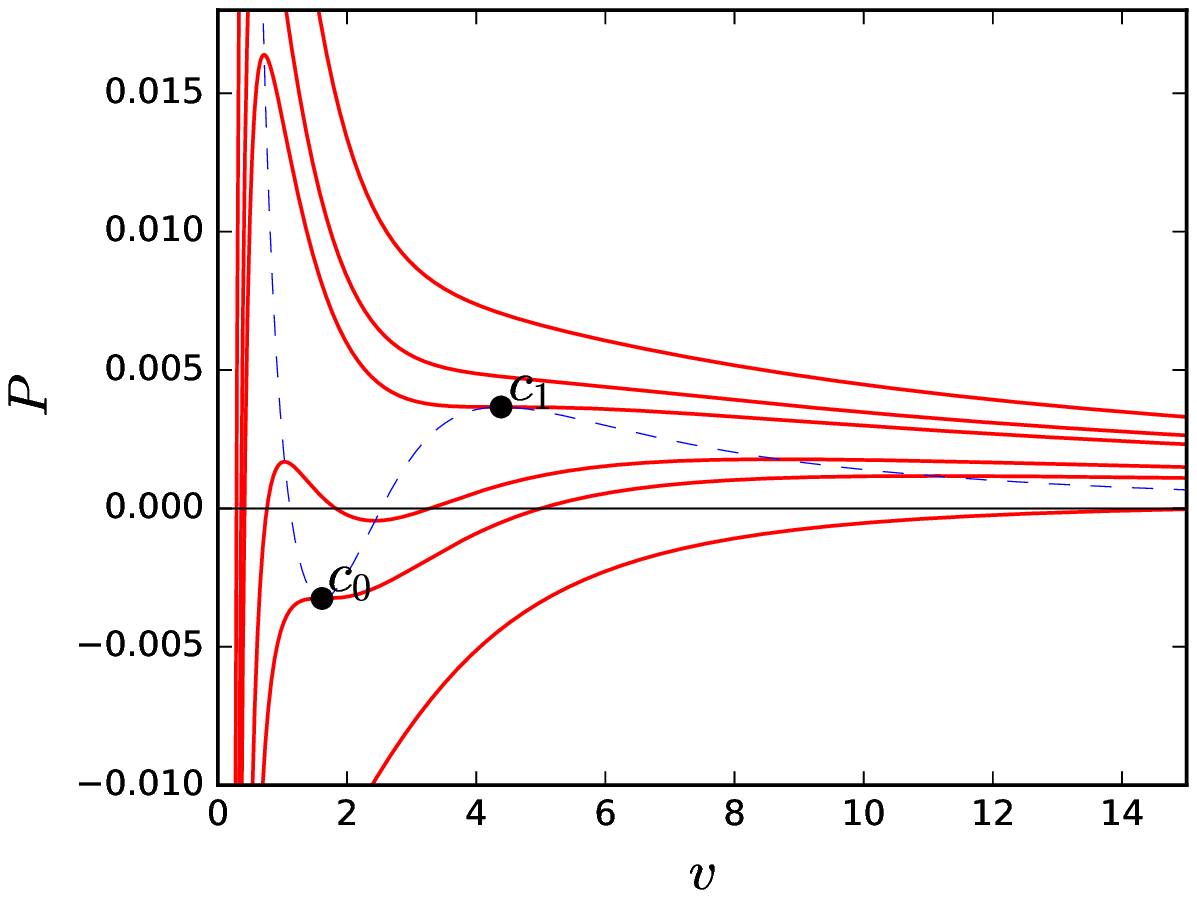}}
\subfigure[Isobaric curves] {\includegraphics[width=0.38\textwidth]{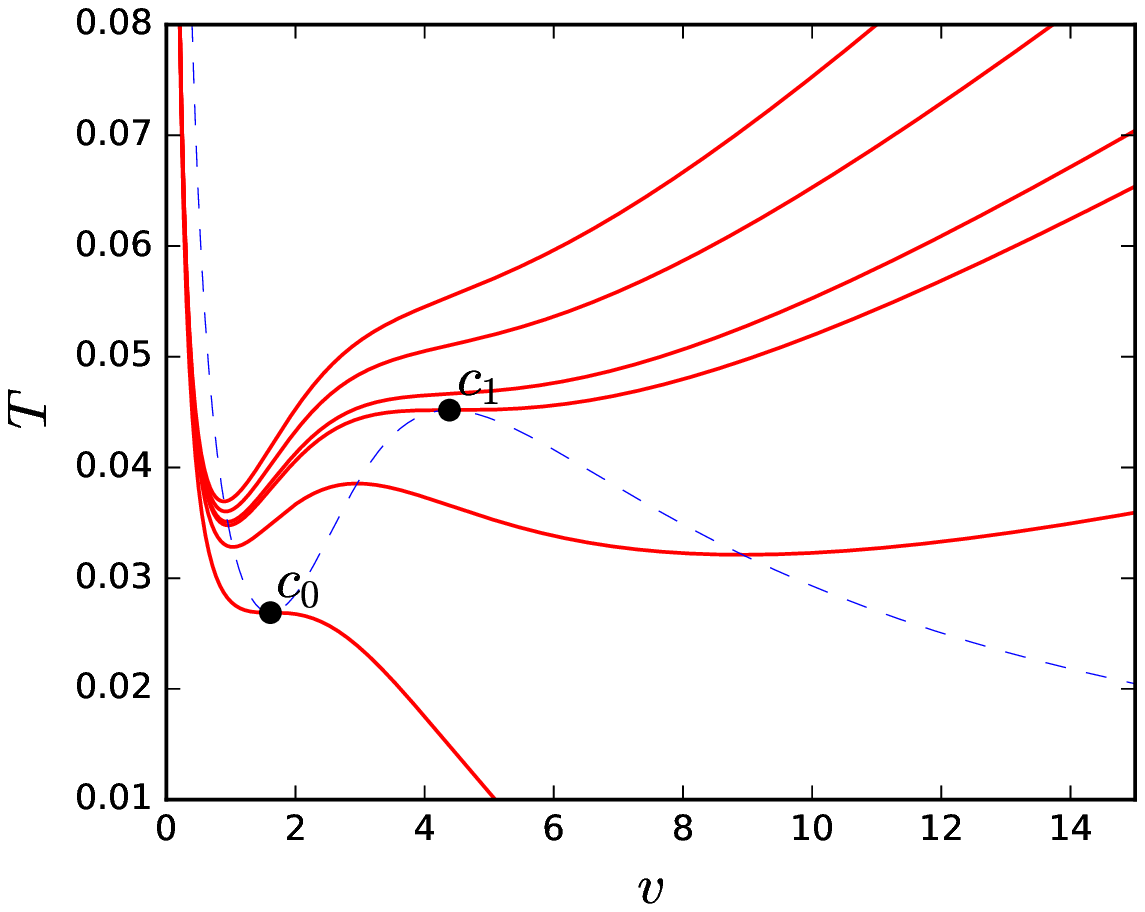}}
\caption{Case III: Isothermal and isobaric curves of the BI-AdS black holes for $b$=0.45 $\in$ ($b_1$, $b_2$). (a) Isothermal curves with the temperature $T$=0.01, 0.026885 ($T_{\rm c0}$), 0.03, 0.045170 ($T_{\rm c1}$), 0.05 and 0.06 from bottom to top. (b) Isobaric curves with the pressure $P$=-0.003253 ($P_{\rm c0}$), 0.001,0.002, 0.003, 0.003664 ($P_{\rm c1}$), and 0.006 from bottom to top. There are two critical points. The second one has a negative pressure while positive temperature, and thus it is unphysical.}\label{figPvTv_b12}
\end{figure}

\begin{figure}
\centering
\subfigure[Isothermal curves] {\label{fig:pv_b1}\includegraphics[width=0.38\textwidth]{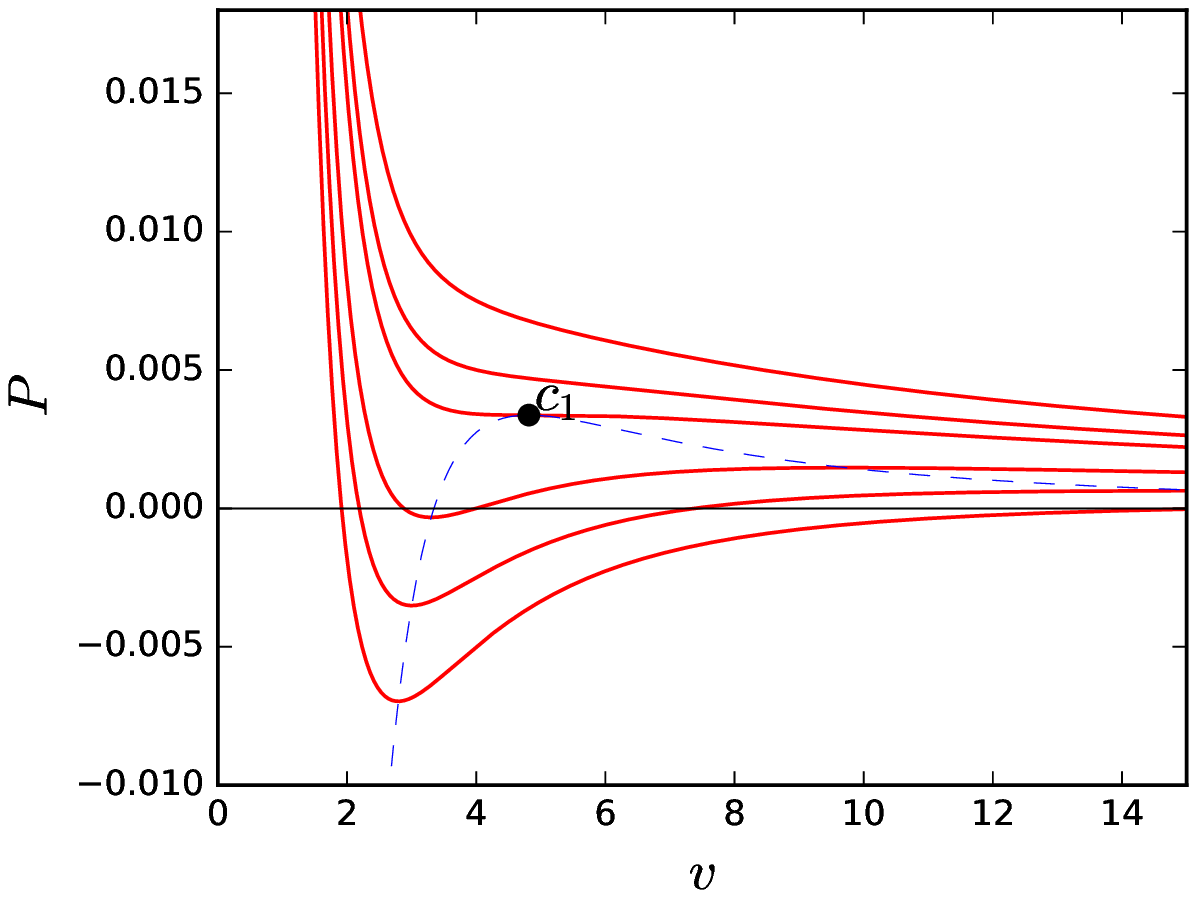}}
\subfigure[Isobaric curves] {\label{fig:tv_b1}\includegraphics[width=0.38\textwidth]{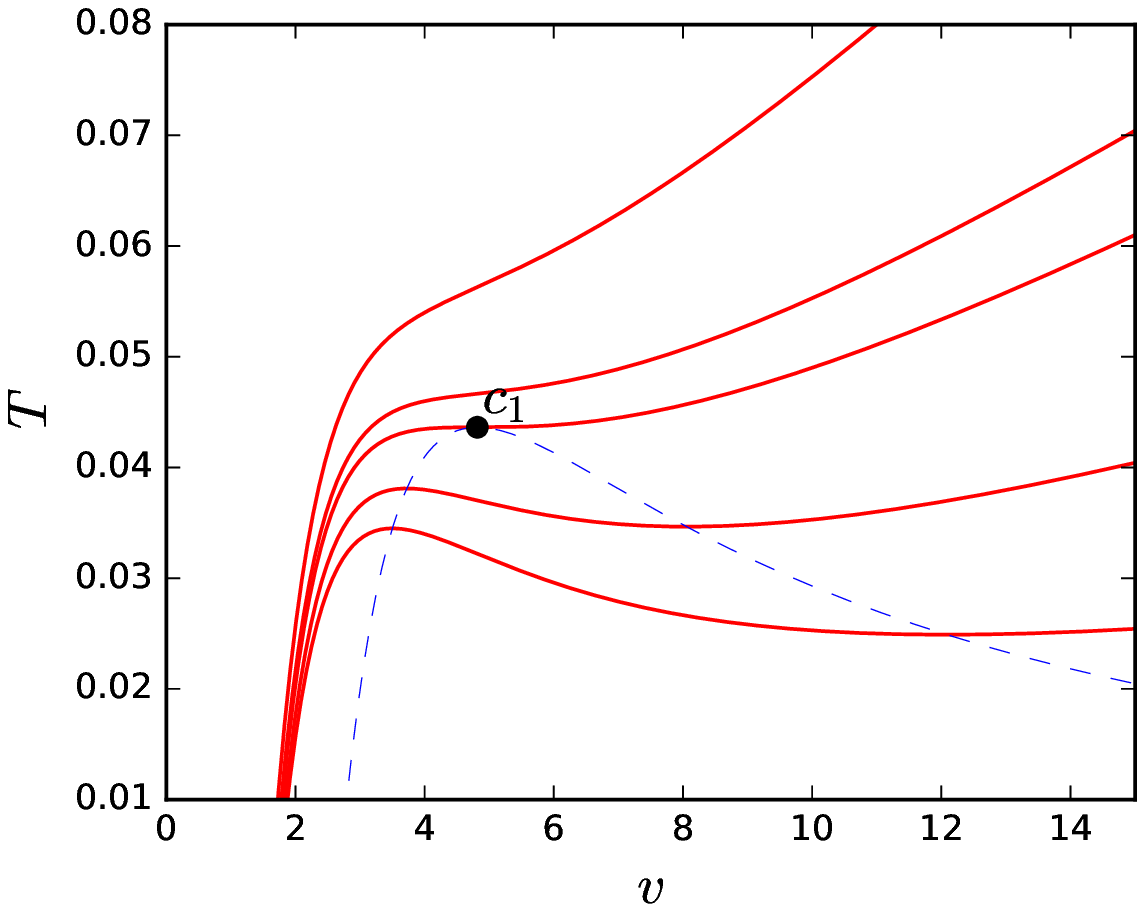}}
\caption{Case IV: Isothermal and isobaric curves of the BI-AdS black holes for $b=$1$>b_2$. (a) Isothermal curves with the pressure $P$=0.01, 0.02, 0.03, 0.043620 ($T_{\rm c1}$), 0.05 and 0.06 from bottom to top. (b) Isobaric curves with the temperature $T$=0.001, 0.002, 0.003372 ($P_{\rm c1}$), 0.004 and 0.006 from bottom to top. For this case, only one critical point is presented, and the typical VdW-like phase transition is displayed.}\label{figPvTv_b2}
\end{figure}

These values of the parameters $b_i$ are given by
\begin{equation}
b_0=\frac{1}{\sqrt{8}Q}\approx\frac{0.3536}{Q},\quad
b_1=\frac{\sqrt{3+2\sqrt{3}}}{6Q}\approx\frac{0.4237}{Q},\quad
b_2=\frac{1}{2Q}=\frac{0.5}{Q}.
\end{equation}
In the followings, we take $Q$=1 for simplicity. For these four cases, we plot the behaviors of the isothermal and isobaric curves in Figs. \ref{figPvTv_b0}-\ref{figPvTv_b2}. All the different properties can be found. In particular, the blue dashed lines are for the metastable curves, which are defined by
\begin{equation}
(\partial_v P)_{T} = 0, \quad\text{or} \quad (\partial_v T)_{P} = 0.
\end{equation}
From these figures, one finds that the metastable curves have no extremal point for $b<b_{0}$, two extremal points for $b\in(b_{0}, b_{2})$, and one point for $b>b_{2}$. Actually, the extremal points are exactly the critical point of the thermodynamic phase transition. In summary, we can find the reentrant phase transition for $b\in(b_{0}, \;b_{2})$, and the VdW-like phase transition for $b>b_{2}$.

%%%%%%%%%%%%%%%
%%%%%%%%%%%%%

%%%%%%%%%%%%%%
%%%%%%%%%%%%%%%%%%

%%%%%%%%%%%%%%%%%%%
%%%%%%%%%%%%%%%%

%%%%%%%%%%%%%%%%%%%
%%%%%%%%%%%%%%%%%%

%%%%%%%%%%%%%%%%%%%
\begin{figure}
\centering
\subfigure[] {\label{fig:reentrant}\includegraphics[width=0.38\textwidth]{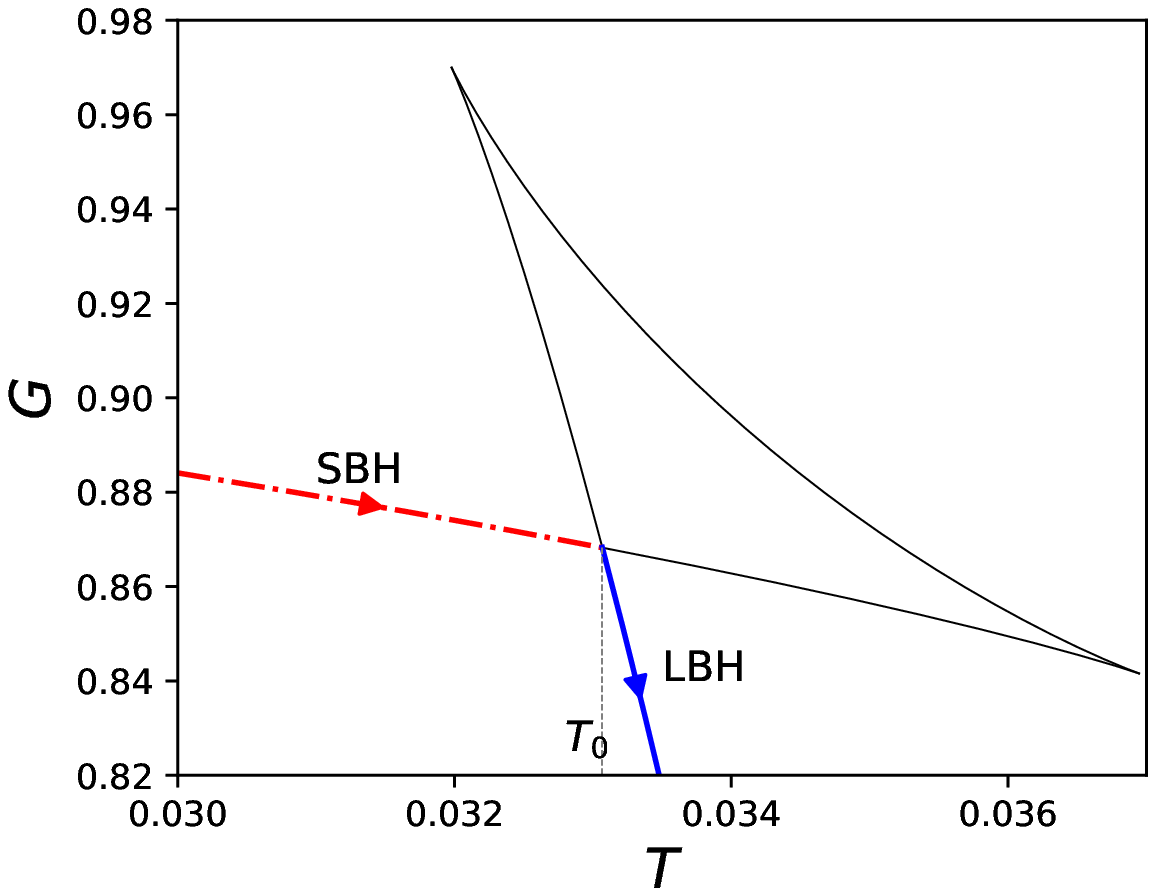}}
\subfigure[] {\label{fig:reentrantNO}\includegraphics[width=0.38\textwidth]{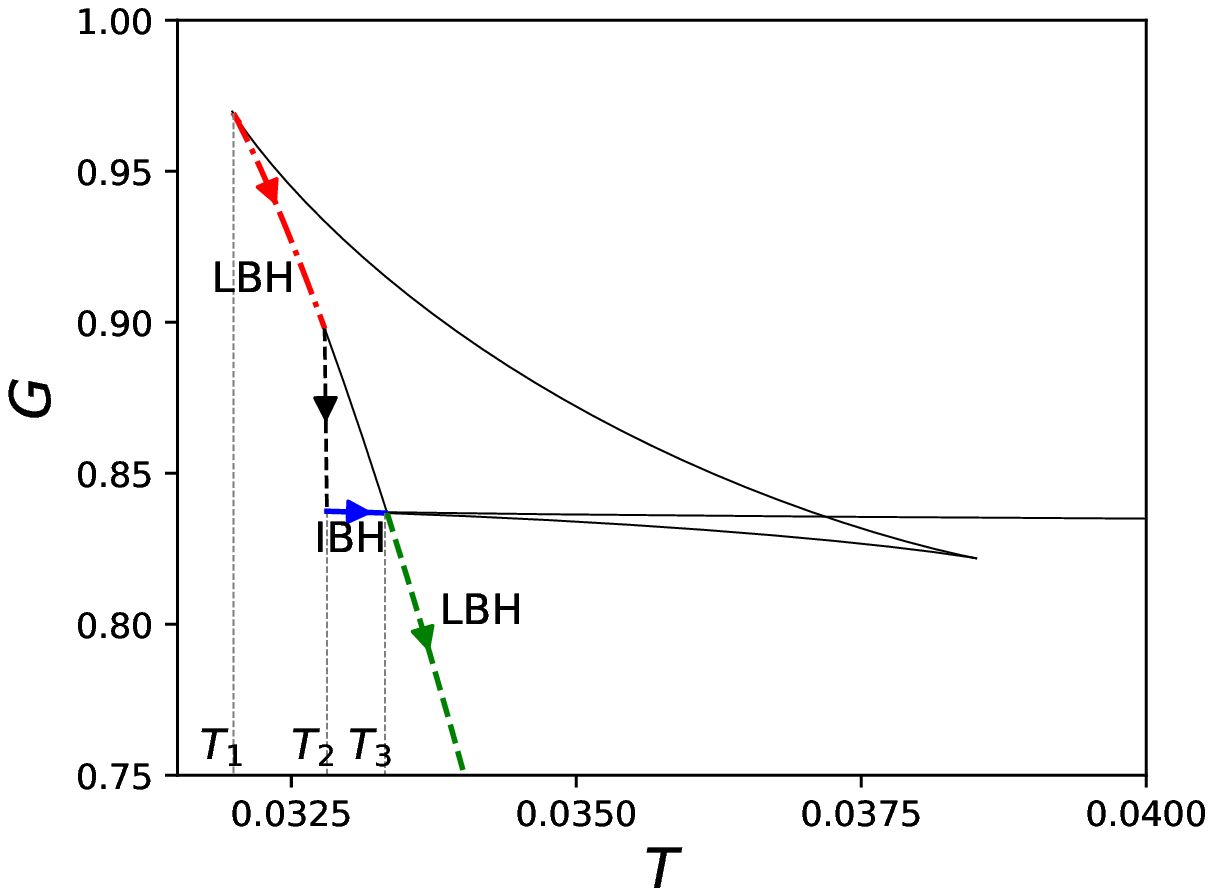}}
\caption{The behavior of the Gibbs free energy. (a) VdW-like phase transiton with $P=0.001685$ and $b=1$. (b) Reentrant phase transition with $P=0.001685$ and $b=0.45$. The temperatures are $T_0=0.033067$, $T_1 =0.031992$, $T_2=0.032812$, and $T_3=0.033322$.}\label{figreentrantNO}
\end{figure}
%%%%%%%%%%%%%%%%%%

%%%%%%%%%%%%%%%%%%%%
\begin{figure}
\centering
\subfigure[$P$-$T$ phase diagram]{\includegraphics[width=0.38\textwidth]{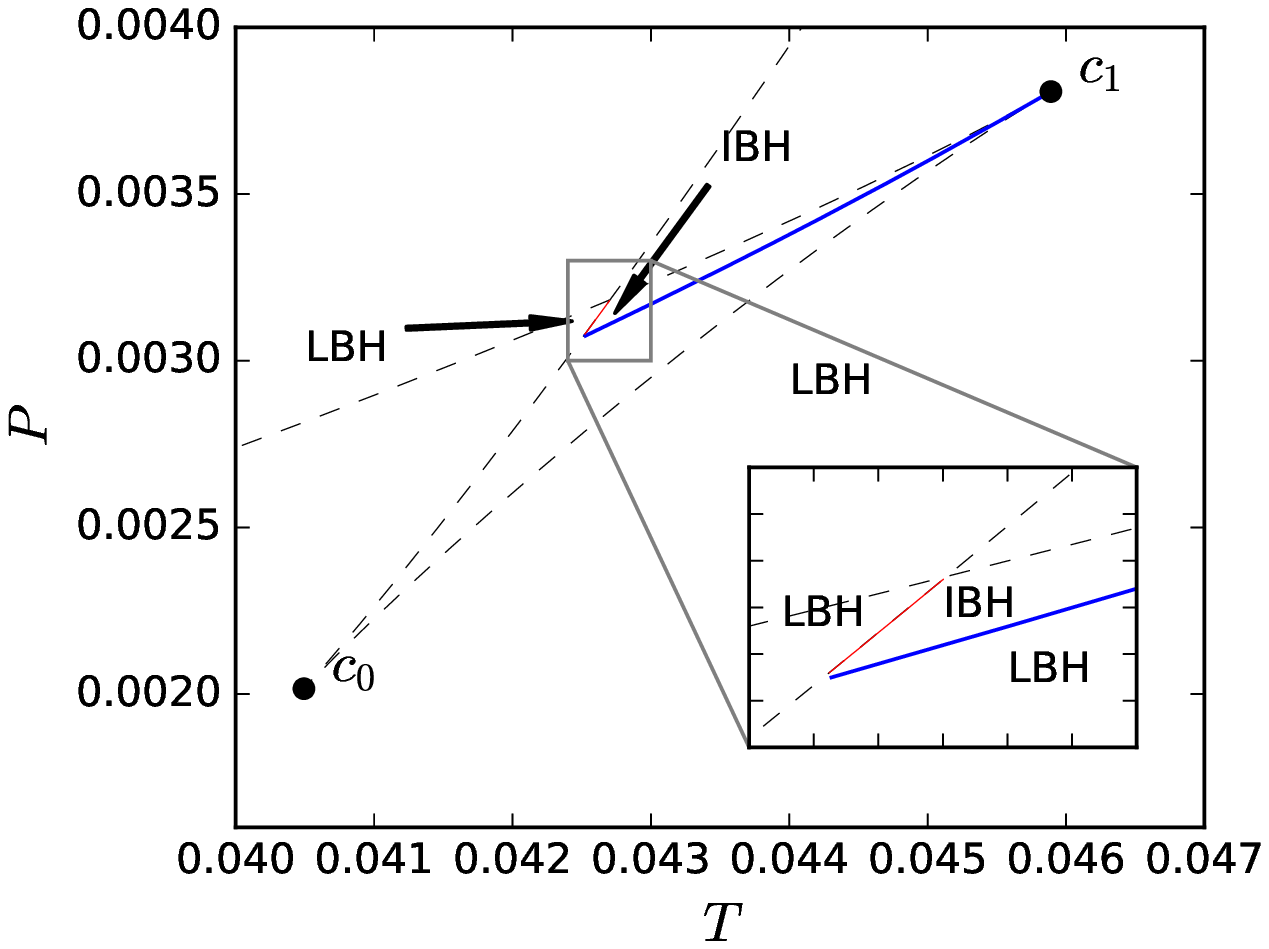}\label{pphaa1}}
\subfigure[$T$-$v$ phase diagram] {\includegraphics[width=0.38\textwidth]{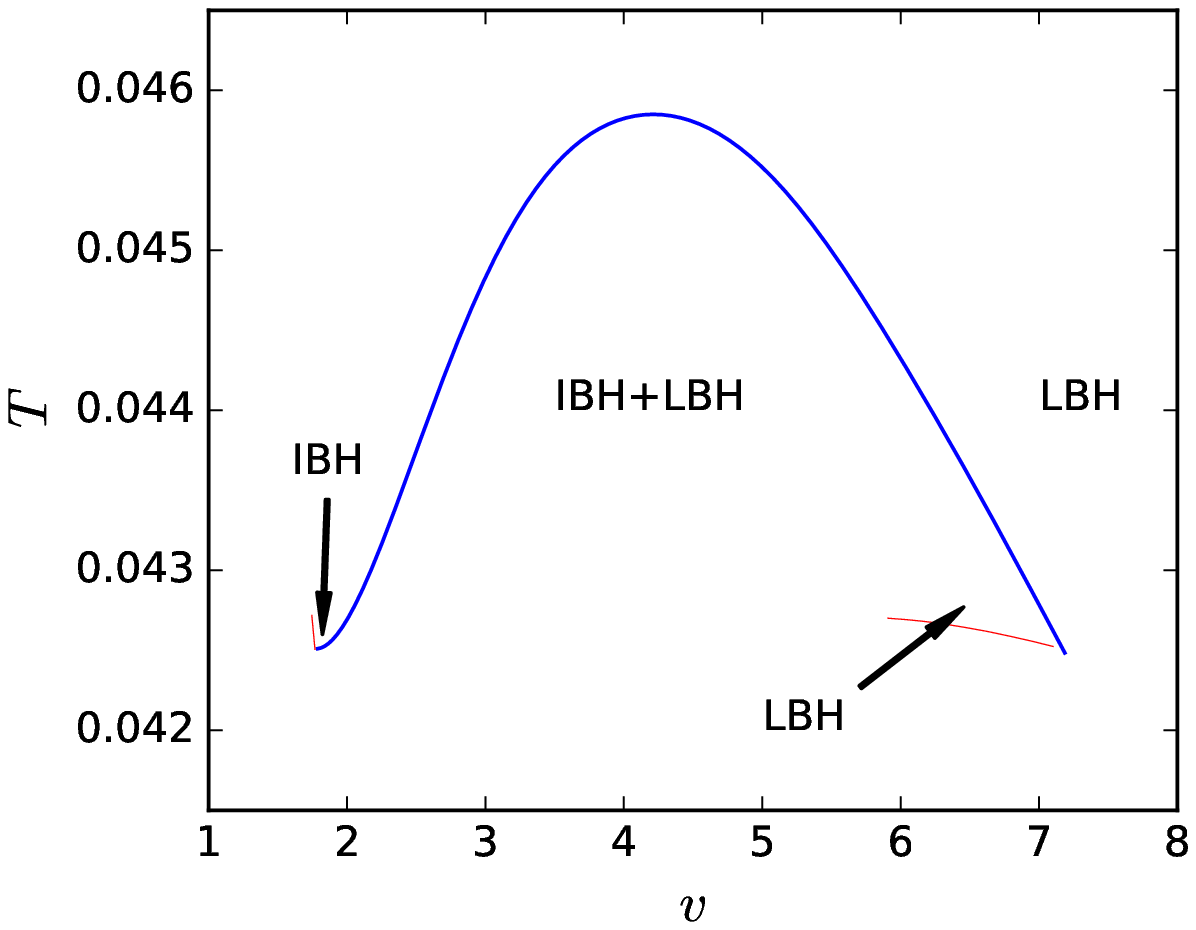}\label{ppha2}}
\caption{Phase structures of the BI-AdS black holes for $b=$0.4$\in (b_0, b_1)$. (a) $P$-$T$ phase diagram. (b) $T$-$v$ phase diagram. The blue solid curves are for the coexistence curves. The red thin solid curves are the `zeroth'-order phase transition curves. The dot dashed curves are the extremal point curves of the isothermal and isobaric curves of the BI-AdS black holes. Black dots denote the critical points.}\label{ppha}
\end{figure}
%%%%%%%%%%%%%%%%%%%%%%

%%%%%%%%%%%%%%%%%%
\begin{figure}
\centering
\subfigure[$P$-$T$ phase diagram]{\label{pphb1}\includegraphics[width=0.38\textwidth]{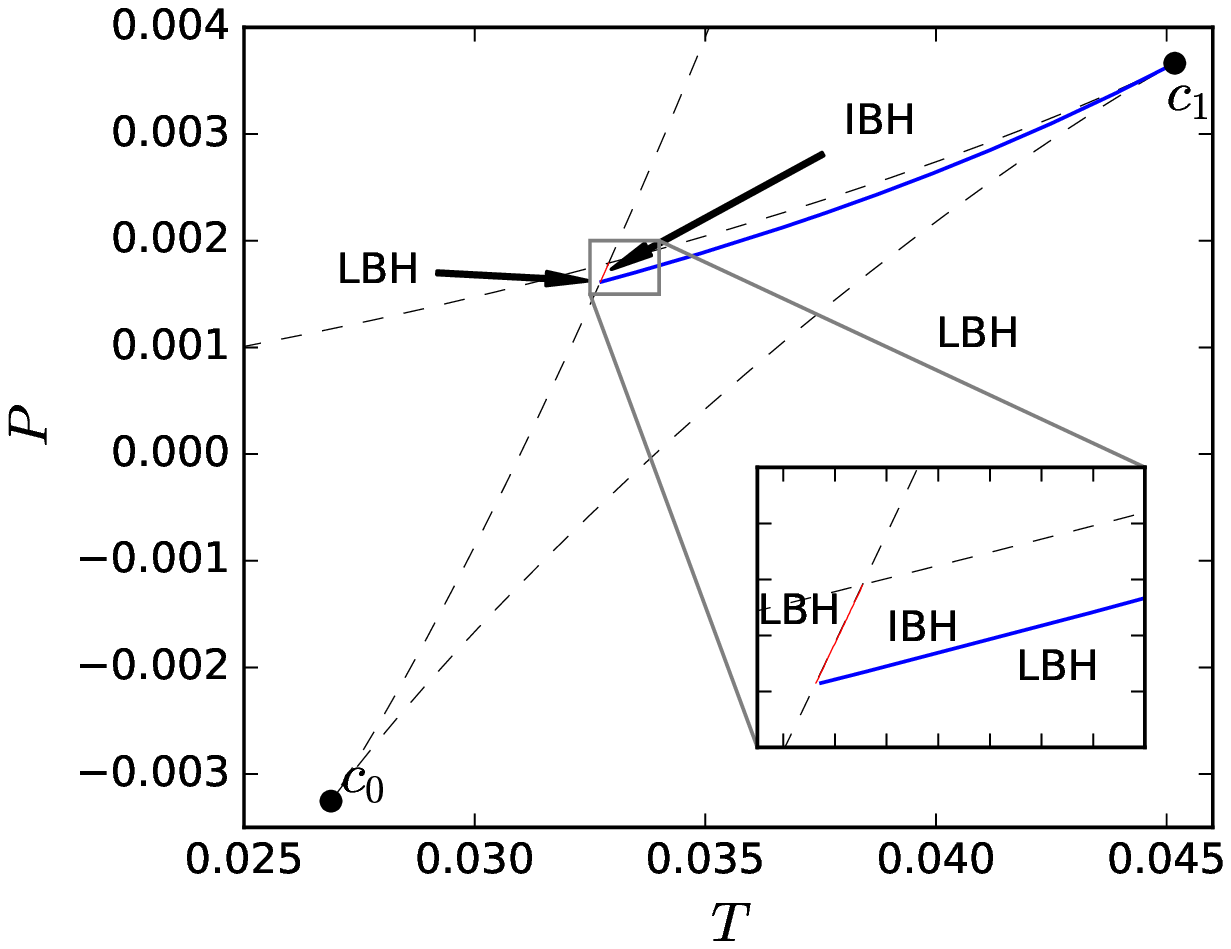}}
\subfigure[$T$-$v$ phase diagram]{\label{pphb2}\includegraphics[width=0.38\textwidth]{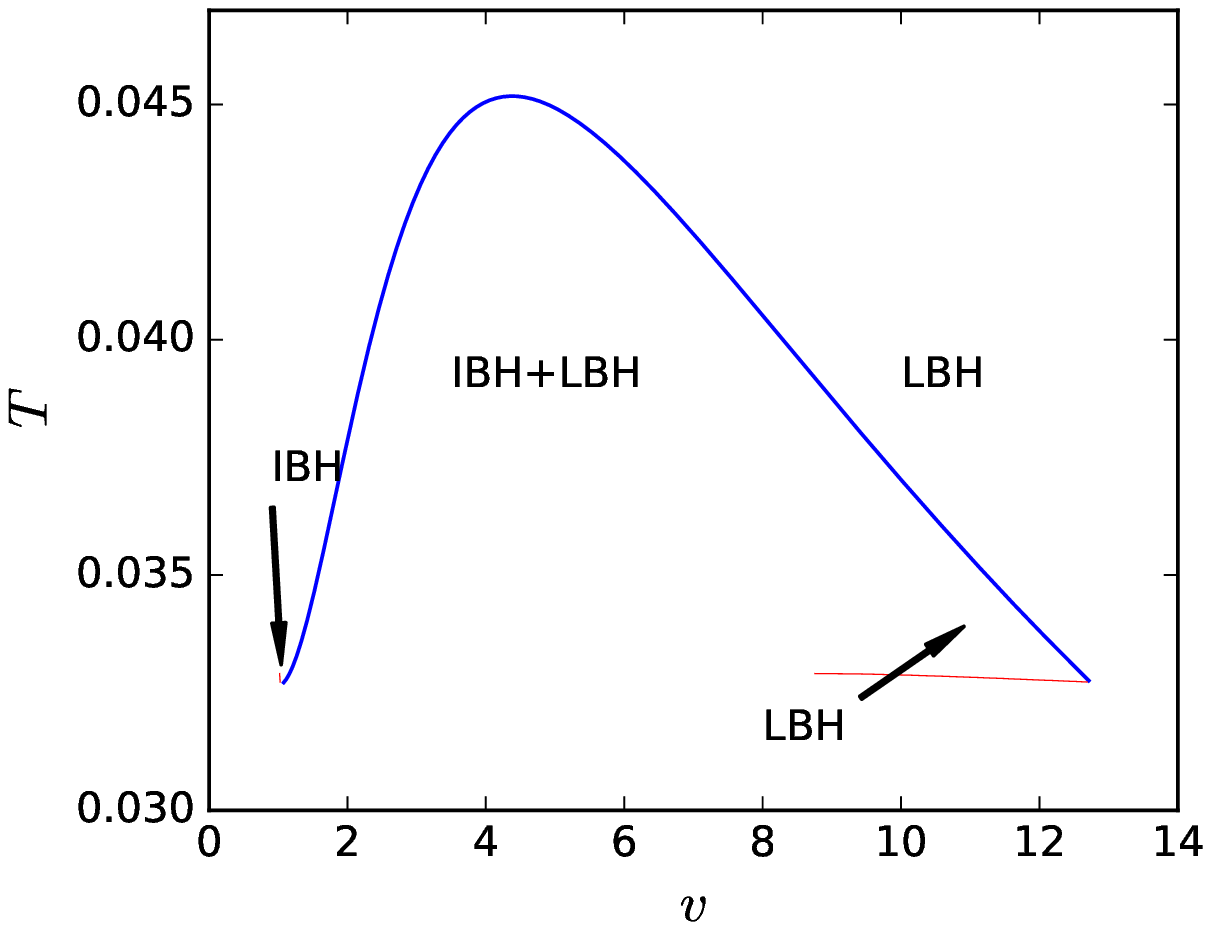}}
\caption{Phase structures of the BI-AdS black holes for $b=$0.45$\in (b_1,b_2)$. (a) $P$-$T$ phase diagram. (b) $T$-$v$ phase diagram. The blue solid curves are for the coexistence curves. The red thin solid curves are the `zeroth'-order phase transition curves. The dot dashed curves are the extremal point curves of the isothermal and isobaric curves of the BI-AdS black holes. Black dots denote the critical points.}\label{pphb}
\end{figure}
%%%%%%%%%%%%%%%%%%%

%%%%%%%%%%%%%%%%%%
\begin{figure}
\centering
\subfigure[$P$-$T$ diagram] {\label{pphc11}\includegraphics[width=0.38\textwidth]{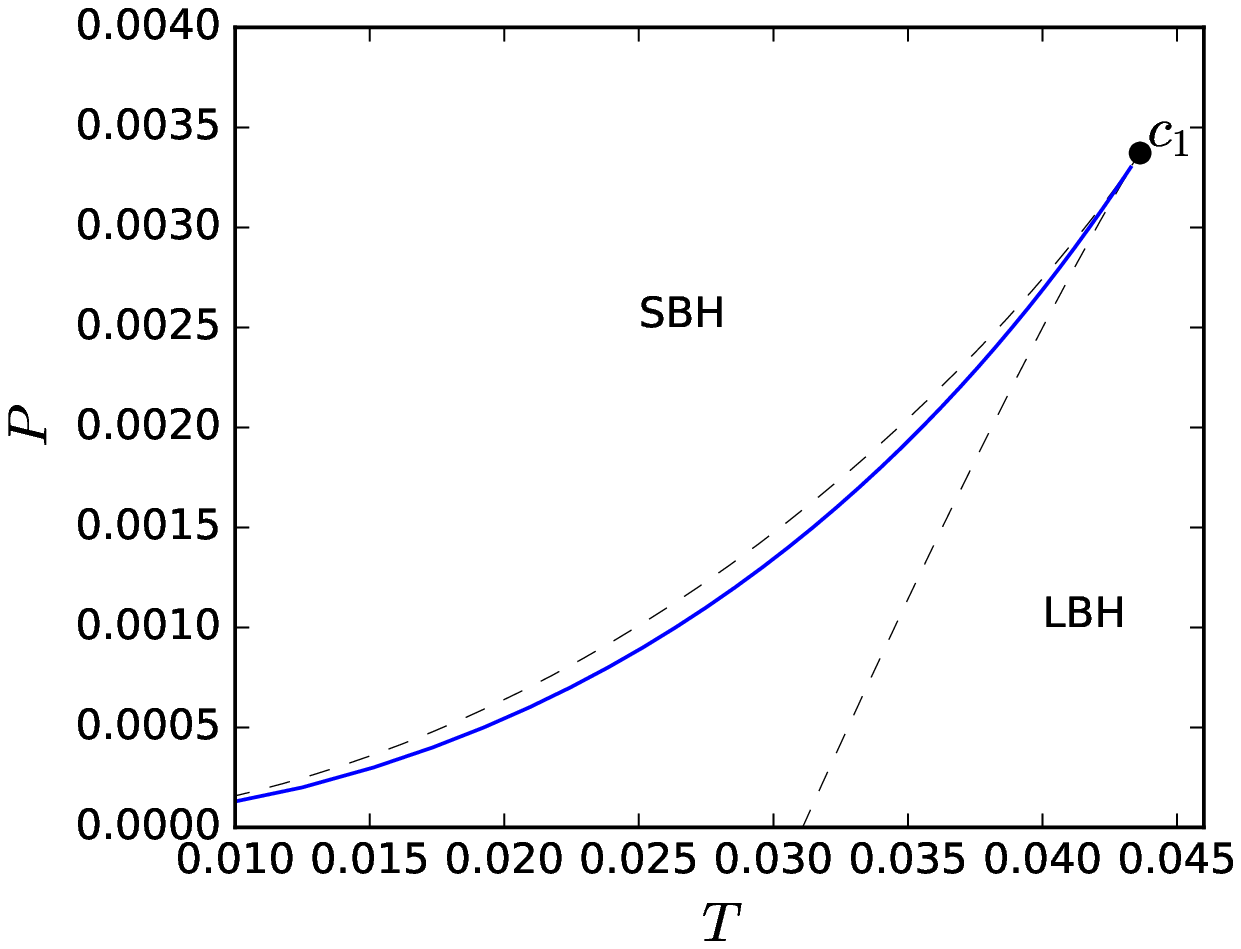}}
\subfigure[$T$-$v$ diagram] {\label{pphc22}\includegraphics[width=0.38\textwidth]{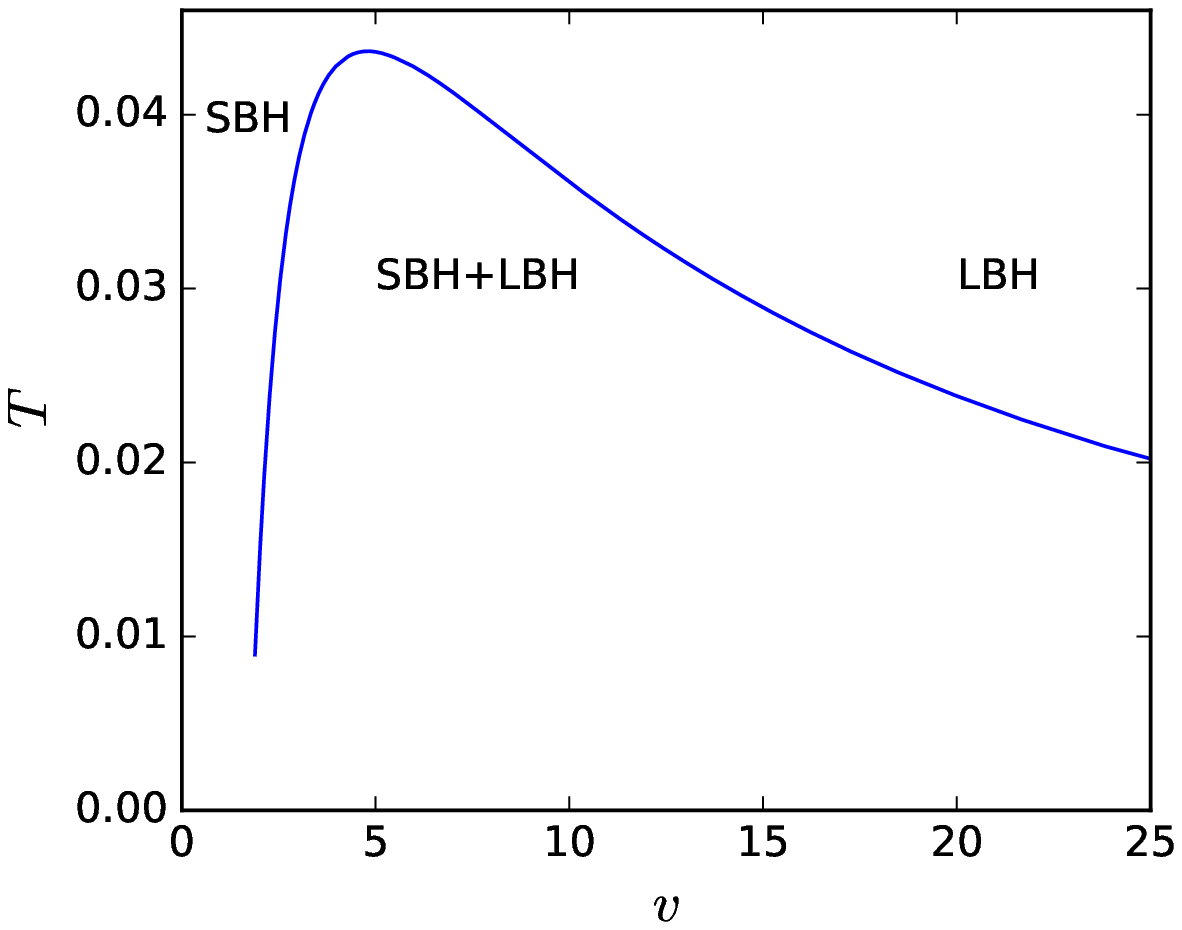}}
\caption{Phase structures of the BI-AdS black holes for $b=$1$> b_2$ in the $P$-$v$ and $T$-$v$ diagrams.}\label{pphc}
\end{figure}
%%%%%%%%%%%%%%%%

The new type phase transition, the reentrant phase transition, can be well understood from the Gibbs free energy. Here we show the the VdW phase transition and the reentrant phase transition in Fig. \ref{fig:reentrant} and \ref{fig:reentrantNO}, respectively. From Fig. \ref{fig:reentrant}, we can find that there exists a typical swallow tail behavior, which indicates a small-large black hole phase transition at $T_0=0.033067$. Actually, with the increase of the temperature, the system prefers the small black hole phase for it has low Gibbs free energy. Until the temperature $T_0$ is approached, the large black hole will have the low Gibbs free energy, so the system will prefer the large black hole phase. Thus at $T_0$, the small-large black hole phase transition takes place, which is similar to the liquid-gas phase transition of VdW fluid. The case of the reentrant phase transition becomes complicated. Taking $P=0.001685$ and $b=0.45$, we show the behavior of Gibbs free energy in Fig. \ref{fig:reentrantNO}. When $T<T_1=0.031992$, there is no black hole phase. With the increase of the temperature and according to the fact that a thermodynamic system always prefers a low Gibbs free energy phase, the spacetime will take a large black hole phase when $T_1<T<T_2$. At $T_2$, there occurs a phase transition between the large black hole and the intermediate black hole, which is `zeroth'-order. With further increasing the temperature, another first-order phase transition between the intermediate black hole and the large black hole takes place at $T_3$. This behavior of the Gibbs free energy admits a typical reentrant phase transition.

Next, we would like to study the phase structure for the BI-AdS black holes. The results are described in Figs. \ref{ppha}-\ref{pphc} for $b>b_{0}$. While for $b<b_0$, only a stable large black hole phase exists, and thus no phase transition occurs at constant charge.

As mentioned above, for $b\in$ ($b_0$, $b_1$), there exists a reentrant phase transition, which is different from the VdW-like phase transition. Taking $b=$0.4 as an example, we plot the phase structure in Fig. \ref{ppha}. In Fig. \ref{pphaa1}, we show the phase structure in the $P$-$T$ diagram, which is a typical reentrant phase diagram. The blue solid line denotes the coexistence curve of the intermediate and large black holes, which is a first-order phase transition. The short red dashed curve describes the `zero'-order phase transition between the large black hole and the intermediate black hole. It is worthwhile noting that this curve is not the coexistence curve of the large and intermediate black holes. It belongs to the intermediate black hole phase because at that point, the intermediate black hole has lower Gibbs free energy than the large black hole. The dot dashed curves are for the extremal points of the isothermal and isobaric curves of the BI-AdS black holes. The intersection points of these extremal point curves marked with two black dots are the two critical points. It is clear that these critical points have positive temperature. On the other hand, the upper left corners bounded by these two extremal point curves are the regions without any black holes.

In order to clearly show the coexistence region of the intermediate and large black holes, we plot the phase structure in the $T$-$v$ diagram, see Fig. \ref{ppha2}. Below the blue solid curve, it is the region of the coexistence intermediate and large black holes. The intermediate black hole region is on the left, and the large one is on the right. Since the small black holes are thermodynamic unstable, we do not show them. Of particular interest is that there also has a region for the large black hole, which locates in the lower right corner of the coexistence region. Although it overlaps with the coexistence region, it is a large black hole region. The reason for this is that we have degeneracy in the $T$-$v$ diagram. So if the phase structure is displayed in other diagram, such degeneracy could be eliminated.

For the BI parameter $b\in$ ($b_1$, $b_2$), there is also a reentrant phase transition. We take $b=$0.45, and plot the phase structures in the $P$-$T$ and $T$-$v$ diagrams in Fig. \ref{pphb}. These phase diagrams are quite similar to the case $b=$0.4. The tiny difference is that one of the critical points has negative pressure, see Fig. \ref{pphb1}, where the critical points are marked with the black dots.

When $b>b_{2}$, the reentrant phase transition disappears, and only the standard VdW-like phase transition is left. The phase diagram is described in Fig. \ref{pphc} for $b=$1. The coexistence curves are plotted in blue solid curve, and the extremal point curves are in dot dashed curves. In Fig. \ref{pphc11}, the phase diagram is described in the $P$-$T$ diagram. Above the coexistence curve is the small black hole phase region and below it is the large one. The $T$-$v$ phase diagram is shown in Fig. \ref{pphc22}. The coexistence phase of small and large black holes is below the coexistence curve. The small black hole phase locates at the left and the large black hole phase locates at the right.

In summary, VdW-like or reentrant phase transition can be found in different parameter region of $b$. The $T$-$v$ phase diagrams are quite different for these two type phase transitions.

\section{Null geodesics and phase transition}\label{sec:null_geodesic_and_critical}

In this section, we would like to study the null geodesics, especially the photon sphere, of the BI-AdS black holes. Then we will examine the relation between the parameter properties of the photon sphere and the thermodynamic phase transition.

\subsection{Null geodesics and photon sphere}
\label{subsec:null_geodesic}

Now, we consider the motion of a free photon orbiting around a BI-AdS black hole. Since the black hole is spherically symmetrical, we fix $\theta=\frac{\pi}{2}$ without loss of generality. Then the Lagrangian of a free photon in the background of a BI-AdS black hole reads
\begin{equation}
\mathcal{L}=\frac{1}{2}g_{\mu\nu}\dot{x}^{\mu}\dot{x}^{\nu}=\frac{1}{2}(-f(r)\dot{t}^{2}+\dot{r}^{2}/f(r)+r^{2}\dot{\phi}^{2}).
\label{lag}
\end{equation}
The dot over the symbol denotes the ordinary differentiation with respect to an affine parameter. Using the Lagrangian (\ref{lag}), the generalized momentum can be calculated by
\begin{equation}
p_{\mu}=\frac{\partial\mathcal{L}}{\partial\dot{x}^{\mu}}
=g_{\mu\nu}\dot{x}^{\nu}.
\end{equation}
For this spacetime, it has two Killing vectors $\partial_{t}$ and $\partial_{\phi}$. Thus for each geodesics, there are two constants. One is the energy $E$ and another is the orbital angular momentum $L$ of the photon. Considering them, the generalized momenta can be expressed as
\begin{equation}
p_{t}=-f(r)\dot{t}=-E,\quad p_{r}=\frac{\dot{r}}{f(r)},\quad p_{\phi}=r^{2}\dot{\phi}=L.
\end{equation}
For a photon, it is required to satisfy $g_{\mu\nu}\dot{x^{\mu}}\dot{x^{\nu}}=0$. Thus, the radial $r$-motion is
\begin{equation}
\dot{r}^{2}+V_{\rm eff}=0,
\end{equation}
where the effective potential is given by
\begin{eqnarray}
V_{\rm eff}=\frac{L^{2}}{r^{2}}f(r)-E^{2}.\label{eff}
\end{eqnarray}
Using the effective potential (\ref{eff}). The unstable photon sphere can be obtained by solving
\begin{eqnarray}
V_{\rm eff}=0,\quad \partial_{r}V_{\rm eff}=0,\quad \partial_{r,r}V_{\rm eff}<0.
\end{eqnarray}
The first equation gives the minimum impact parameter
\begin{eqnarray}
u_{\rm ps}=\frac{L}{E}\bigg|_{r_{\rm ps}}=\frac{r}{\sqrt{f(r)}}\bigg|_{r_{\rm ps}}.\label{upss}
\end{eqnarray}
Then using the second equation, one can find that the radius of the photon sphere is determined by the following equation
\begin{eqnarray}
2f(r_{\rm ps})-r_{\rm ps}\partial_{r}f(r_{\rm ps})=0.\label{rpss}
\end{eqnarray}
Plugging the metric function (\ref{metricf}) into this equation, we will obtain
\begin{eqnarray}
r_{\rm ps}^{2}-3Mr_{\rm ps}+2Q^{2}{}_2 F_1\left(\frac{1}{4},\frac{1}{2}; \frac{5}{4};-\frac{Q^2}{b^2 r_{\rm ps}^4}\right)=0.\label{spherps}
\end{eqnarray}
For $b=0$, one gets $r_{\rm ps}=\frac{1}{2}(3M+\sqrt{9M^{2}-8Q^{2}})$, which is the radius of the photon sphere for the Reissner-Nordstr\"om AdS black hole. However, when $b\neq0$, there is no analytic result.

\subsection{Phase transition and photon sphere}

For given $b$ and $Q$, we can solve the radius of the photon sphere from (\ref{spherps}). Substituting the radius into (\ref{upss}), the minimum impact parameter will be obtained. In order to examine the relation between the photon sphere and the phase transition, we would like to study the behavior of the radius and the minimum impact parameter along the isothermal and isobaric curves of the BI-AdS black holes.

\begin{figure*}[!htbp]
\centering
\subfigure[] {\label{fig:t:b_b0}\includegraphics[width=0.38\textwidth]{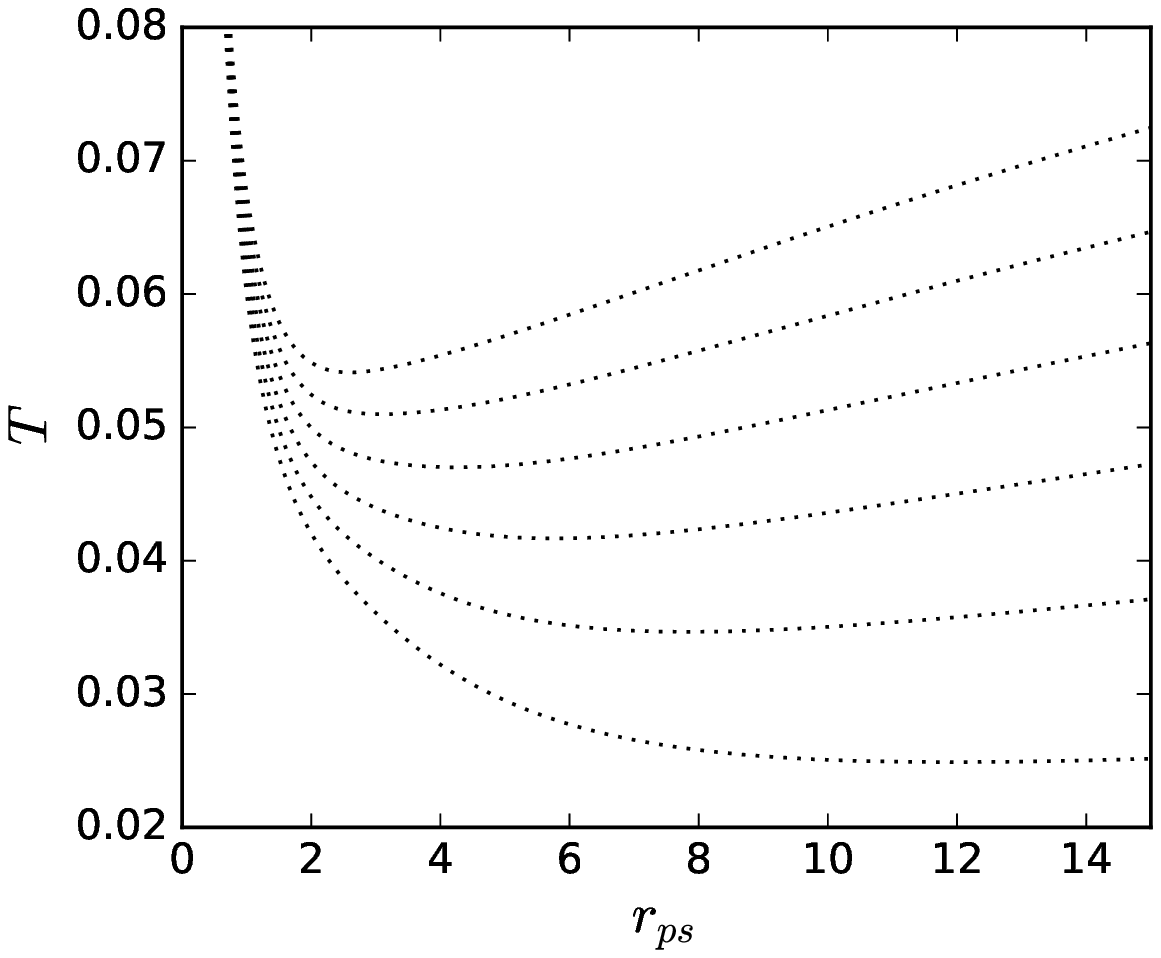}}
\subfigure[] {\label{fig:t:b0_b_b1}\includegraphics[width=0.38\textwidth]{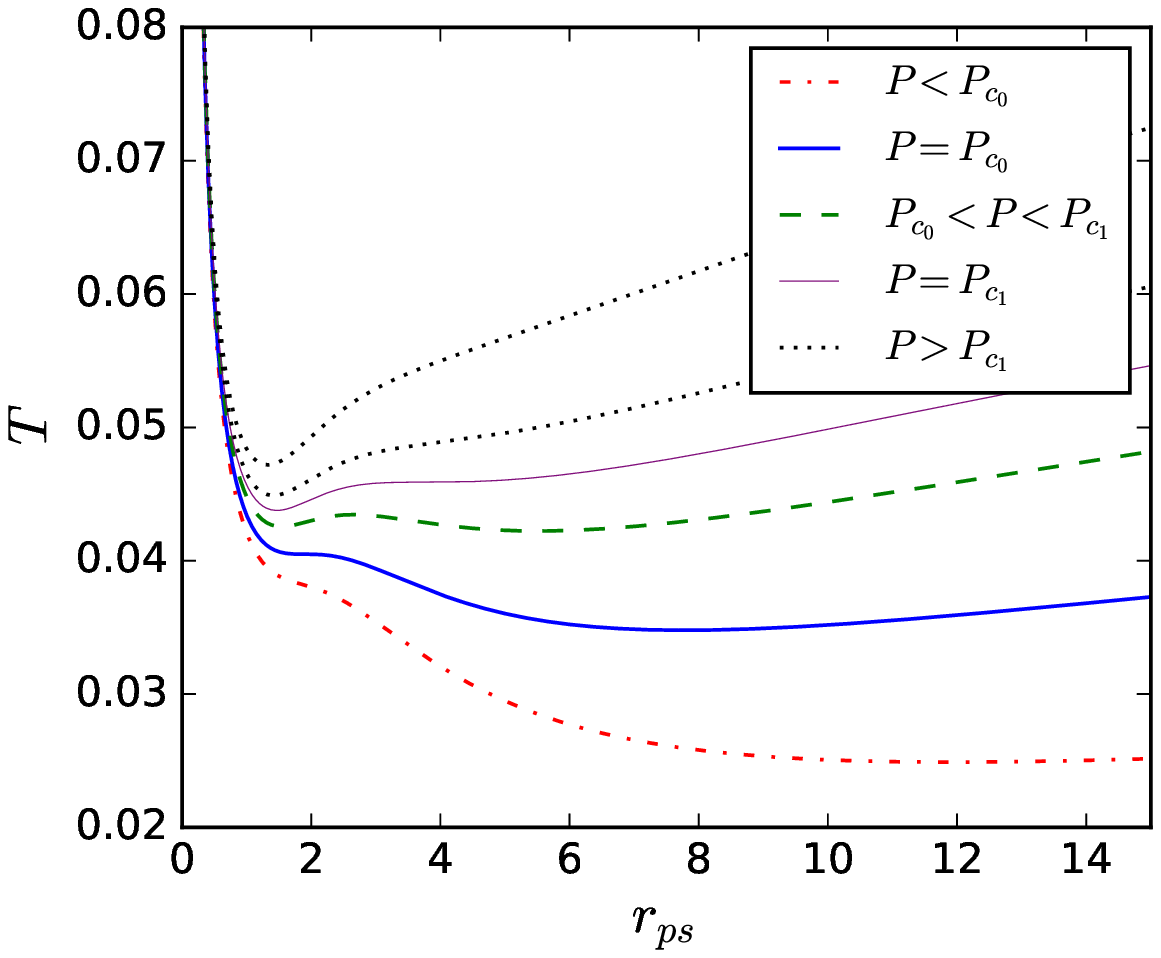}}
\subfigure[] {\label{fig:t:b1_b_b2}\includegraphics[width=0.38\textwidth]{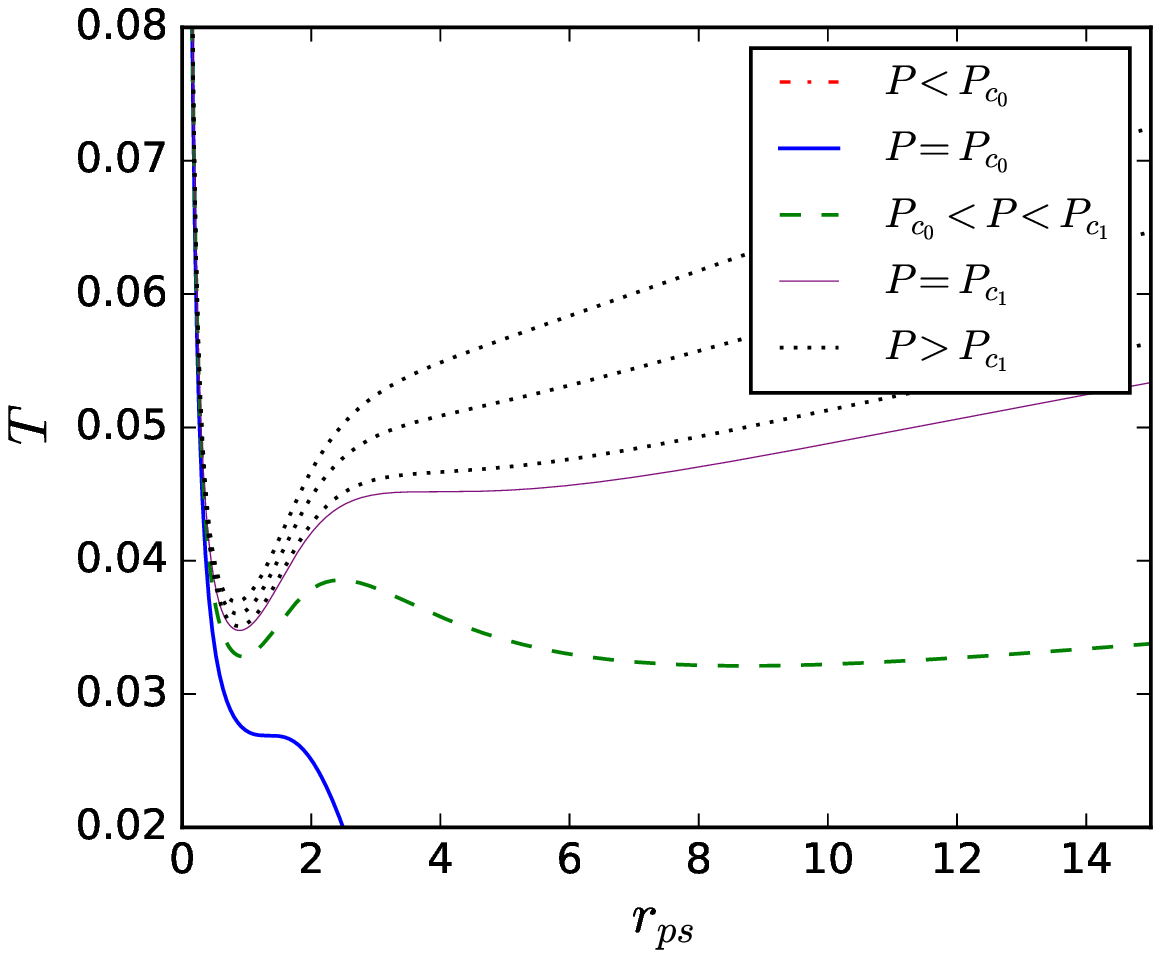}}
\subfigure[] {\label{fig:t:b2_b_b3}\includegraphics[width=0.38\textwidth]{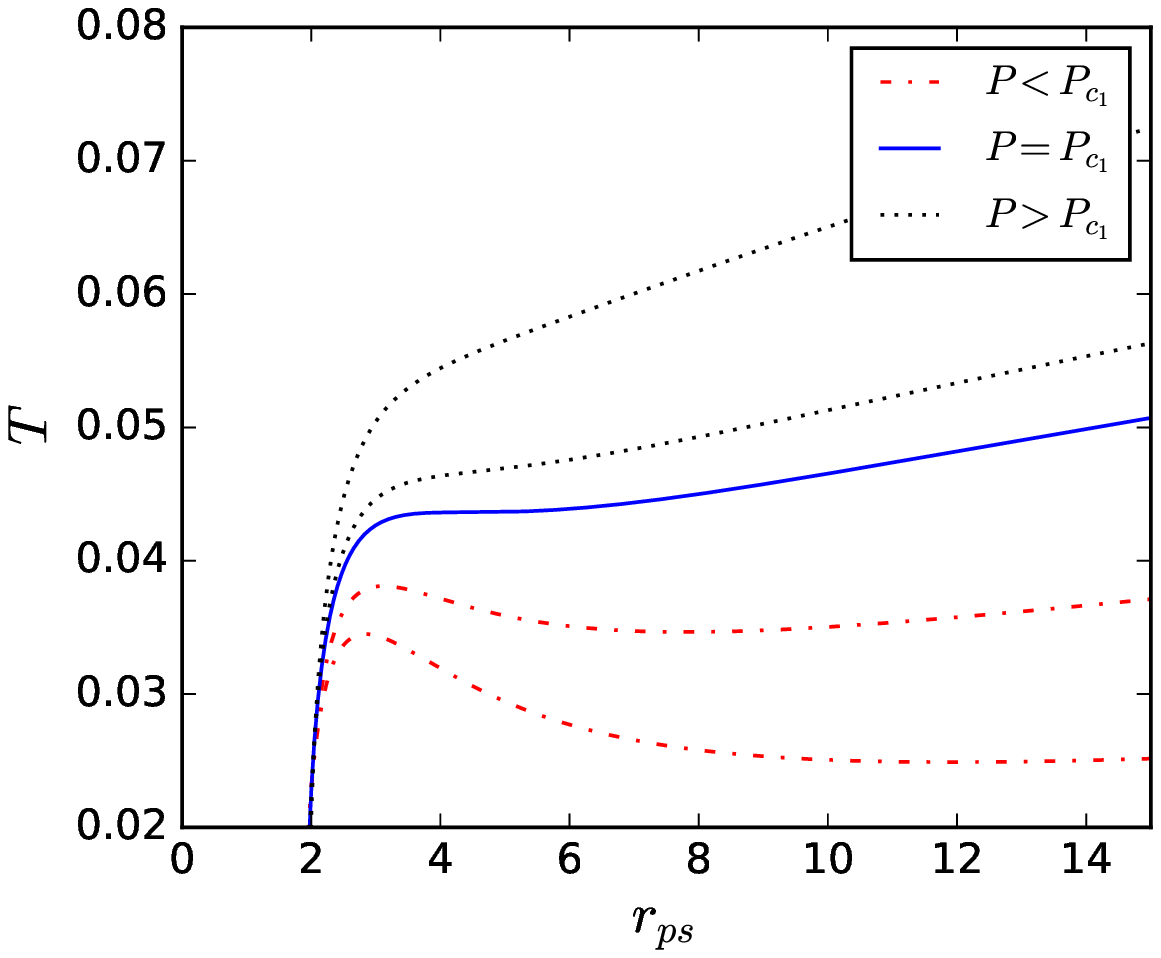}}
\caption{The temperature $T$ as a function of the radius $r_{\rm ps}$ of the photon sphere for different values of the BI parameter $b$ with fixed pressure $P$. (a) $b=0.3<b_{0}$ with $P$ = 0.001, 0.002, 0.003, 0.004, 0.005, and 0.006 from bottom to top. (b) $b=0.4\in(b_{0}, b_{1})$ with $P$ = 0.001, 0.002016 ($P_{\rm c0}$), 0.003, 0.003807 ($P_{\rm c1}$), 0.0045 and 0.006 from bottom to top. (c) $b=0.45\in (b_{1}, b_{2})$ with $P$ = -0.003253 ($P_{\rm c0}$), 0.001,0.002, 0.003, 0.003664 ($P_{\rm c1}$), and 0.006 from bottom to top. (d) $b=1>b_{2}$ with $P$ = 0.01, 0.02, 0.03, 0.043620 ($T_{\rm c1}$), 0.05 and 0.06 from bottom to top.}\label{trpss}
\end{figure*}

In Fig. \ref{trpss}, we show the temperature $T$ as a function of the radius $r_{\rm ps}$ of the photon sphere at constant pressures for different values of the parameter $b$. In Fig. \ref{fig:t:b_b0}, the pressure $P$=0.001, 0.002,  0.003, 0.004, 0.005, and 0.006 from bottom to top, and the BI parameter $b$=0.3$<b_{0}$, which corresponds to the first case. For this case, $T$ firstly decreases with $r_{\rm ps}$ and then increases with it. As shown in Refs. \cite{WeiLiuLiu,WeiLiuLiu2}, such behavior of $T$ reveals that no VdW-like phase transition exists. When $b$=0.4 and 0.45, the results are described in Figs. \ref{fig:t:b0_b_b1} and \ref{fig:t:b1_b_b2}, which are related with the case II and case III given above. For these two cases, the BI-AdS black holes demonstrate the reentrant phase transition. From these figures, we find that the temperature has a complicated behavior as the radius of the photon sphere varies. For example, when the pressure $P_{\rm c0}<P<P_{\rm c1}$, the temperature has a decrease-increase-decrease-increase behavior, which is quite different from the case of $b$=0.3, and can be used to be a sign of the reentrant phase transition. Further increasing the value of $b$ such that $b=1$, the behavior is presented in Fig. \ref{fig:t:b2_b_b3}. For this case, only one critical point exists, and the VdW-like phase transition can be found.
When the pressure is lower than the critical values, the temperature firstly increases, then decreases, and finally increases with the radius of the photon sphere, which is a typical behavior implying the existence of the VdW-like phase transition.

%%%%%%%%%%
%%%%%%%%%%%

%%%%%%%%%%%
\begin{figure*}[!htbp]
\centering
\subfigure[] {\label{fig:p:b_b0}\includegraphics[width=0.38\textwidth]{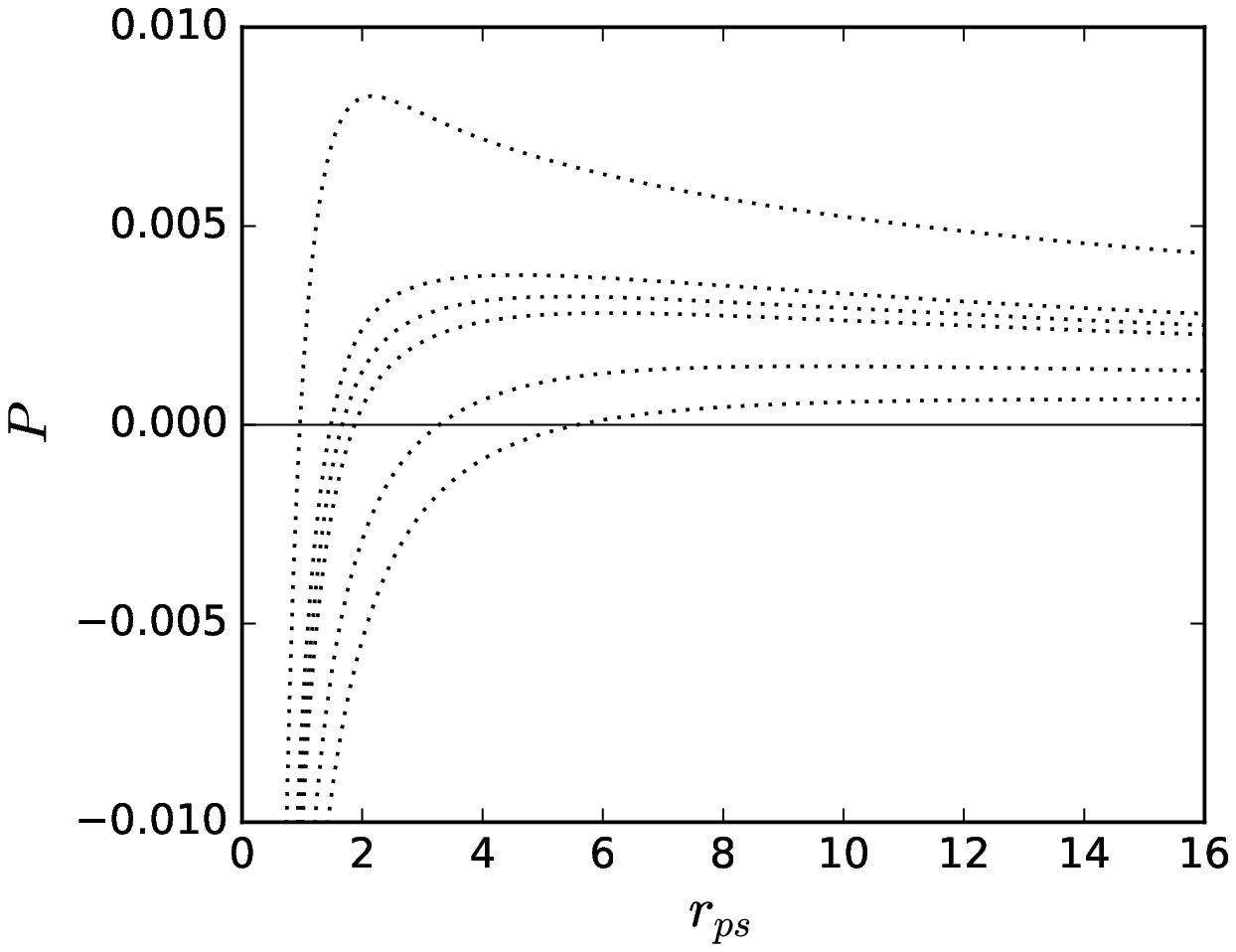}}
\subfigure[] {\label{fig:p:b0_b_b1}\includegraphics[width=0.38\textwidth]{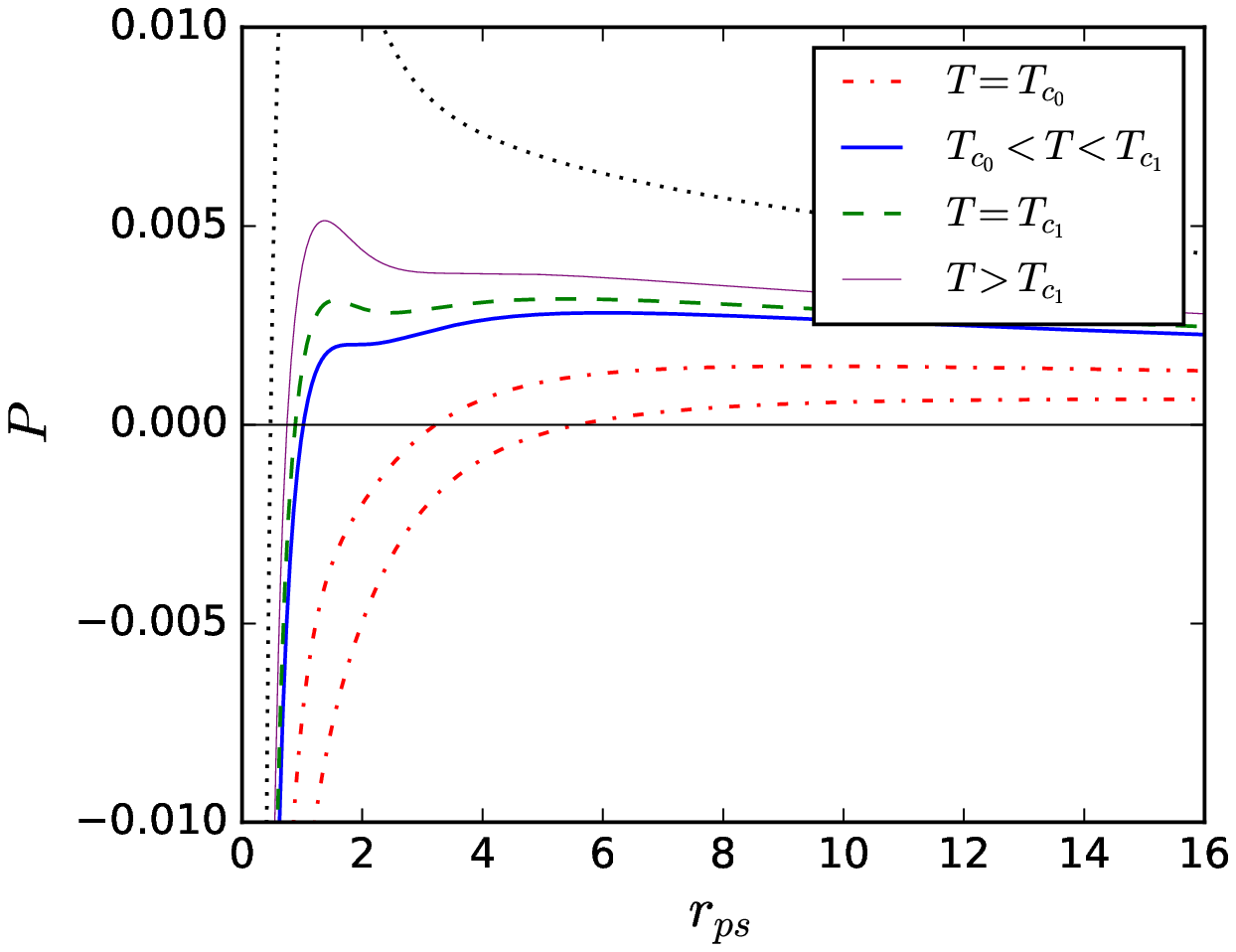}}
\subfigure[] {\label{fig:p:b1_b_b2}\includegraphics[width=0.38\textwidth]{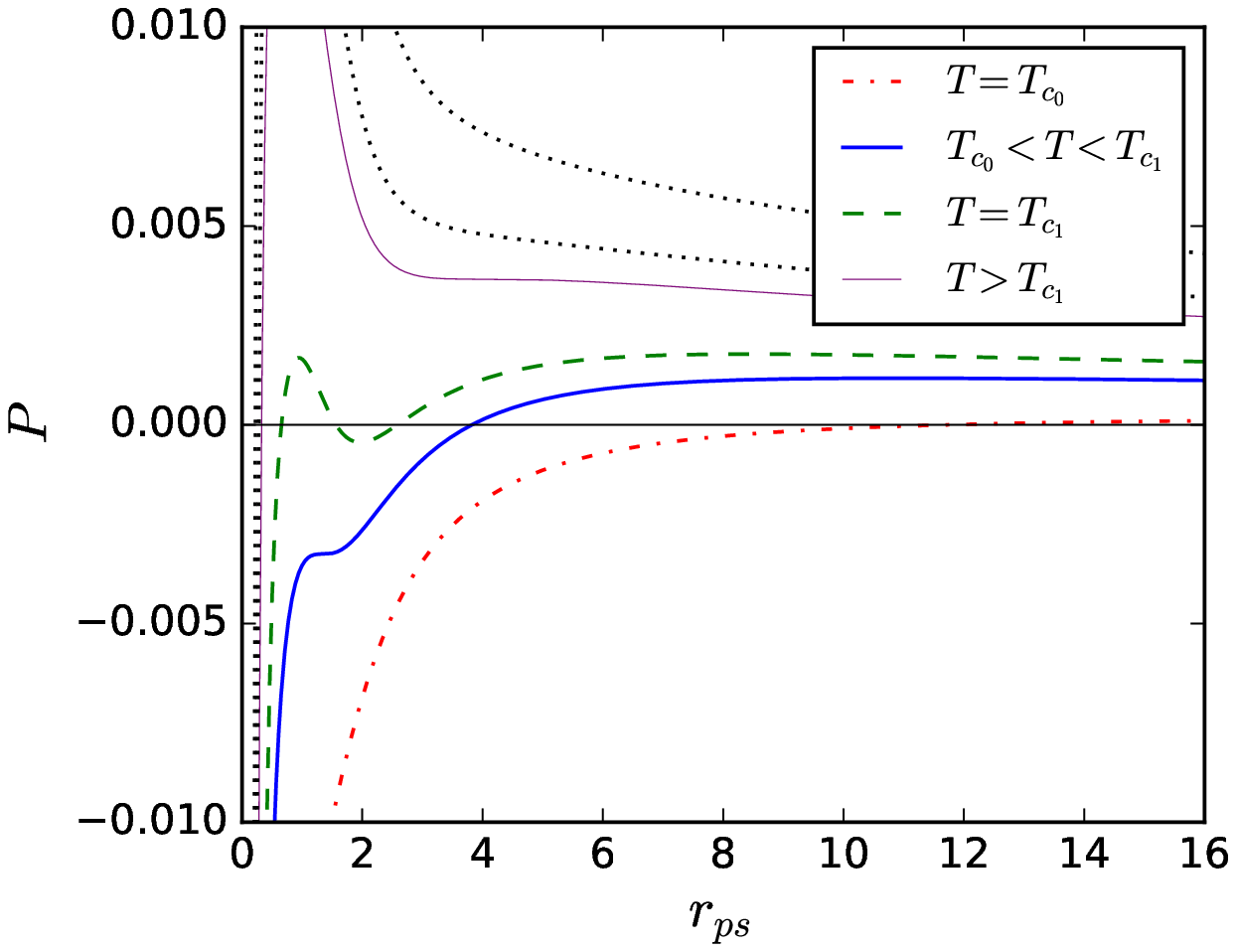}}
\subfigure[] {\label{fig:p:b2_b_b3}\includegraphics[width=0.38\textwidth]{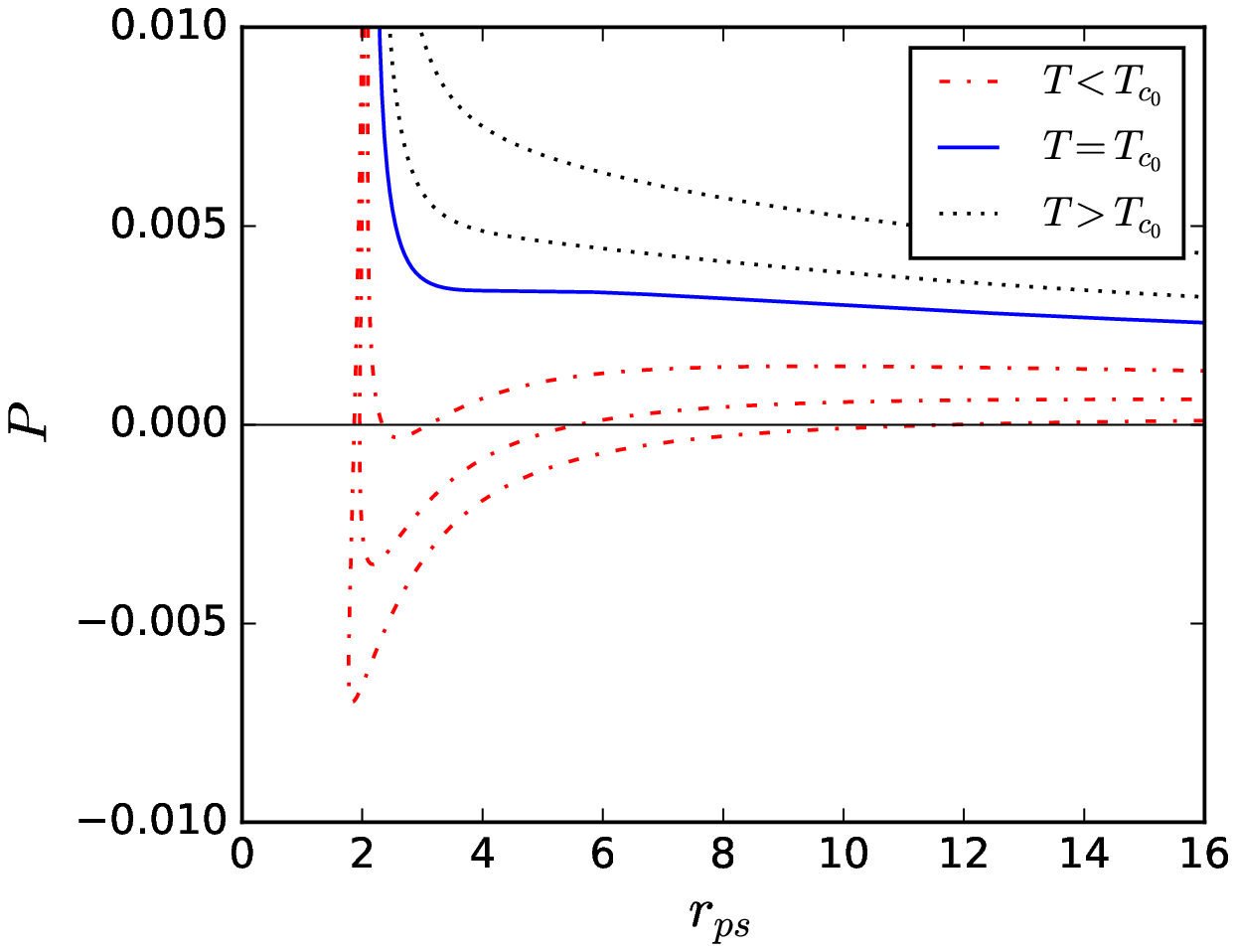}}
\caption{The pressure $P$ as a function of the radius $r_{\rm ps}$ of the photon sphere for different values of the BI parameter $b$ with fixed temperature $T$. (a) $b=0.3<b_{0}$ with $T$ = 0.02, 0.03, 0.04, 0.043, and 0.06 from bottom to top. (b) $b=0.4\in(b_{0}, b_{1})$ with $T$ = 0.02, 0.03, 0.040492 ($T_{\rm c0}$), 0.043, 0.045890 ($T_{\rm c1}$) and 0.06 from bottom to top. (c) $b=0.45\in (b_{1}, b_{2})$ with $T$ = 0.01, 0.026885 ($T_{\rm c0}$), 0.03, 0.045170 ($T_{\rm c1}$), 0.05 and 0.06 from bottom to top. (d) $b=1>b_{2}$ with $T$ = 0.001, 0.002, 0.003372 ($P_{\rm c1}$), 0.004 and 0.006 from bottom to top.}\label{prpss}
\end{figure*}
%%%%%%%%%%%%

%%%%%%%%%%%%
\begin{figure}
\centering
\subfigure[]{\label{tupsaa}\includegraphics[width=0.38\textwidth]{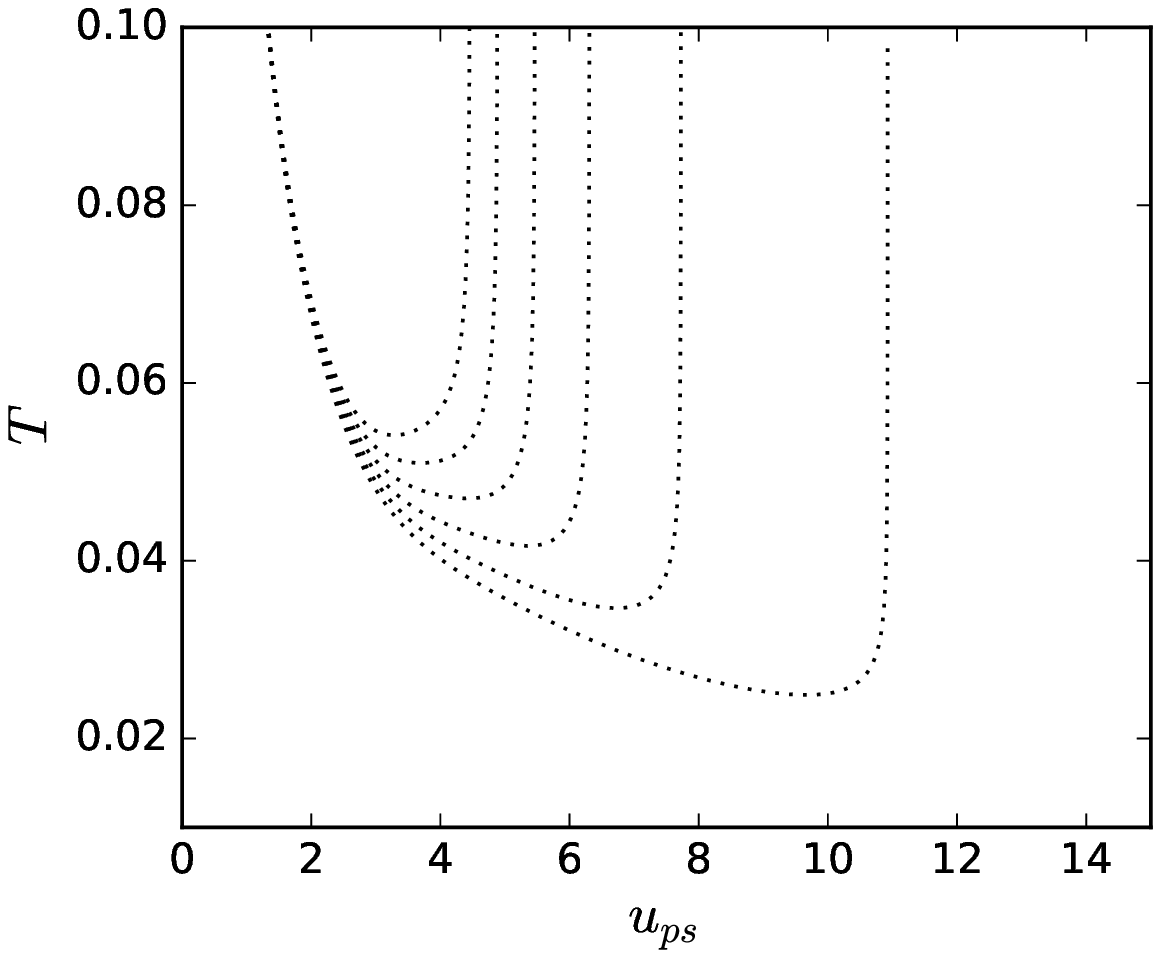}}
\subfigure[]{\label{tupsab}\includegraphics[width=0.38\textwidth]{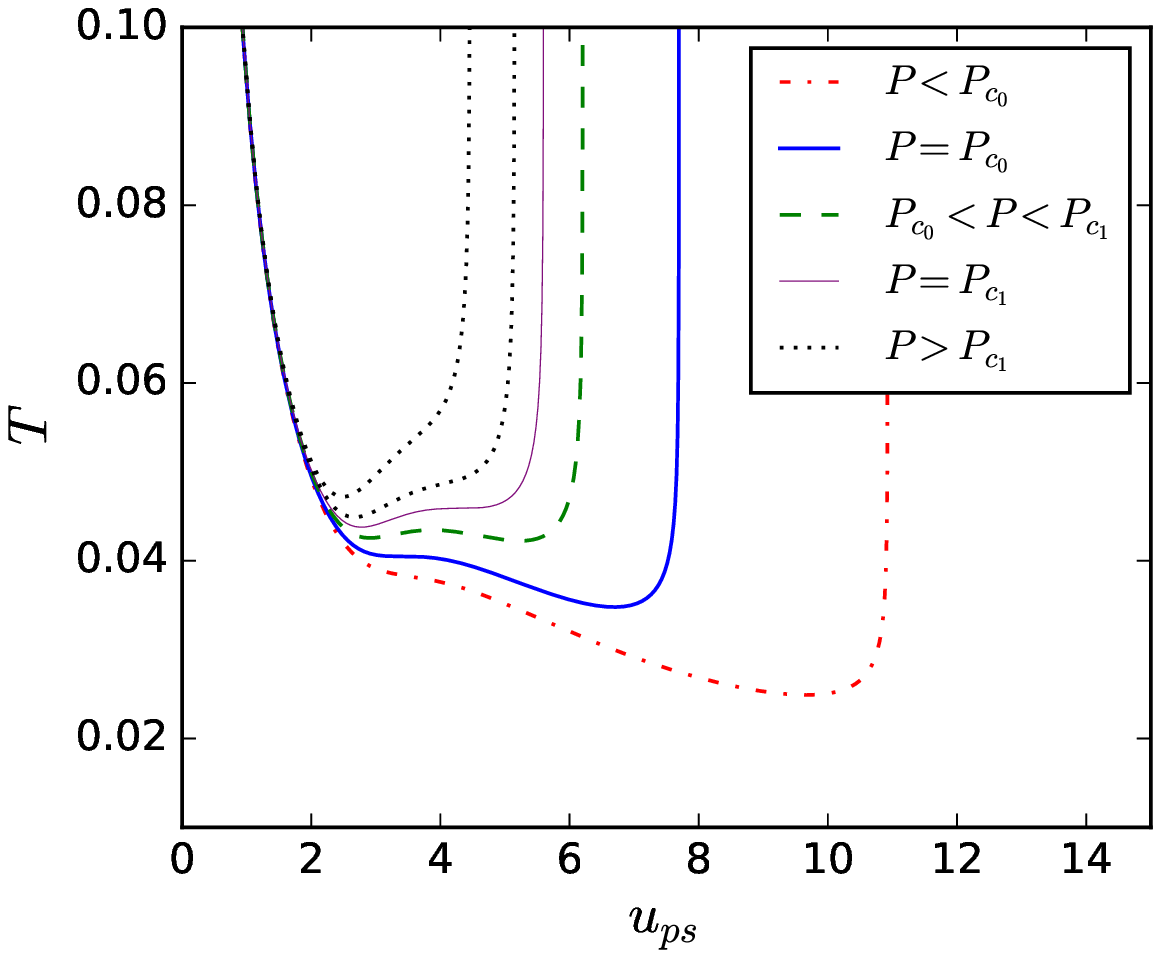}}
\subfigure[]{\label{tupsac}\includegraphics[width=0.38\textwidth]{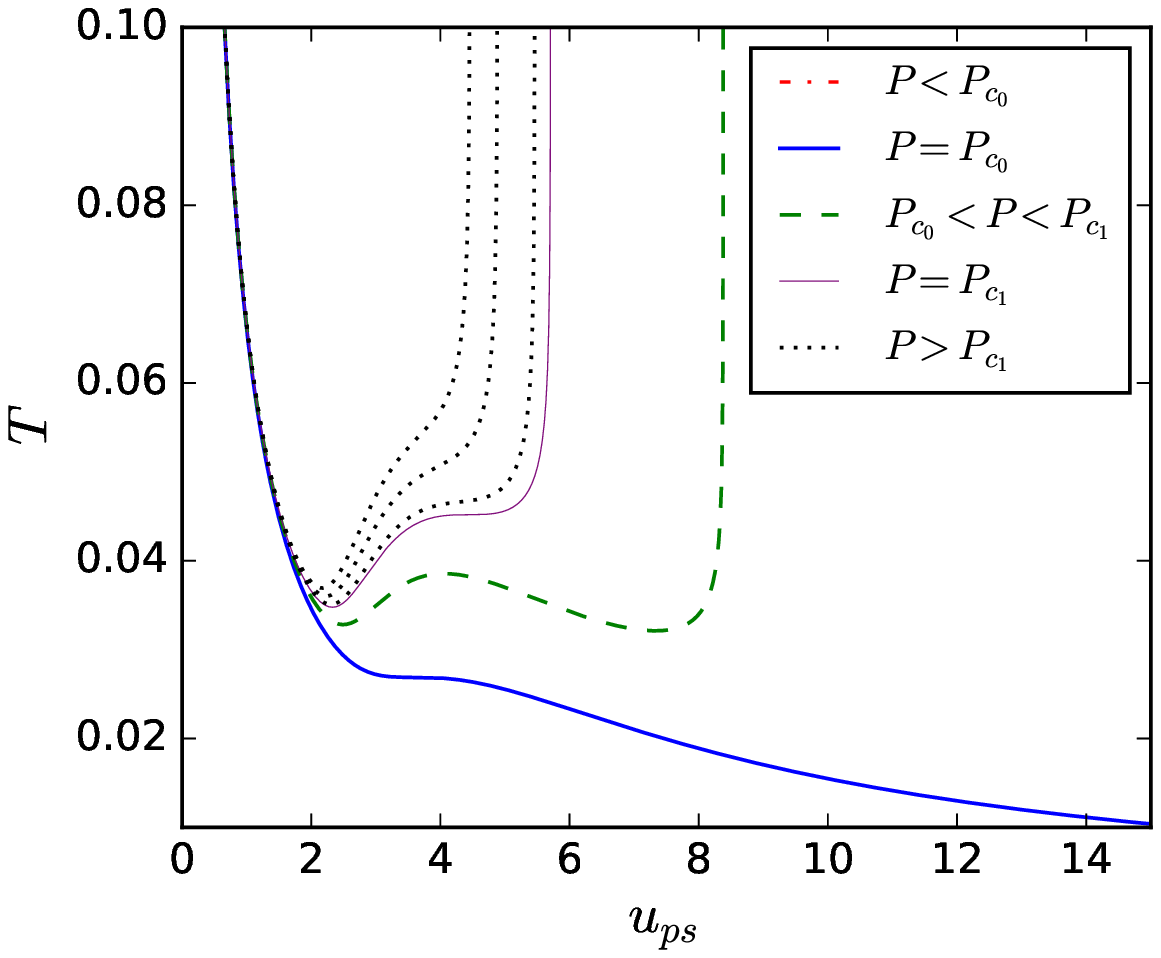}}
\subfigure[]{\label{tupsad}\includegraphics[width=0.38\textwidth]{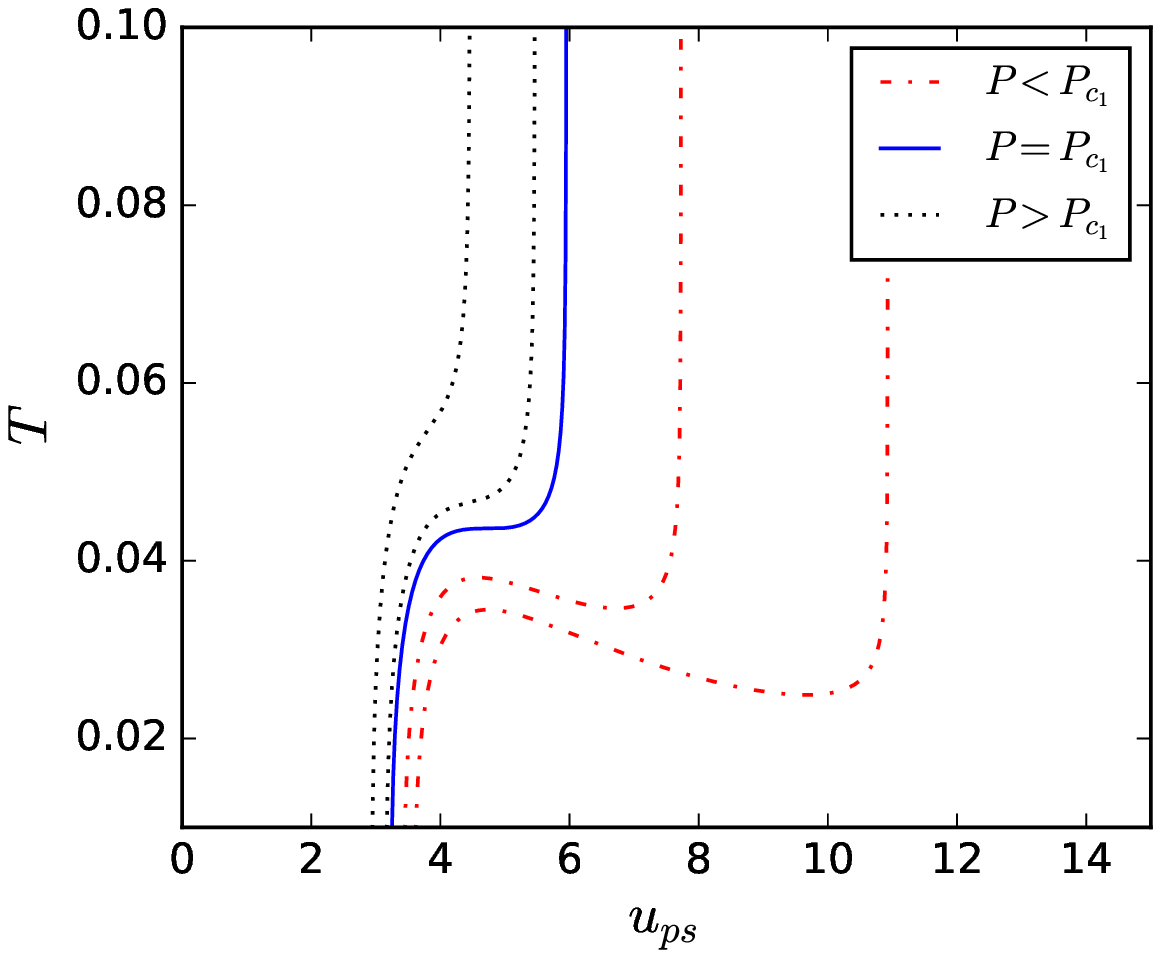}}
\caption{The temperature $T$ as a function of the minimum impact parameter $u_{\rm ps}$ of the photon sphere for different values of the BI parameter $b$ with fixed pressure $P$. (a) $b=0.3<b_{0}$ with $P$ = 0.001, 0.002, 0.003, 0.004, 0.005, and 0.006 from bottom to top. (b) $b=0.4\in(b_{0}, b_{1})$ with $P$ = 0.001, 0.002016 ($P_{\rm c0}$), 0.003, 0.003807 ($P_{\rm c1}$), 0.0045 and 0.006 from bottom to top. (c) $b=0.45\in (b_{1}, b_{2})$ with $P$=-0.003253 ($P_{\rm c0}$), 0.001,0.002, 0.003, 0.003664 ($P_{\rm c1}$), and 0.006 from bottom to top. (d) $b=1>b_{2}$ with $P$ = 0.01, 0.02, 0.03, 0.043620 ($T_{\rm c1}$), 0.05 and 0.06 from bottom to top.}\label{tupss}
\end{figure}
%%%%%%%%%%%%

%%%%%%%%%%%%
\begin{figure*}[!htbp]
\centering
\subfigure[]{\includegraphics[width=0.38\textwidth]{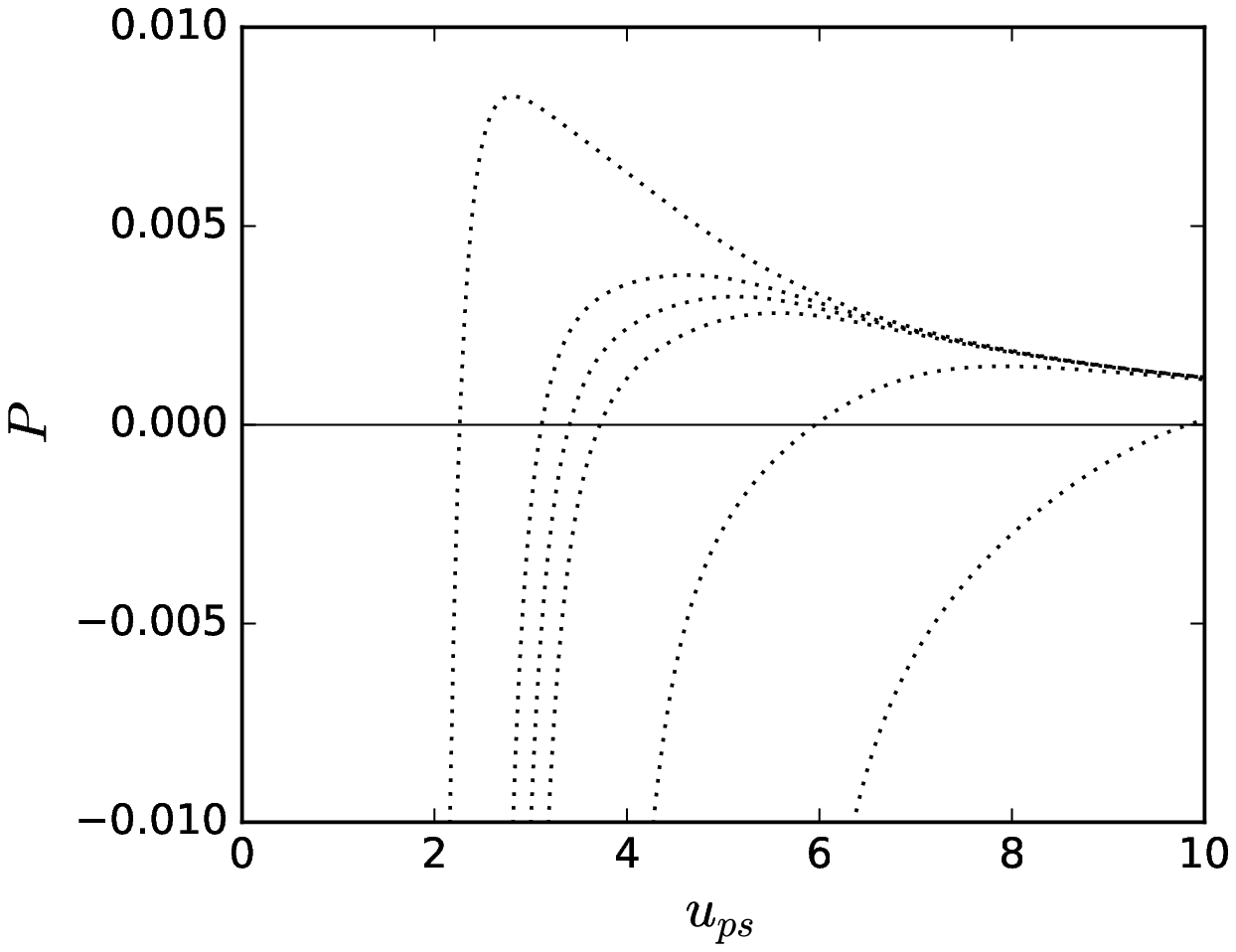}}
\subfigure[]{\includegraphics[width=0.38\textwidth]{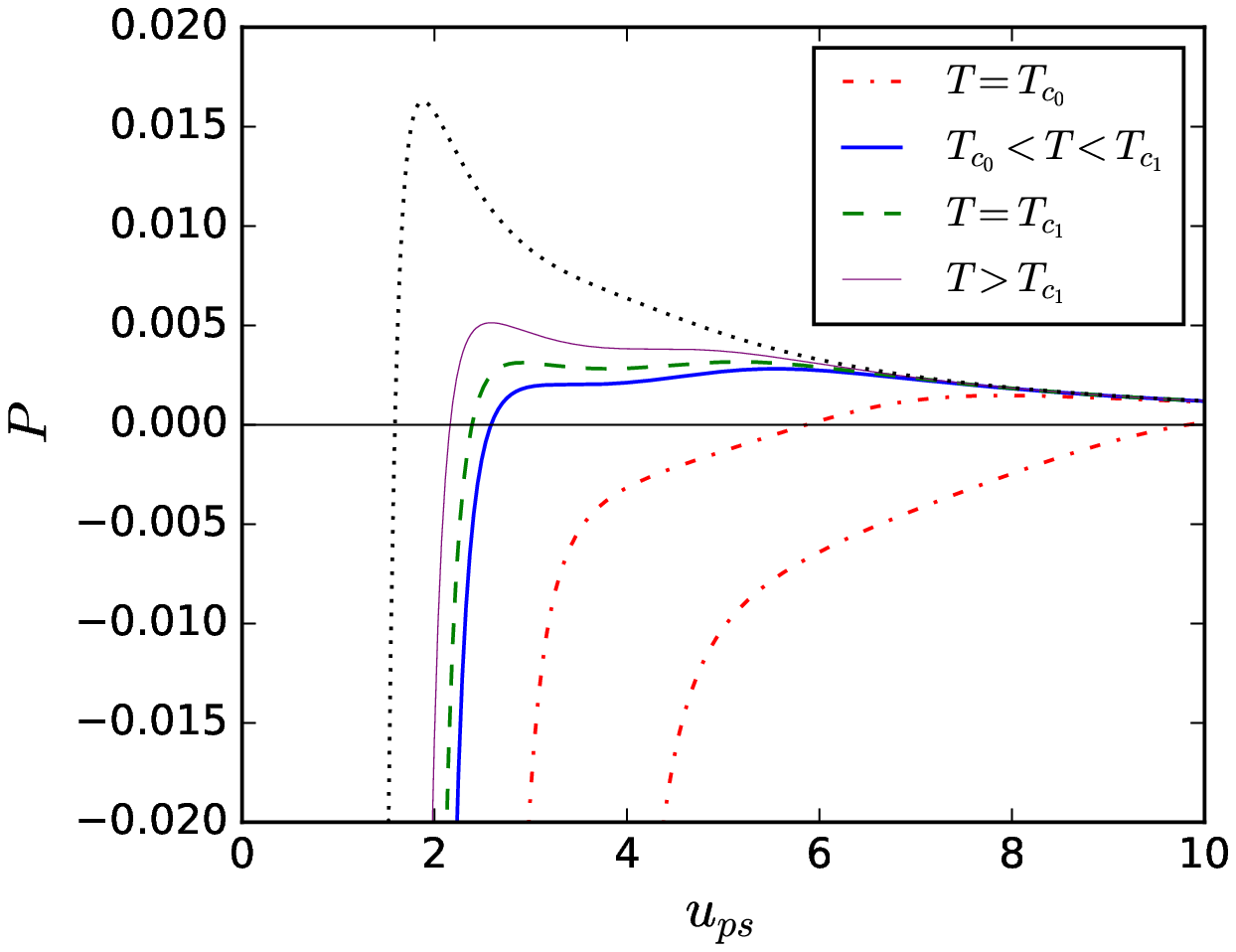}}
\subfigure[]{\includegraphics[width=0.38\textwidth]{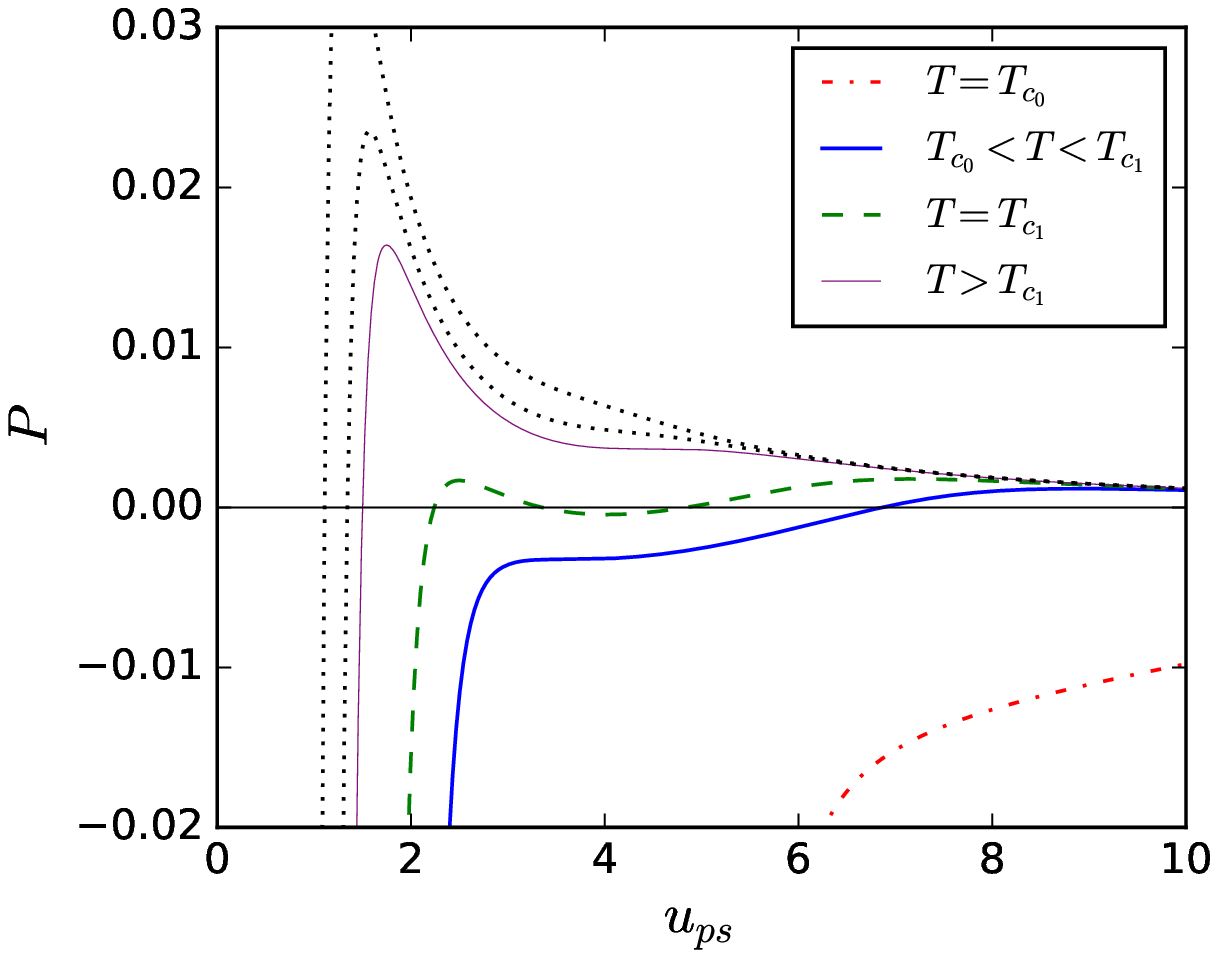}}
\subfigure[]{\includegraphics[width=0.38\textwidth]{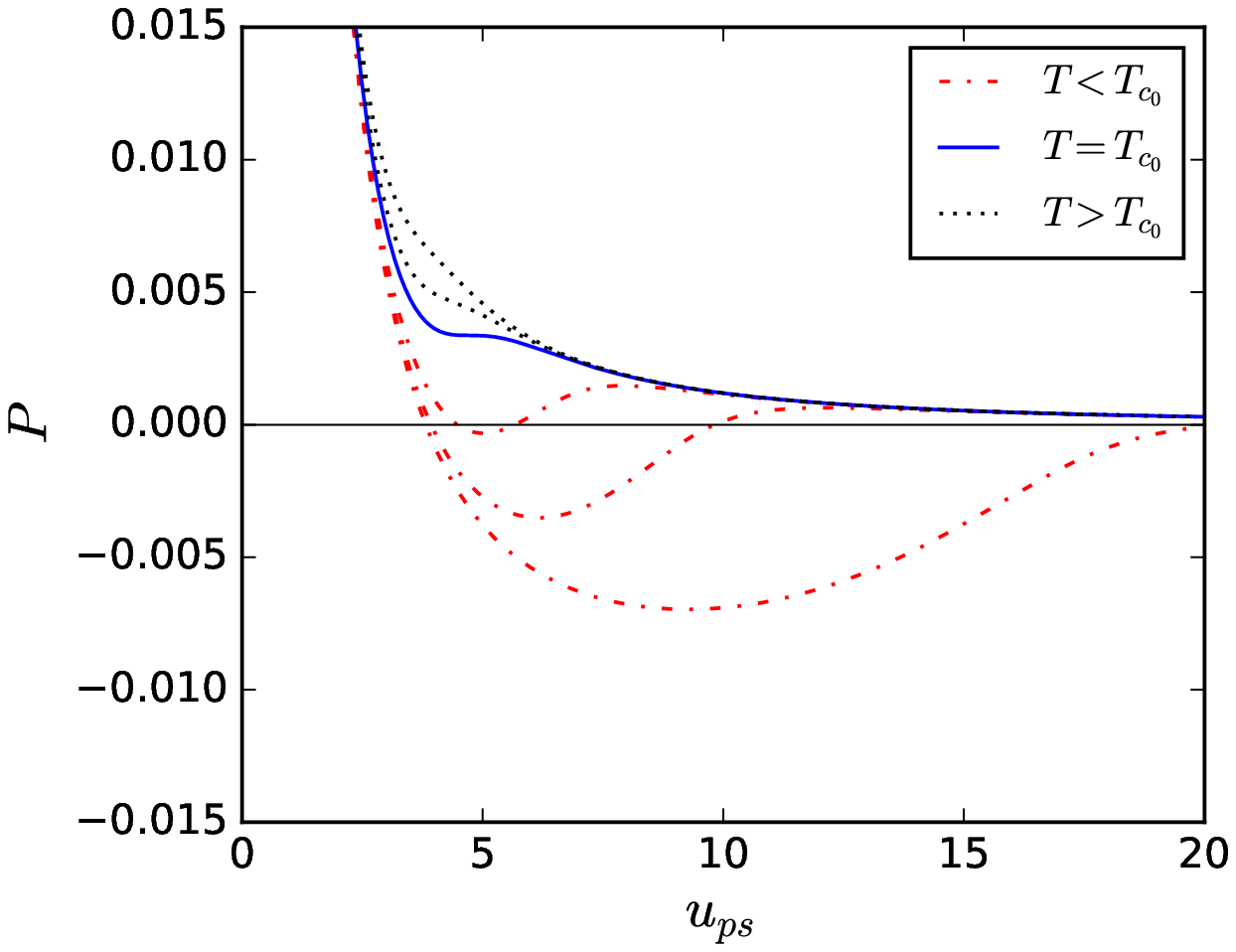}}
\caption{The pressure $P$ as a function of the minimum impact parameter $u_{\rm ps}$ of the photon sphere for different values of the BI parameter $b$ with fixed temperature $T$. (a) $b=0.3<b_{0}$ with $T$ = 0.02, 0.03, 0.04, 0.043, and 0.06 from bottom to top. (b) $b=0.4\in(b_{0}, b_{1})$ with $T$ = 0.02, 0.03, 0.040492 ($T_{\rm c0}$), 0.043, 0.045890 ($T_{\rm c1}$) and 0.06 from bottom to top. (c) $b=0.45\in (b_{1}, b_{2})$ with $T$ = 0.01, 0.026885 ($T_{\rm c0}$), 0.03, 0.045170 ($T_{\rm c1}$), 0.05 and 0.06 from bottom to top. (d) $b=1>b_{2}$ with $T$ = 0.001, 0.002, 0.003372 ($P_{\rm c1}$), 0.004 and 0.006 from bottom to top.}\label{pupss}
\end{figure*}
%%%%%%%%%%%%

We also plot the pressure $P$ as a function of the radius $r_{\rm ps}$ of the photon sphere at constant temperatures for different values of the parameter $b$, see Fig. \ref{prpss}. For different values of $b$, we can find that the behaviors are quite similar. The significant difference is that $T$ and $P$ have a reverse behavior with $r_{\rm ps}$. For example, when one increases with $r_{\rm ps}$, the other one will decreases. However, the number of the extremal point is the same.

Next, let us turn to the behavior of the minimum impact parameter $u_{\rm ps}$ of the photon sphere. At constant pressure, we display the temperature $T$
as a function of $u_{\rm ps}$ for different values of $b$ in Fig. \ref{tupss}. When $b$=0.3 corresponding to case I, see Fig. \ref{tupsaa}, we find that at small $u_{\rm ps}$, the temperature $T$ decreases rapidly. Then $T$ increases and approaches to a high value at a certain finite $u_{\rm ps}$. However, this behavior does not indicate the VdW-like phase transition. For $b$=0.4 and 0.45 described in Figs. \ref{tupsab} and \ref{tupsac}, the behaviors of $T$ become more complicated. In particular, for fixed pressure $P_{\rm c0}<P<P_{\rm c1}$, with the increase of $u_{\rm ps}$, the temperature $T$ demonstrates a decrease-increase-decrease-increase behavior, which implies that there exists a reentrant phase transition in the corresponding BI-AdS black hole background. While when $b=1>b_{2}$ shown in Fig. \ref{tupsad}, the black hole system admits a VdW-like phase transition. We observe that the temperature shows an increase-decrease-increase behavior for the constant pressure $P$ below its critical value. While when the pressure is above the critical value, the temperature is only a monotonic increasing function of $u_{\rm ps}$, and thus no phase transition exists for this case.

Moreover, we also plot the pressure $P$ as a function of the minimum impact parameter $u_{\rm ps}$ in Fig. \ref{pupss}. From it, we can find that both the reentrant phase transition and the VdW-like phase transition can be reflected from the behavior of the pressure $P$.

In summary, by observing the behavior of the photon sphere of the BI-AdS black hole, both the reentrant phase transition and the VdW-like phase transition can be reflected. And these two types phase transitions can be clearly distinguished by the photon sphere. Thus, we confirm our previous conjecture that there indeed exists a relation between the photon sphere and thermodynamic phase transition.

\subsection{Critical behavior of photon sphere}\label{sec:critial_exponent}

As discussed above, when $b>b_{0}$, there will be one or two critical points for the BI-AdS black hole systems. However one of the points has negative pressure or higher Gibbs free energy, which makes it unstable or unphysical. Nevertheless, one physical critical point always exists for $b>b_{0}$. Meanwhile, there exist the first-order phase transition between the intermediate black hole and large black hole for $b_{0}<b<b_{2}$, and the first-order phase transition between the small black hole and large black hole for $b>b_{2}$. With the increase of the pressure and temperature, these first-order phase transition terminates at the critical point. Thus it is worth to examine the changes of the radius and the minimum impact parameter of the photon sphere along the coexistence curve and to consider their critical behavior.

\begin{figure*}[!htbp]
\centering
\subfigure[$\Delta r_{\rm ps} - T/T_{\rm c} $] {\includegraphics[width=0.38\textwidth]{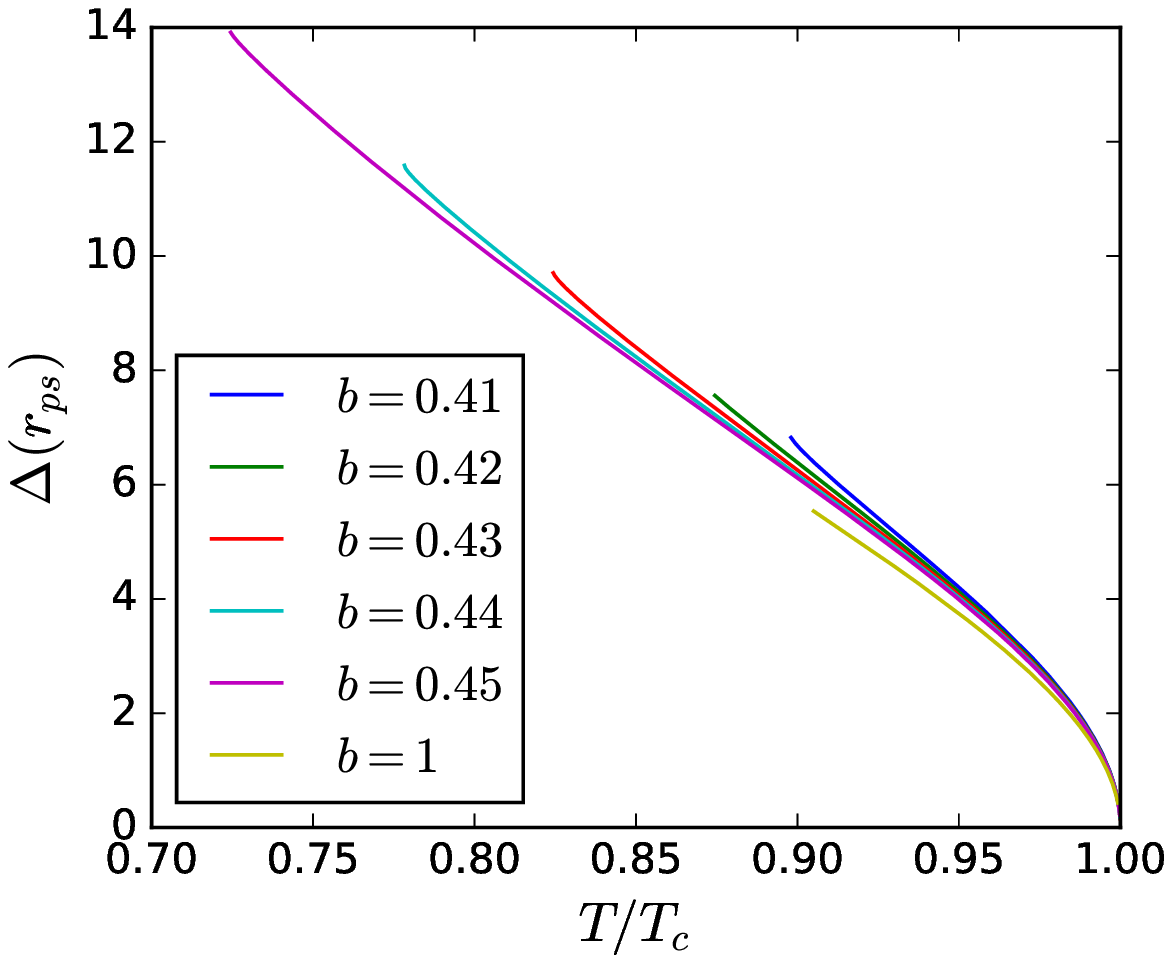}}
\subfigure[$\Delta u_{\rm ps} - T/T_{\rm c} $] {\includegraphics[width=0.38\textwidth]{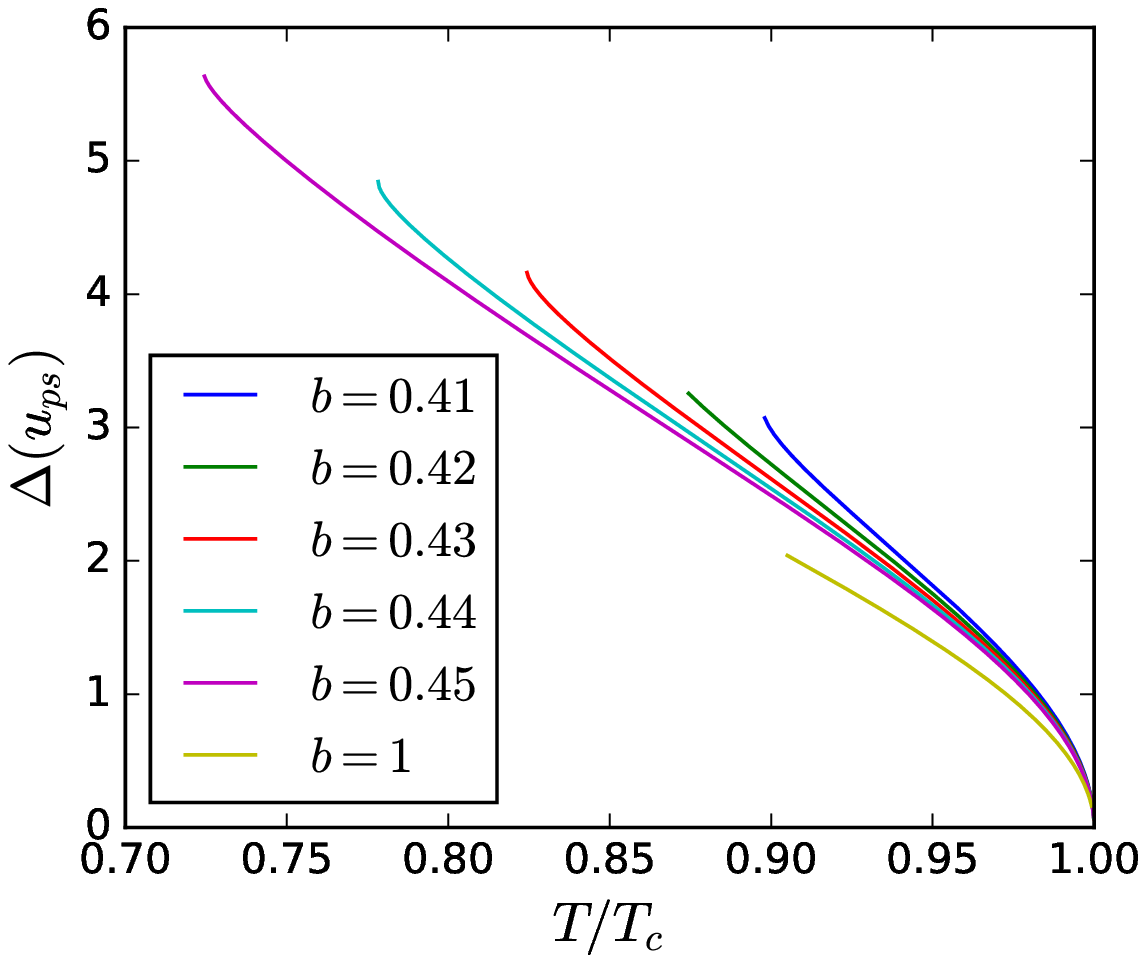}}
\caption{Behaviors of $\Delta r_{\rm ps}$ and $\Delta u_{\rm ps}$ as a function of the phase transition temperature $T/T_{\rm c}$ for $b$=0.41$\sim$1 from top to bottom.}
\label{fig:p:drps_dups_T}
\end{figure*}

Following Refs. \cite{WeiLiuLiu,WeiLiuLiu2}, we plot the changes of $r_{\rm ps}$ and $u_{\rm ps}$ when the first-order phase transition occurs. From Fig. \ref{fig:p:drps_dups_T}, we find that with the increase of the phase transition temperature, both $\Delta r_{\rm ps}$ and $\Delta u_{\rm ps}$ decrease, and when the critical temperature is approached, both of them vanish. This behavior holds for $b>b_{0}$. Such behaviors are also consistent with that observed in Refs. \cite{WeiLiuLiu,WeiLiuLiu2}.

%%%%%%%%%%%
%%%%%%%%%%%%%

Next, we would like to calculate the critical exponent of $\Delta r_{\rm ps}$ and $\Delta u_{\rm ps}$ for different values of the BI parameter $b$. Since there is no the analytic result, we use the numerical data to fit them with the following form \cite{WeiLiuLiu}
\begin{equation}
\Delta r_{\rm ps}, \Delta u_{\rm ps} \sim a \times (1 - \tilde{T})^\delta,\label{fiteq}
\end{equation}
where $a$ and $\delta$ are two fitting coefficients. If $\Delta r_{\rm ps}$ and $\Delta u_{\rm ps}$ have an universal critical behavior, then the coefficient $\delta$ must keep constant for different values of $b$.

The fitting results are listed in Table \ref{tb:critical_value}. One can notice that the fitting coefficient $a$ of $\Delta r_{\rm ps}$ and $\Delta u_{\rm ps}$ decreases with $b$. While the coefficient $\delta$ is always around $\frac{1}{2}$ with numerical error no more than 1.73\% for different values of $b$. So even when the reentrant phase transition is included in, the universal critical exponent does not change.

\begin{figure*}[!htbp]
\centering
\subfigure[$b< b_0$] {\label{fig:PointOnPT:b_b0}\includegraphics[width=0.38\textwidth]{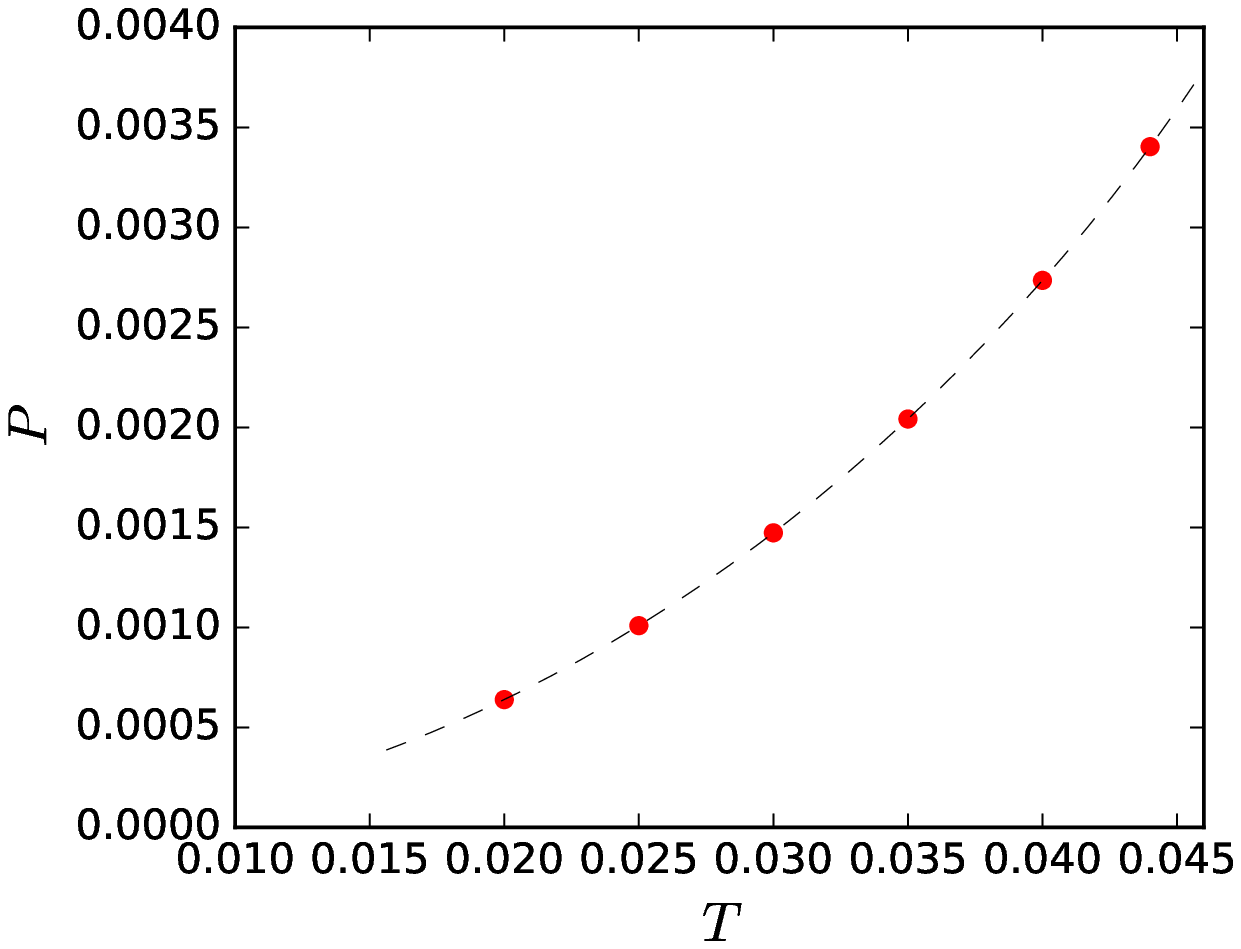}}
\subfigure[$b\in (b_0,b_1)$] {\label{fig:PointOnPT:b1_b_b2}\includegraphics[width=0.38\textwidth]{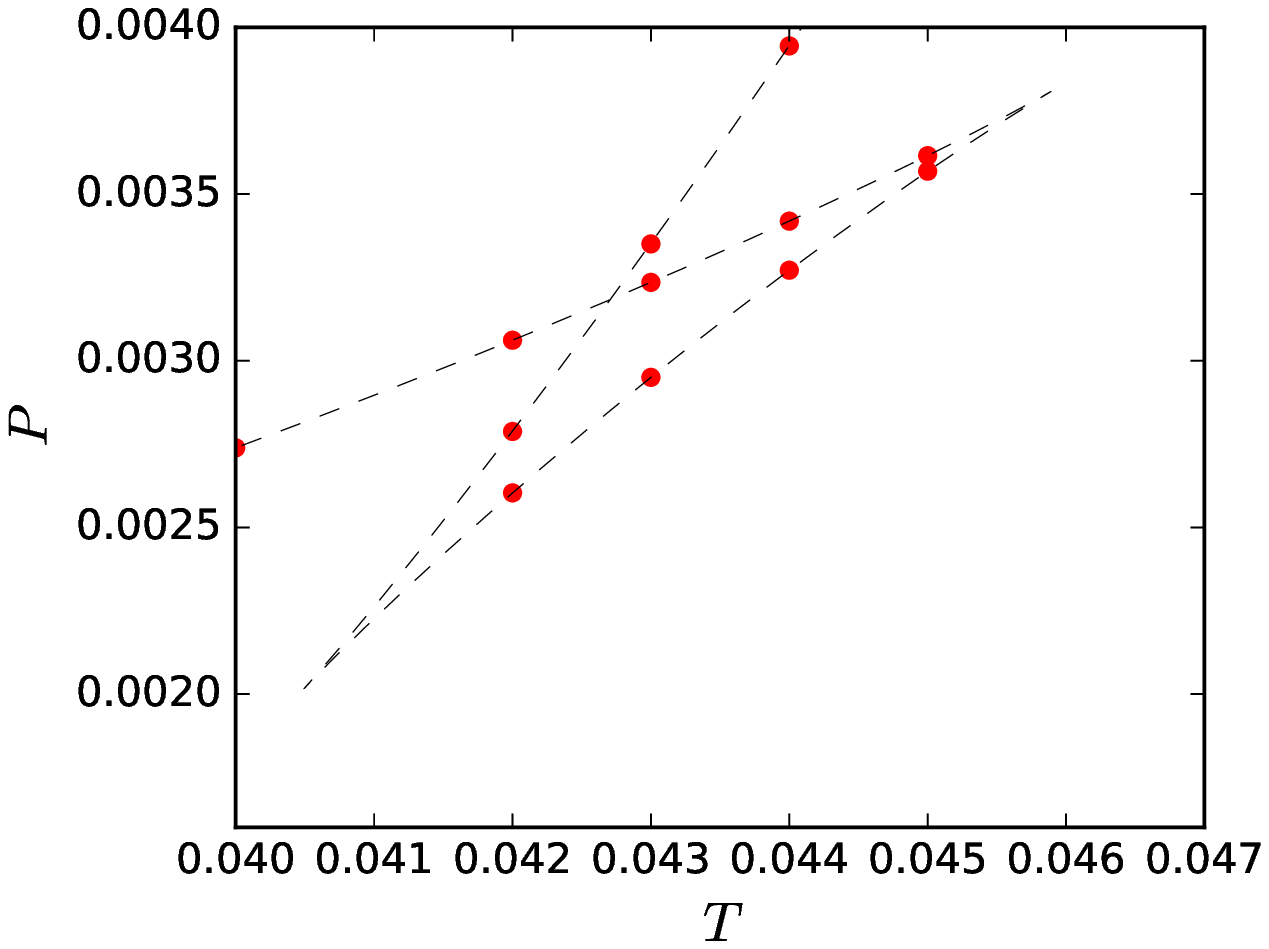}}
\subfigure[$b\in (b_1,b_2)$] {\label{fig:PointOnPT:b2_b_b3}\includegraphics[width=0.38\textwidth]{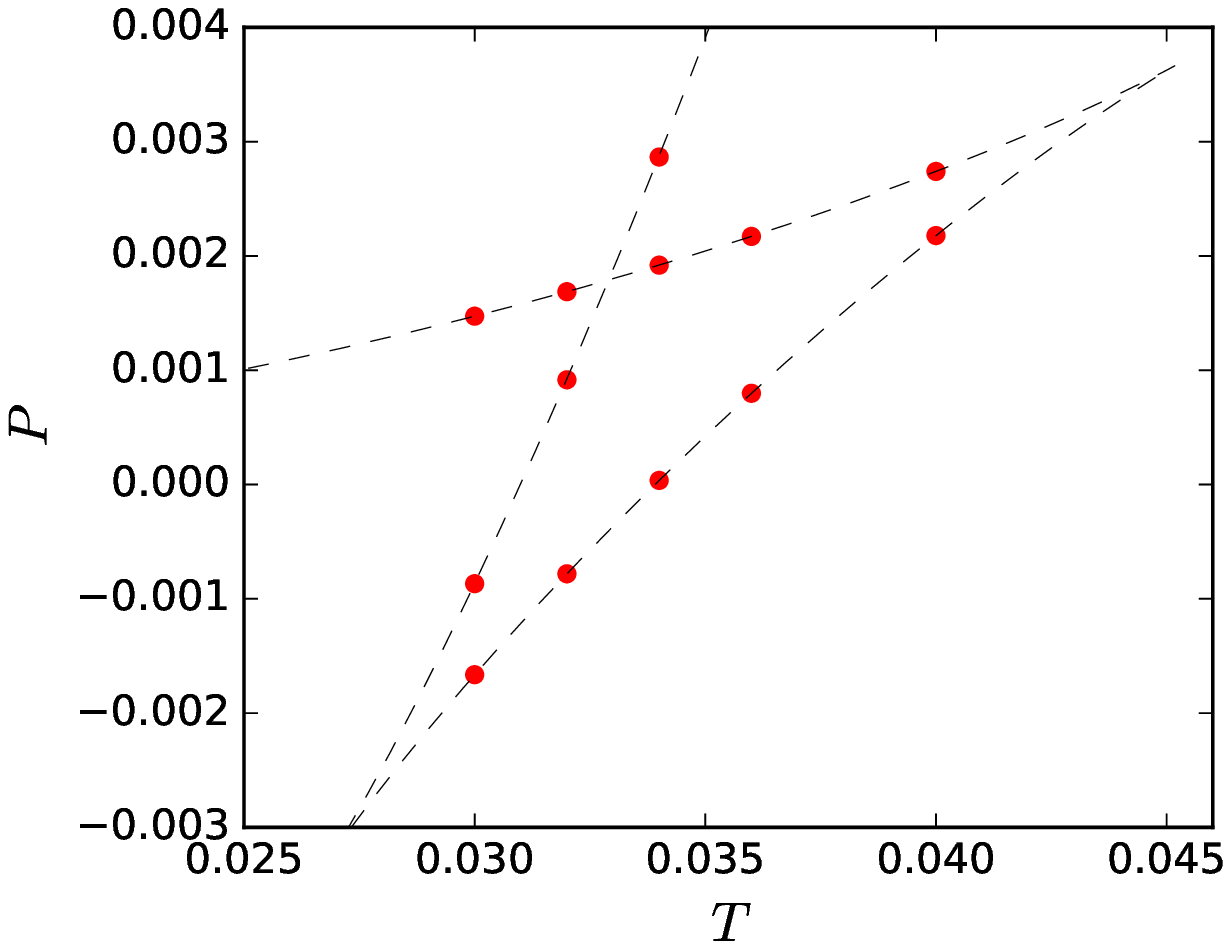}}
\subfigure[$b> b_2$] {\label{fig:PointOnPT:b_b2}\includegraphics[width=0.38\textwidth]{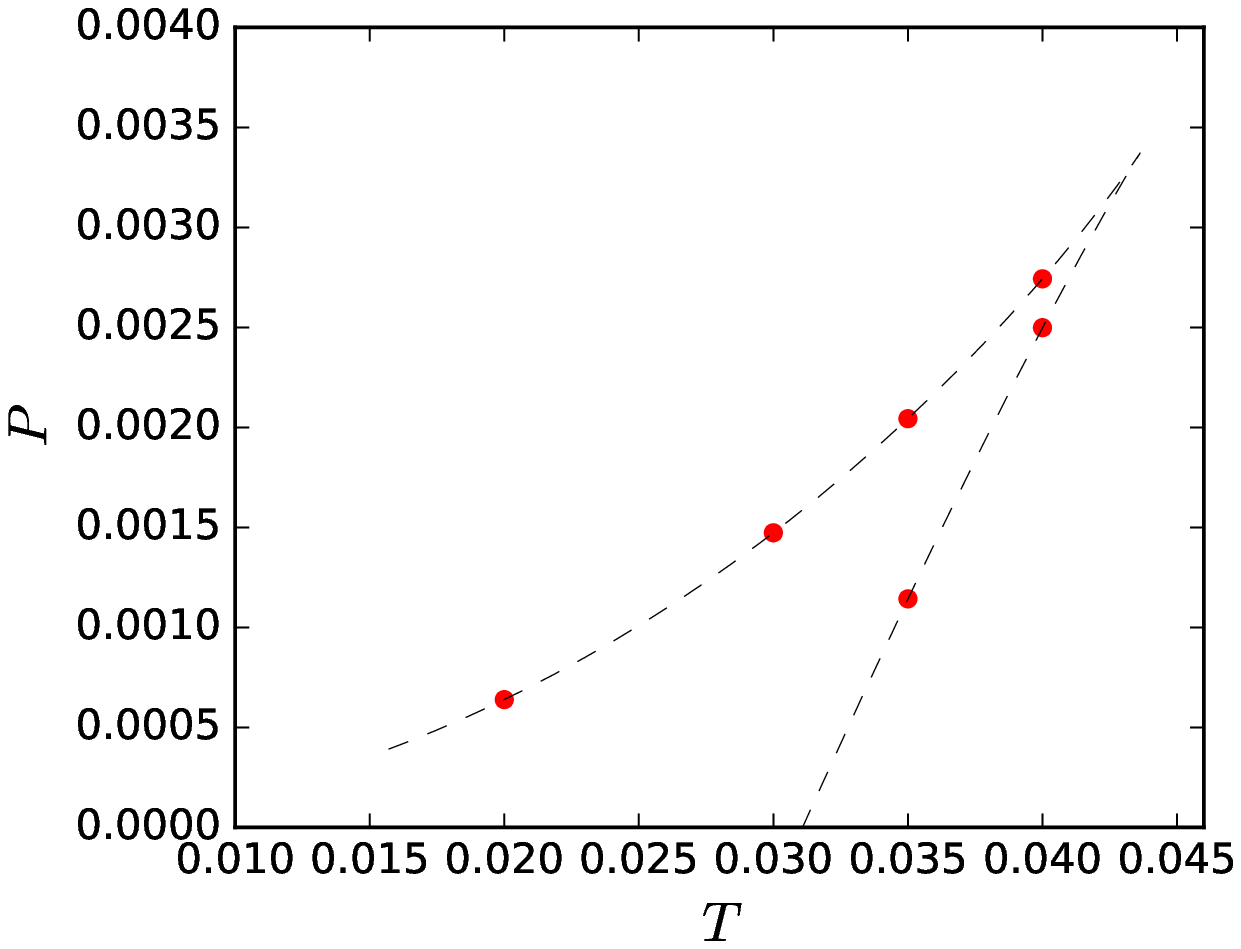}}
\caption{Extremal points of the radius $r_{\rm ps}$ and the minimal impact parameter $u_{\rm ps}$ of the photon sphere (black dots) and the thermodynamic metastable curves (dashed lines) of the BI-AdS black holes shown in the $P$-$T$ diagram. (a) $b$=0.3. (b) $b$=0.4. (c) $b$=0.45. (d) $b$=1.0.}\label{fig:PointOnPT}
\end{figure*}

At last, we would like to examine the consistency between the extremal points of the radius and the minimum impact parameter of the photon sphere along the constant temperature or pressure curve and the metastable curve from the thermodynamic side. For the Kerr-AdS black hole, we have shown that they are consistent with each other when the VdW-like phase transition is presented \cite{WeiLiuLiu2}. However, for the case of the reentrant phase transition, the metastable curves became rather interesting, see Figs. \ref{pphaa1} and \ref{pphb1}. We display in Fig. \ref{fig:PointOnPT} the metastable curves with the dashed lines and the extremal point of the radius $r_{\rm ps}$ of the photon sphere marked with the black dots in the $P$-$T$ diagram for different values of $b$. When $b>b_{2}$ described in Fig. \ref{fig:PointOnPT:b_b2}, it is the case that the typical VdW phase transition takes place. From it, one can find that these results are highly consistent with each other, which also supports our result for the Kerr-AdS black holes \cite{WeiLiuLiu2}. When $b_{0}<b<b_{2}$, we have the reentrant phase transition, see Figs. \ref{fig:PointOnPT:b1_b_b2} and \ref{fig:PointOnPT:b2_b_b3}. It is worth to note that there are three metastable curves. Despite that, the extremal points are still in good agreement with the metastable curves. For $b<b_{0}$, we can also find that they well agree with each other even when there is no any phase transition. Thus we confirm that in all the range of the BI parameter $b$, the extremal points of $r_{\rm ps}$ always coincide with the thermodynamic metastable curves for the BI-AdS black holes, and it is independent of the type of the phase transition.

%%%%%%%%%%%%%%%%%%%%%
\begin{table}[!htbp]%[H] add [H] placement to break table across pages
\centering
\begin{tabular}{cccccccc}
\hline\hline
&b &        0.41 &         0.42 &         0.43 &         0.44 &         0.45 &         1.00\\
\hline
$\Delta r_{\rm ps}$ &a &   18.059025 &    17.761526 &    17.419893 &    17.173444 &    17.072752 &    16.346841\\
&$\delta$&    0.508291 &     0.507801 &     0.505974 &     0.504958 &     0.505207 &     0.508663\\
\hline
$\Delta u_{\rm ps}$ &a &    7.737585 &     7.508655 &     7.313941 &     7.143246 &     7.041705 &     6.083478\\
&$\delta$&    0.506776 &     0.506175 &     0.505248 &     0.504228 &     0.504426 &     0.506008 \\
\hline\hline
\end{tabular}
\caption{Values of the fitting coefficients $a$ and $\delta$ near the critical point following the fitting equation (\ref{fiteq}) for different values of $b$.}\label{tb:critical_value}
\end{table}
%%%%%%%%%%%%

%%%%%%%%%%%%
%%%%%%%%%

\section{Null geodesics and phase transition for higher dimensional black holes}\label{sec:higher-dim}

In the higher dimensional BI-AdS black hole cases, i.e., $d\geq5$, it is found that there is no the reentrant phase transition \cite{ZouZouZou}, and only the VdW-like phase transition is presented. Here we would like to extend the above treatment of the null geodesics to the higher dimensional cases and to examine the behavior of the temperature in terms of $r_{\rm ps}$ and $u_{\rm ps}$.

The metric for $d$-dimensional BI-AdS black holes is
\begin{equation}
d s^{2}=-f(r) d t^{2}+\frac{1}{f(r)} d r^{2}+r^{2} d\Omega_{d-2}^{2},
\end{equation}
where $d\Omega_{(d-2)}^{2}$ is the line element on the unit
$(d-2)$-dimensional sphere $S^{(d-2)}$, and the metric function is given by
\begin{eqnarray}
f(r) &=& 1 + \frac{r^2}{l^2} -\frac{16 \pi  M r^{3-d}}{(d-2) \omega_{d-2}} + \frac{4 b^2 r^2}{(d-2)(d-1)}\left(1-\sqrt{\frac{16 \pi ^2 Q^2 r^{4-2 d}}{b^2 \omega_{d-2} ^2}+1}\right)\\
& & + \frac{64 \pi ^2 Q^2 r^{6-2 d} }{(d-3) (d-1) \omega^2_{d-2}}\, _2F_1\left(\frac{1}{2},\frac{d-3}{2 d-4};\frac{3 d-7}{2 d-4};-\frac{16 \pi ^2 Q^2 r^{4-2 d}}{b^2 \omega_{d-2} ^2}\right),
\end{eqnarray}
with $\omega_{d-2}=\frac{2 \pi ^{\frac{d-1}{2}}}{\Gamma \left(\frac{d-1}{2}\right)}$ is the volume of the unit sphere $S^{(d-2)}$. The temperature of the black hole can be calculated as
\begin{eqnarray}
 T=\frac{\partial_{r}f(r_{+})}{4\pi}=\frac{(4\pi P+b^{2})r_+}{(d-2)\pi}
  -\frac{b\sqrt{br_{+}^{2d}\omega^2_{d-2}+16\pi^{2}Q^{2}r_{+}^{4}}}{(d-2)\pi \omega_{d-2}r_{+}^{d-1}}.
\end{eqnarray}
Solving the pressure, we can get the equation of state for the black holes
\begin{equation}
P=\frac{(d-2)T}{4r_{+}}-\frac{(d-2)(d-3)}{16\pi r_{+}^2}-\frac{b^2\left(1-\sqrt{\frac{16\pi^2Q^2}{b^2\omega^2r_{+}^{2 d-4}}+1}\right)}{4\pi}.
\end{equation}
It is found that this equation of state describes a small-large black hole phase transition. The critical point can be obtained by solving
\begin{equation}
 (\partial_v P)_{T}= 0,\quad (\partial^2_v P)_{T}= 0.
\end{equation}
Since there is no the analytical result, we list the critical values of the thermodynamical quantities with fixed $b=0.1$ and $Q=1$ for $d$=5-8 in Table \ref{tb:critical_params_nd}. From it, we can find that, with the increase of the dimension number $d$, the value of the critical specific volume $v_{\rm c}$ increases, while the values of $T_{\rm c}$ and $P_{\rm c}$ first decrease, and then increase. Moreover the dimensionless parameter $\frac{P_{\rm c} v_{\rm c}}{T_{\rm c}}$ is larger than $1/8$ of the four dimensional charged AdS black hole.

\begin{figure*}[!htbp]
  \centering
  \subfigure[$d=5$] {\label{fig:prps:5d}\includegraphics[width=0.38\textwidth]{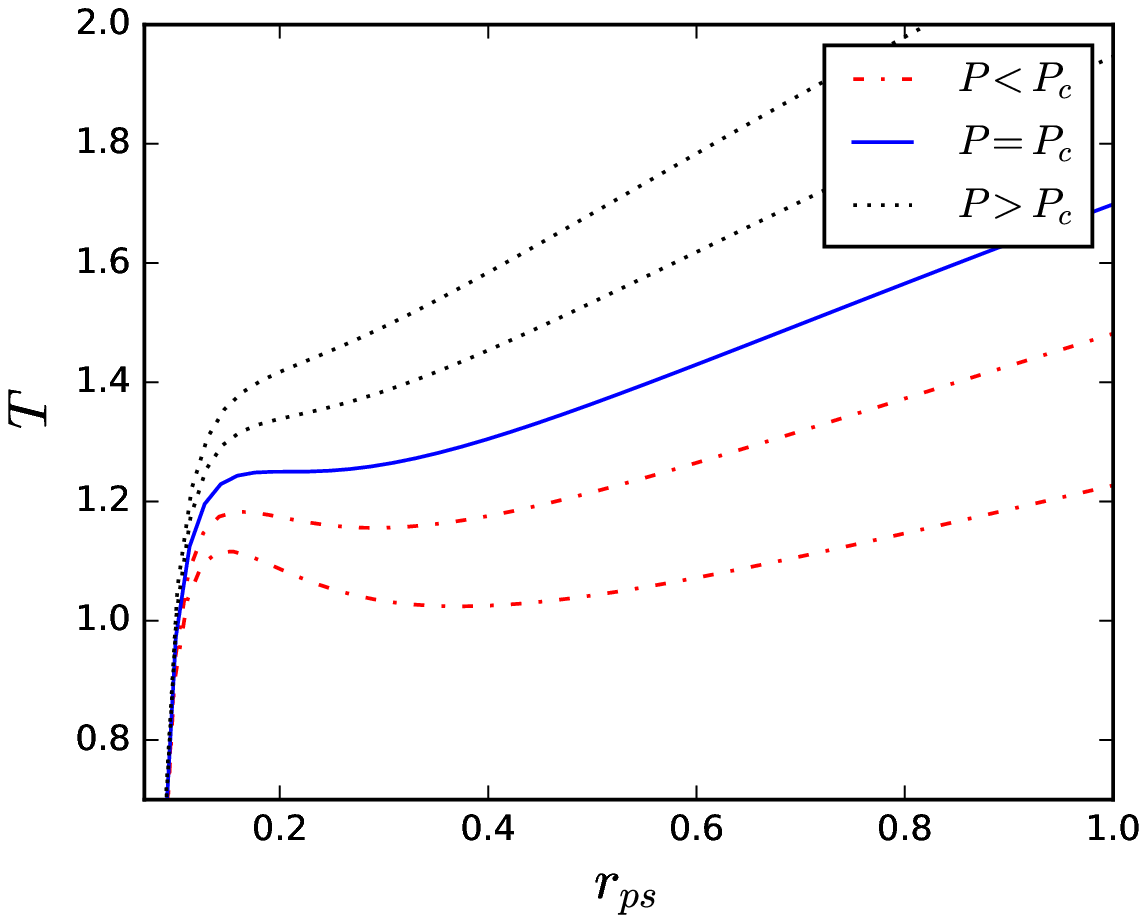}}
  \subfigure[$d=6$] {\label{fig:prps:6d}\includegraphics[width=0.38\textwidth]{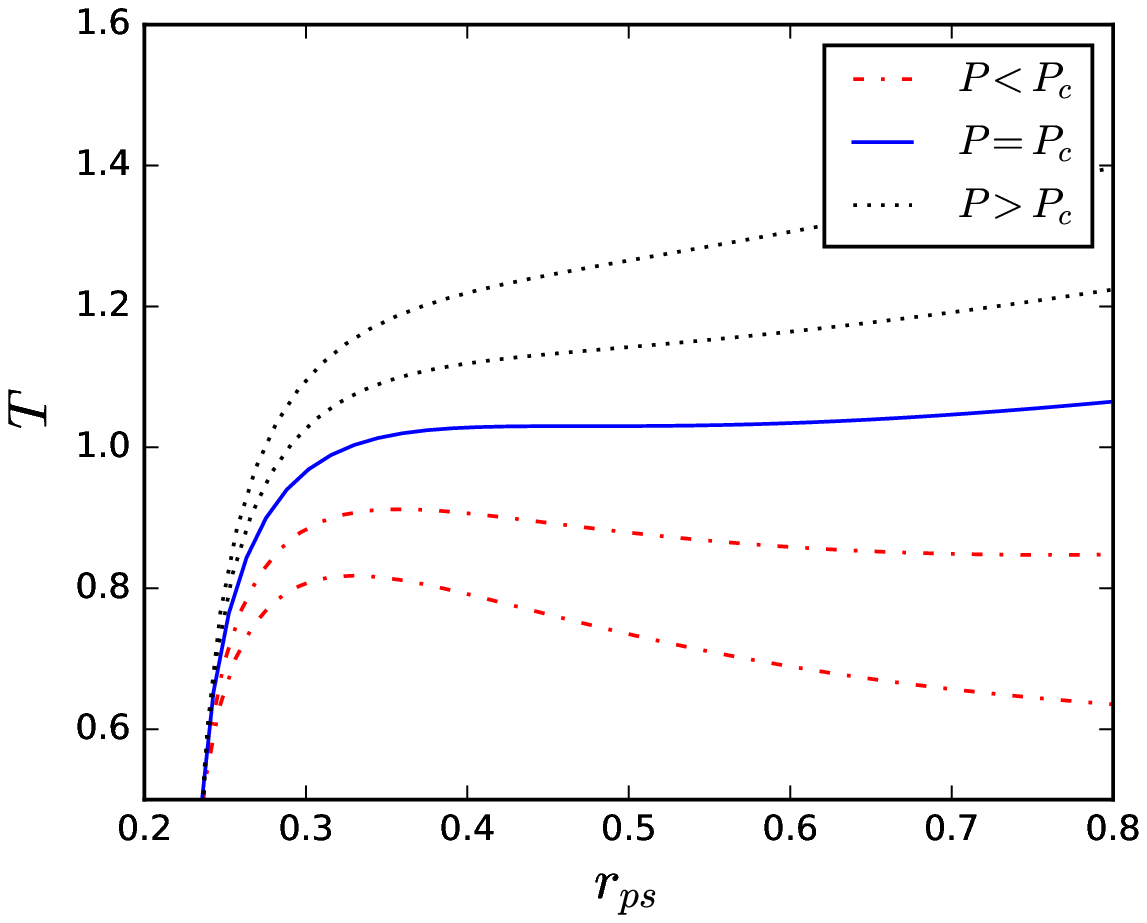}}
  \subfigure[$d=7$] {\label{fig:prps:7d}\includegraphics[width=0.38\textwidth]{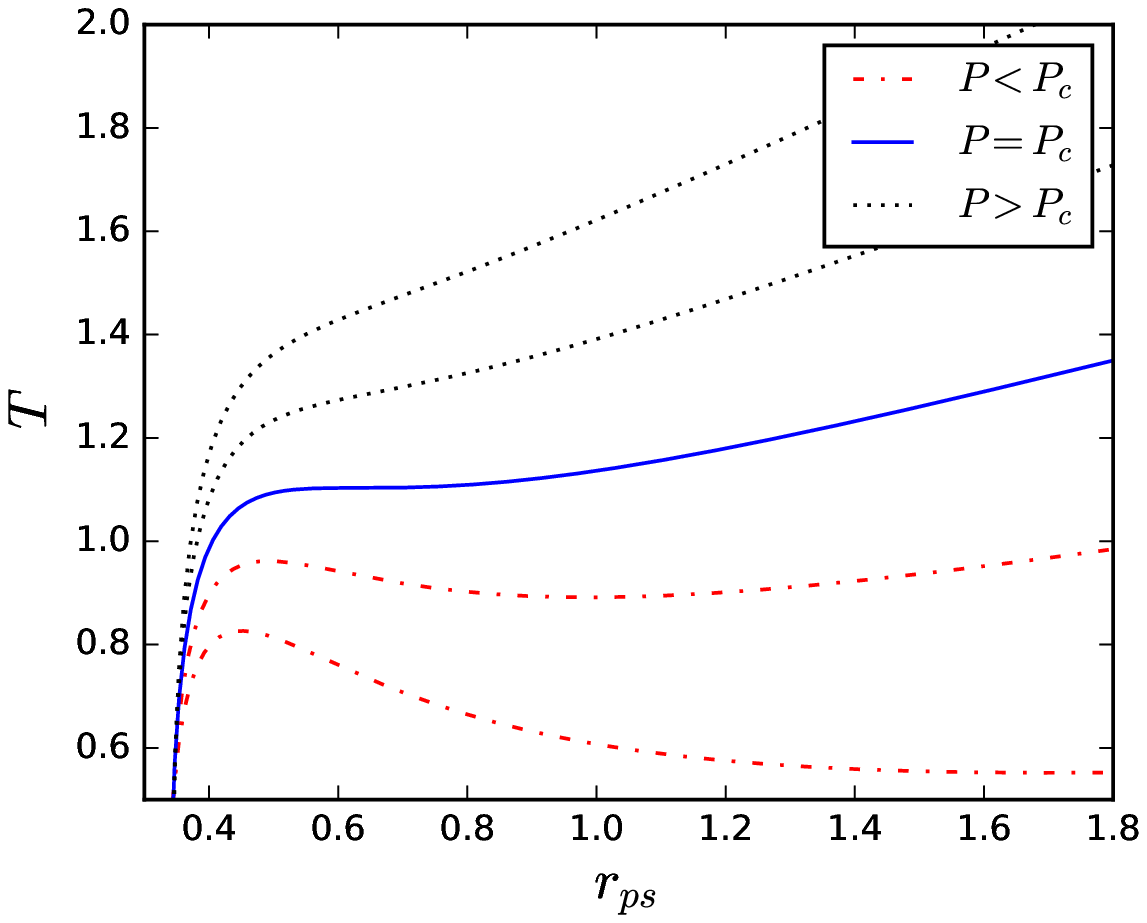}}
  \subfigure[$d=8$] {\label{fig:prps:8d}\includegraphics[width=0.38\textwidth]{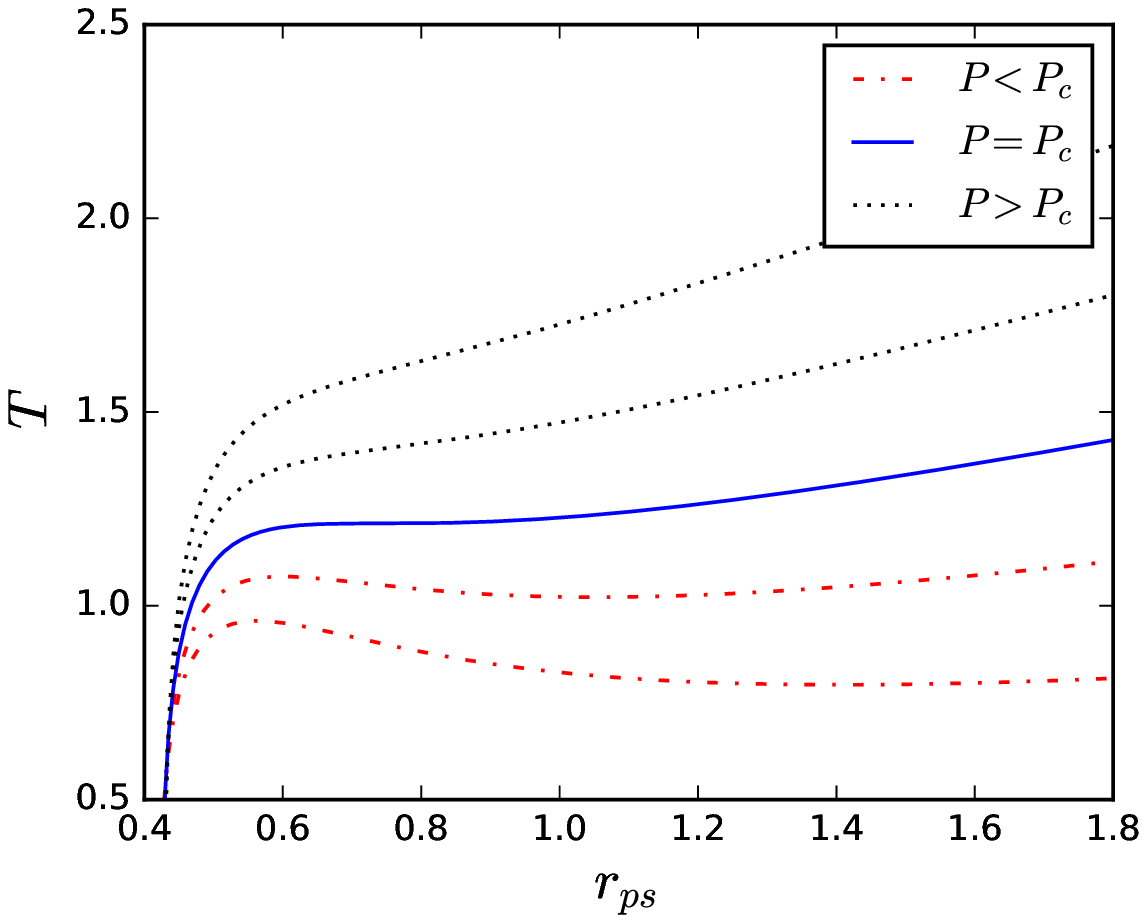}}
\caption{The temperature $T$ as a function of the radius $r_{\rm ps}$ of the photon sphere with $b=0.1$, $Q=1$. (a) $d=5$, and the pressure $P$=1.5, 2.0, 2.45357 ($P_{\rm c}$), 3.0, and 3.5 from bottom to top. (b) $d=6$, and the pressure $P$=0.4, 0.8, 1.24921 ($P_{\rm c}$), 1.6, and 2.0 from bottom to top. (c) $d=7$, and the pressure $P$=0.3, 0.8, 1.27384 ($P_{\rm c}$), 1.8, and 2.3 from bottom to top. (d) $d=8$, and the pressure $P$=0.6, 1.0, 1.4426 ($P_{\rm c}$, 2.0, and 2.6 from bottom to top.}\label{fig:nd-prps-photon-sphere}
\end{figure*}

\begin{figure*}[!htbp]
  \centering
  \subfigure[$d=5$] {\label{fig:prps:5d}\includegraphics[width=0.38\textwidth]{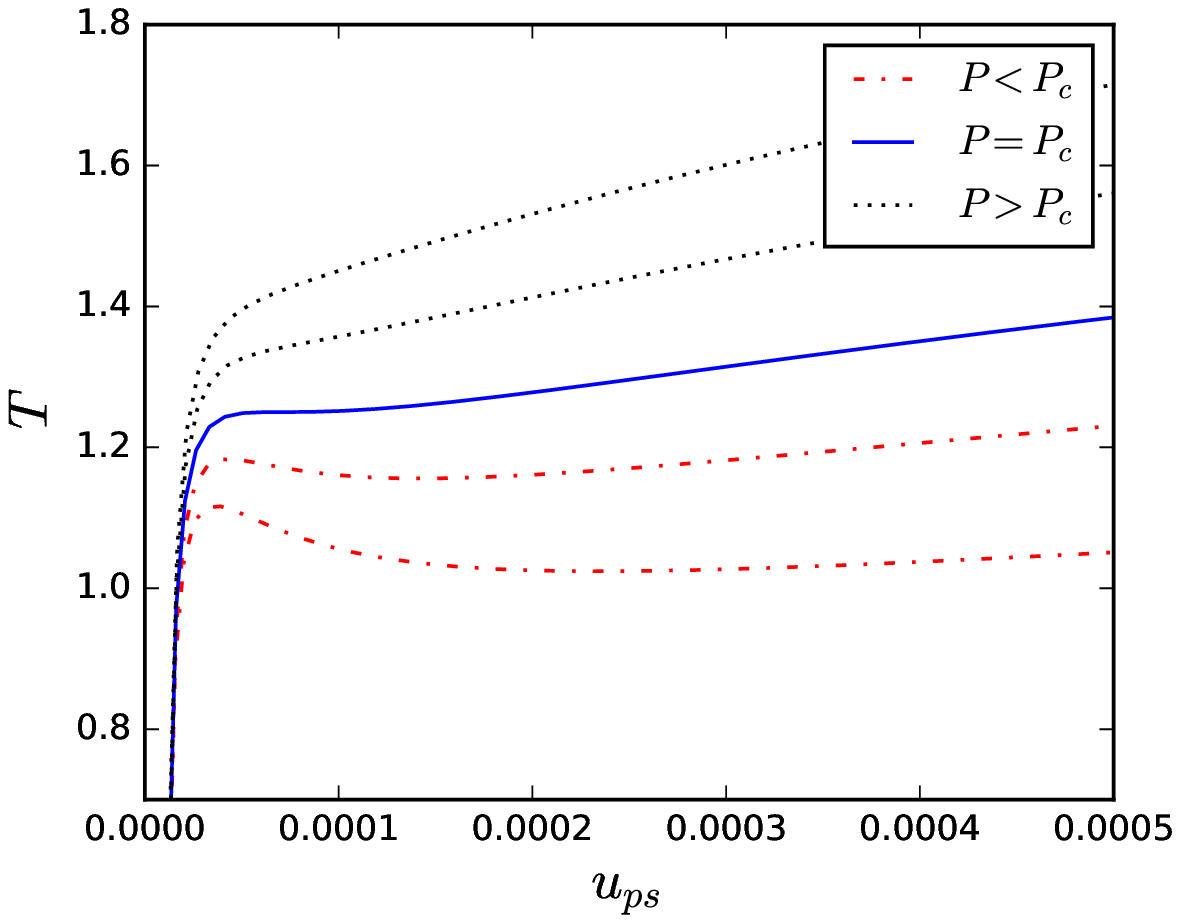}}
  \subfigure[$d=6$] {\label{fig:prps:6d}\includegraphics[width=0.38\textwidth]{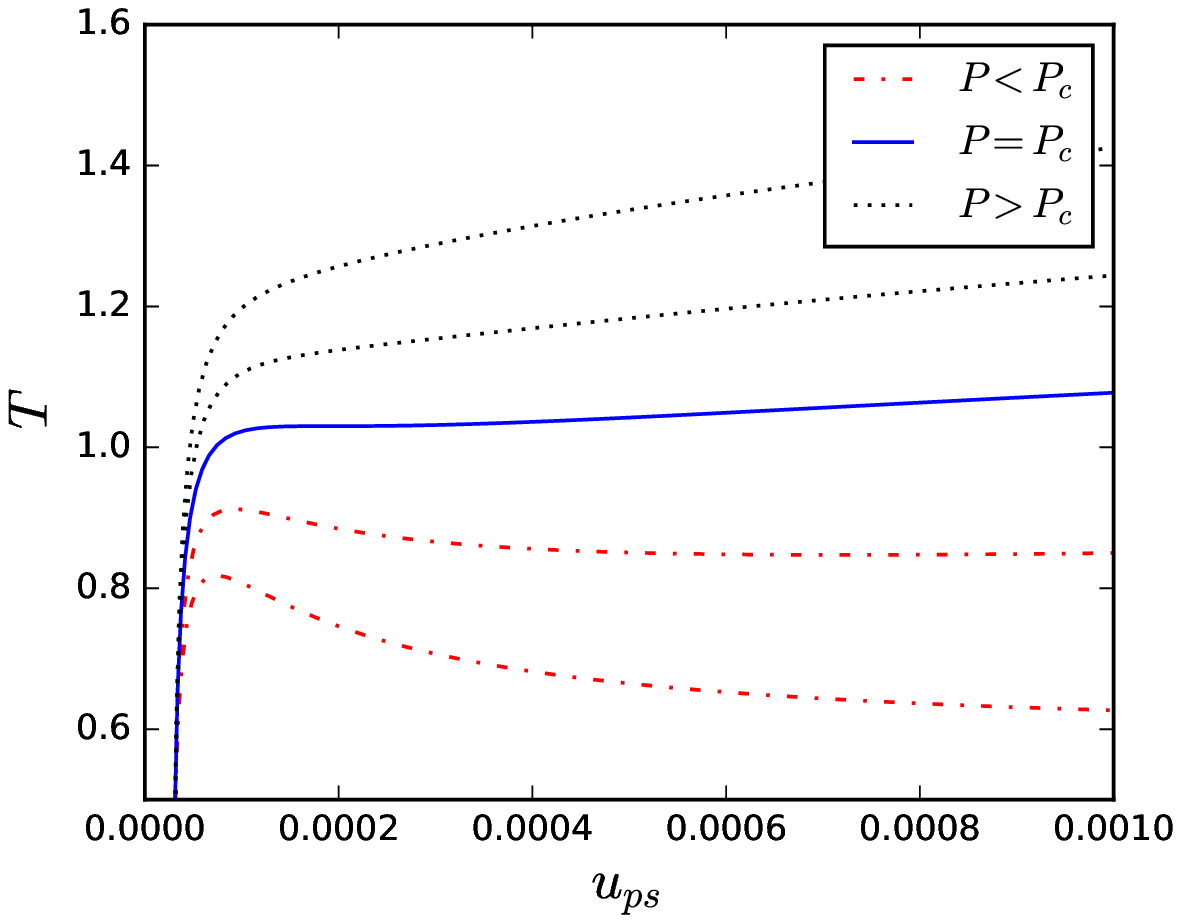}}
  \subfigure[$d=7$] {\label{fig:prps:7d}\includegraphics[width=0.38\textwidth]{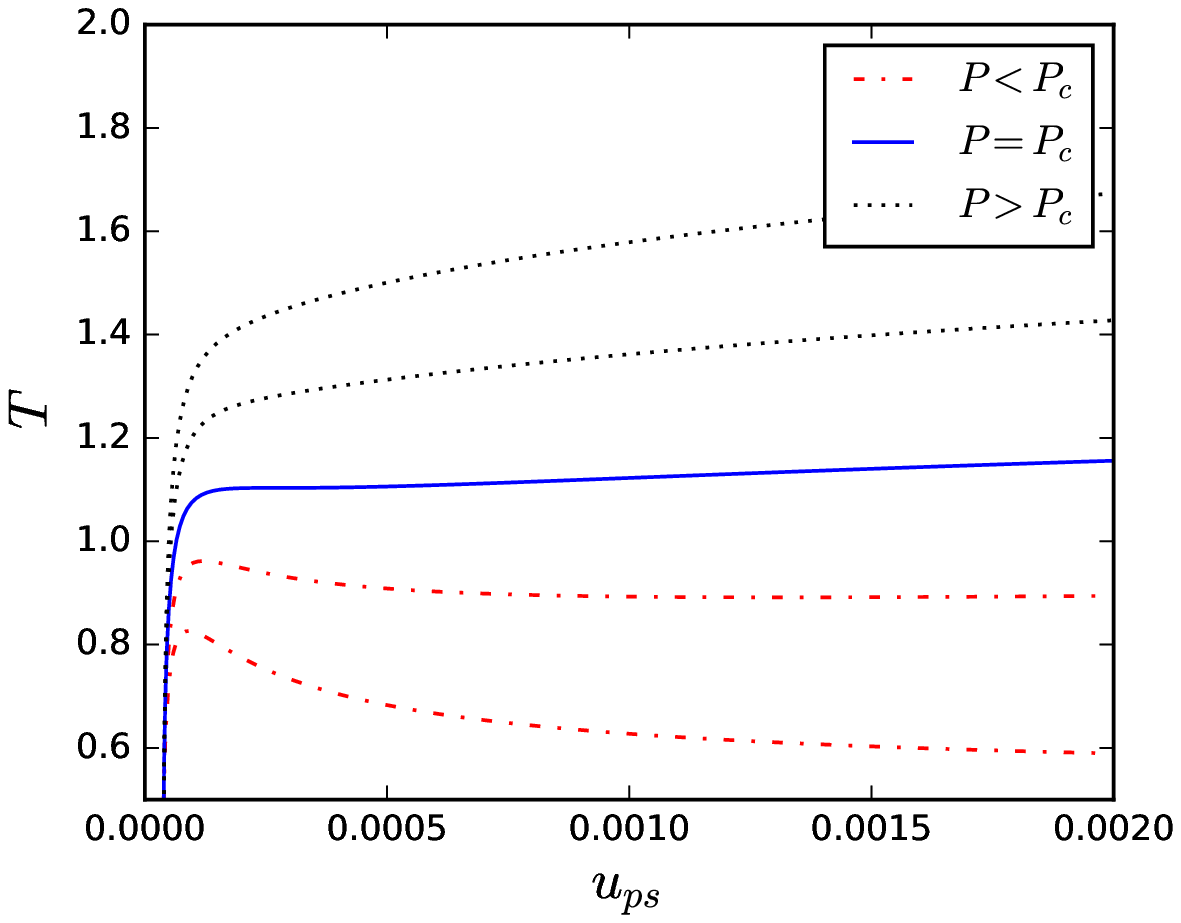}}
  \subfigure[$d=8$] {\label{fig:prps:8d}\includegraphics[width=0.38\textwidth]{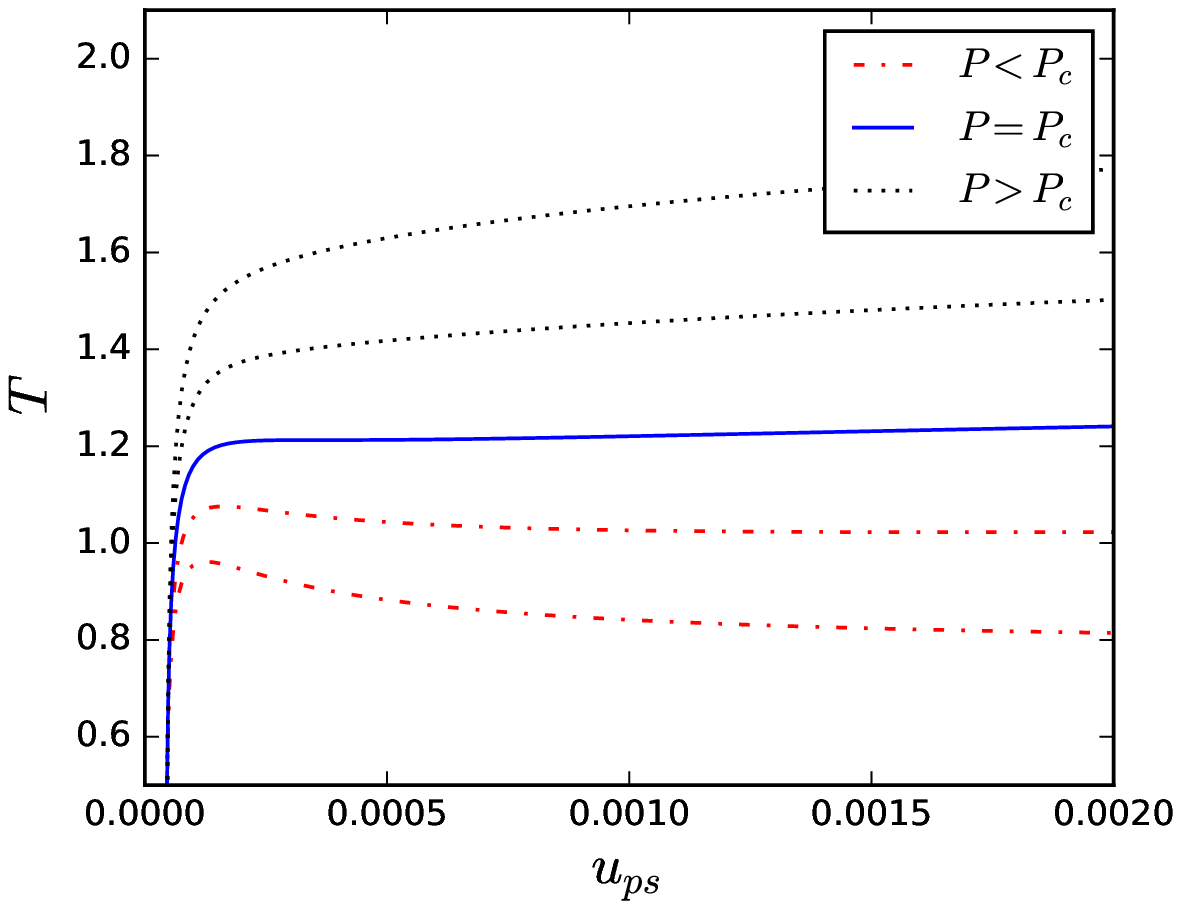}}
  \caption{The temperature $T$ as a function of the minimum impact parameter $u_{\rm ps}$ of the photon sphere with $b=0.1$, $Q=1$. (a) $d=5$, and the pressure $P$=1.5, 2.0, 2.45357 ($P_{\rm c}$), 3.0, and 3.5 from bottom to top. (b) $d=6$, and the pressure $P$=0.4, 0.8, 1.24921 ($P_{\rm c}$), 1.6, and 2.0 from bottom to top. (c) $d=7$, and the pressure $P$=0.3, 0.8, 1.27384 ($P_{\rm c}$), 1.8, and 2.3 from bottom to top. (d) $d=8$, and the pressure $P$=0.6, 1.0, 1.4426 ($P_{\rm c}$, 2.0, and 2.6 from bottom to top. }\label{fig:nd-pups-photon-sphere}
\end{figure*}

For the higher dimensional black holes, it also possesses photon sphere. The equations determine it is also the same as (\ref{upss}) and (\ref{rpss}) of four dimensional case. Here for simplicity, we plot the temperature $T$ as function of $r_{\rm ps}$ and $u_{\rm ps}$ in Figs. \ref{fig:nd-prps-photon-sphere} and \ref{fig:nd-pups-photon-sphere}, respectively.

In Fig. \ref{fig:nd-prps-photon-sphere}, the temperature $T$ is shown against $r_{\rm ps}$ for $d$=5, 6, 7, and 8. It is clear that, when the pressure $T<T_{c}$, the temperature first increases, then decreases, and finally increases with $r_{\rm ps}$. This behavior is a typical behavior indicates that the VdW-like phase transition exits. We can also find that this behavior disappears when $T\geq T_{c}$. So the small-large black hole can be reflected by the radius of the photon sphere. This result is also independent of the dimension number $d$ of the spacetime.

Moreover, we also show the temperature $T$ with the minimum impact parameter $u_{\rm ps}$ in Fig. \ref{fig:nd-pups-photon-sphere} for $d$=5, 6, 7, and 8. For the fixed pressure $P$ below the critical value, with the increase of $u_{\rm ps}$, the temperature has an increase-decrease-increase behavior. Above the critical value, this behavior also disappears.

In summary, with the increase of $r_{\rm ps}$ or $u_{\rm ps}$, the behavior of the temperature is similar to that of the four dimensional black hole case with $b=1$ given in Sec. \ref{sec:null_geodesic_and_critical}. So the small-large black hole phase transition can be revealed even in higher dimensional BI-AdS black holes.

It is worth to examine the critical exponent of $\Delta r_{\rm ps}$ and $\Delta u_{\rm ps}$ at the critical point. We first numerically calculate $\Delta r_{\rm ps}$ and $\Delta u_{\rm ps}$, then fit the data with the formula (\ref{fiteq}). The result is list in Table III. Obviously, the fitting coefficient is always around 0.5 for different $d$. So one can conclude that, $\Delta r_{\rm ps}$ and $\Delta u_{\rm ps}$ have a universal critical exponent $\frac{1}{2}$. This result confirms that of four dimensional BI-AdS black hole and $d$-dimensional charged AdS black hole share the same critical exponent \cite{WeiLiuLiu}.

%%%%%%%%%%%%%%%%%%%%%
\begin{table}[!htbp]%[H] add [H] placement to break table across pages
        \begin{ruledtabular}
            \begin{tabular}{ccccccc}
                    $d$   &       5 &       6  &       7 &         8\\\hline
                $v_c$ &0.169765 & 0.309020 & 0.346222& 0.350027 \\
                $T_c$ &1.250000 & 1.030065 & 1.103263& 1.212527 \\
                $P_c$ &2.453574 & 1.249208 & 1.273844& 1.442595 \\
$\frac{P_c v_c}{T_c}$ &0.333225 & 0.374763 & 0.399754& 0.416442 \\
\end{tabular}
\caption{Critical values of $v_{\rm c}$, $T_{\rm c}$, $P_{\rm c}$, and $\frac{P_{\rm c} v_{\rm c}}{T_{\rm c}}$ with fixed $b=0.1$ and $Q=1$ for $d$=5, 6, 7, and 8.}\label{tb:critical_params_nd}
\end{ruledtabular}
\end{table}
%%%%%%%%%%%%

%%%%%%%%%%%%
%%%%%%%%%

%%%%%%%%%%%%%%%%%%%%%
\begin{table}[!htbp]%[H] add [H] placement to break table across pages
        \begin{ruledtabular}
            \begin{tabular}{lclclclclclclclcl}
                &d &        5 &         6 &         7 &         8\\\hline
$\Delta r_{ps}$ &a &    0.821686 &     1.465679 &     1.702252 &     1.758366\\
          &$\delta$&    0.506893 &     0.497529 &     0.500749 &     0.499735\\\hline
$\Delta u_{ps}$ &a &    0.000711 &     0.001931 &     0.003142 &     0.004750\\
          &$\delta$&    0.548137 &     0.531734 &     0.544612 &     0.514444
\end{tabular}
\caption{Values of the fitting coefficients $a$ and $\delta$ near the critical point following the fitting equation (\ref{fiteq}) for $d$=5, 6, 7, and 8 with $b=0.1$ and $Q=1$}\label{tb:critical_value_nd}
\end{ruledtabular}
\end{table}
%%%%%%%%%%%%

\section{Conclusions and discussions}\label{sec:conclusion}

In this paper, we studied the relationship between the null geodesic and thermodynamic reentrant phase transition of the BI-AdS black holes by examining the radius $r_{\rm ps}$ and the minimal impact parameter $u_{\rm ps}$ of the photon sphere.

At first, we started with the state equation of the BI-AdS black holes. It was shown that the thermodynamic properties are closely dependent of the BI parameter $b$. According to the number of the critical points of the phase transition, the region of $b$ is divided into four cases, case I: $b<b_{0}$, case II: $b\in(b_{0}, b_{1})$, case III: $b\in(b_{1}, b_{2})$, and case IV: $b>b_{2}$. In case I, there is no any phase transition. While in cases II and III, there are two critical points and the typical reentrant phase transitions. The tiny difference between them is that one of the critical points is positive in case II, and negative in case III. However, since they have higher Gibbs free energy, they will not appear in the phase diagram of the black hole system. For case IV, there is only one critical point, and thus only the VdW-like phase transition is shown. For these three different cases of $b>b_{0}$, the reentrant phase transition and VdW-like phase transition are described in the $P$-$T$ phase diagram, see Figs. \ref{ppha}-\ref{pphc}. The extremal point curves were also plotted. It is clear that the critical points always occur at the interaction point of the two extremal point curves. Also, the `zeroth'-order phase transition curve coincides with the extremal point curves. Moreover, the $T$-$v$ phase diagram were also displayed. For the reentrant phase diagram, the coexistence of the intermediate and large black hole cannot be extended to $T$=0, while ends at a certain temperature, where the `zeroth'-order phase transitions start.

Next, we followed the Lagrangian of a free photon to obtain the null geodesics for the black holes. Employing the effective potential of the radial motion, we numerically solved the radius $r_{\rm ps}$ and the minimal impact parameter $u_{\rm ps}$ of the photon sphere for the BI-AdS black holes. The results also indicate that these quantities are closely dependent of the black hole charge and the BI parameter. Then in order to find the relationship between the photon sphere and the phase transition, we plotted the temperature and pressure in terms of $r_{\rm ps}$ and $u_{\rm ps}$, respectively, for different values of $b$.

When fixing the pressure $P$, we showed the temperature $T$ as a function of $r_{\rm ps}$ in Fig. \ref{trpss} and $u_{\rm ps}$ in Fig. \ref{tupss}, respectively. For different types of the phase transition, the temperature behaves very differently, and thus the phase transition can be reflected from the behavior of the temperature. In particular, the reentrant phase transition can be indicated from the decrease-increase-decrease-increase behavior of the temperature with the increase of $r_{\rm ps}$. With the increase of $u_{\rm ps}$, the similar feature of the temperature can also be found for the reentrant phase transition.

On the other hand, when fixed the temperature $T$, the pressure $P$ was plotted with $r_{\rm ps}$ in Fig. \ref{prpss} and with $u_{\rm ps}$ in Fig. \ref{pupss}. For different values of $b$, $P$ also behaves differently. With the increase of $r_{\rm ps}$ or $u_{\rm ps}$, a reentrant phase transition could take place if the pressure has an increase-decrease-increase-decrease behavior.

We also examined the critical behaviors of $r_{\rm ps}$ and $u_{\rm ps}$ near the critical point. Since only one of the critical points is physical when $b>b_{0}$, we fitted the formula (\ref{fiteq}) with the numerical results near the critical point. The results were described in Table \ref{tb:critical_value}. The fitting coefficient $a$ for both $\Delta r_{\rm ps}$ and $\Delta u_{\rm ps}$ decreases with the parameter $b$. While, it is obvious that the fitting coefficient $\delta$ is around $\frac{1}{2}$ with the numerical error no more than 1.73\%. Therefore, $\Delta r_{\rm ps}$ and $\Delta u_{\rm ps}$ have a universal critical exponent $\frac{1}{2}$. These results are exactly consistent with that of the VdW-like phase transition of the charged and rotating AdS black hole given in Refs. \cite{WeiLiuLiu,WeiLiuLiu2}.

Furthermore, we calculated the temperature and pressure corresponding to the extremal points of the radius $r_{\rm ps}$ and the minimal impact parameter $u_{\rm ps}$ of the photon sphere. For different values of $b$, our results show that they completely agree with that of the metastable curves obtained from the thermodynamic side, see Fig. \ref{fig:PointOnPT}. This property also holds even for $b<b_{0}$, where no reentrant phase transition and VdW-like phase transition exist.

Finally, we extended the study to the higher dimensional BI-AdS black holes. The results indicate that the small-large black hole phase transition in higher dimensional cases can also be reflected from the null geodesics. Especially, at the critical point, $\Delta r_{\rm ps}$ and $\Delta u_{\rm ps}$ also have a universal critical exponent $\frac{1}{2}$.

In conclusion, we have investigated the relationship between the photon sphere and the thermodynamic phase transition for the charged BI-AdS black holes. All the results imply that the reentrant phase transition and VdW-like phase transition information of the black holes are encoded in and can be reflected by the property of the photon sphere. The reentrant phase transition can also be distinguished from the VdW-like phase transition through the photon sphere.

\section*{Acknowledgements}
This work was supported by the National Natural Science Foundation of China (Grants No. 11675064, No. 11875151, and No. 11522541). S.-W. Wei was also supported by the Chinese Scholarship Council (CSC) Scholarship (201806185016) to visit the University of Waterloo.


\begin{thebibliography}{99}

\bibitem{Bekenstein}
 J. D. Bekenstein,
 {\em Black holes and entropy},
 Phys. Rev. D \textbf{7}, 2333 (1973).

\bibitem{Bardeen}
 J. M. Bardeen, B. Carter, and S. W. Hawking,
 {\em The four laws of black hole mechanics},
 Comm. Math. Phys. \textbf{31}, 161 (1973)161.


\bibitem{Kastor}
 D. Kastor, S. Ray, and J. Traschen,
  {\em Enthalpy and the Mechanics of AdS Black Holes},
     Class. Quant. Grav. \textbf{26}, 195011 (2009), [arXiv:0904.2765 [hep-th]].

\bibitem{Dolan00}
 B. P. Dolan,
  {\em The cosmological constant and black-hole thermodynamic potentials},
     Class. Quant. Grav. \textbf{28}, 125020 (2011), [arXiv:1008.5023 [gr-qc]].

\bibitem{Cvetic}
 M. Cvetic, G. W. Gibbons, D. Kubiznak, and C. N. Pope,
  {\em Black hole enthalpy and an entropy inequality for the thermodynamic volume},
     Phys. Rev. D \textbf{84}, 024037 (2011), [arXiv:1012.2888 [hep-th]].

\bibitem{Kubiznak}
 D. Kubiznak and R. B. Mann,
  {\em $P$-$V$ criticality of charged AdS black holes},
     J. High Energy Phys. \textbf{1207}, 033 (2012), [arXiv:1205.0559 [hep-th]].

\bibitem{Gunasekaran}
  S. Gunasekaran, D. Kubiznak, and R. B. Mann,
   {\em Extended phase space thermodynamics for charged and rotating black holes and Born-Infeld vacuum polarization},
  J. High Energy Phys. \textbf{1211}, 110 (2012), [arXiv:1208.6251 [hep-th]].

\bibitem{ZouZouZou}
  D.-C. Zou, S.-J. Zhang, and B. Wang,
   {\em Critical behavior of Born-Infeld AdS black holes in the extended phase space thermodynamics},
  Phys. Rev. D \textbf{89}, 044002 (2014), [arXiv:1311.7299 [hep-th]].

\bibitem{Altamirano}
  N. Altamirano, D. Kubiznak, and R. B. Mann,
   {\em Reentrant Phase Transitions in Rotating AdS Black Holes},
  Phys. Rev. D \textbf{88}, 101502(R) (2013), [arXiv:1306.5756 [hep-th]].

\bibitem{Mann}
  N. Altamirano, D. Kubiznak, R. B. Mann, and Z. Sherkatghanad,
   {\em Kerr-AdS analogue of triple point and solid/liquid/gas phase transition},
  Class. Quant. Grav. \textbf{31}, 042001 (2014), [arXiv:1308.2672 [hep-th]].

\bibitem{Frassino}
  A. M. Frassino, D. Kubiznak, R. B. Mann, and F. Simovic,
   {\em Multiple reentrant phase transitions and triple points in lovelock thermodynamics},
  J. High Energy Phys. \textbf{1409}, 080 (2014), [arXiv:1406.7015 [hep-th]].

\bibitem{Wei0}
  S.-W. Wei and Y.-X. Liu,
   {\em Triple points and phase diagrams in the extended phase space of charged Gauss-Bonnet black holes in AdS space},
  Phys. Rev. D \textbf{90}, 044057 (2014), [arXiv:1402.2837 [hep-th]].

\bibitem{Kostouki}
  B. P. Dolan, A. Kostouki, D. Kubiznak, and R. B. Mann,
   {\em Isolated critical point from Lovelock gravity},
  Class. Quant. Grav. \textbf{31}, 242001 (2014), [arXiv:1407.4783 [hep-th].

\bibitem{Wei1}
  S.-W. Wei, P. Cheng, and Y.-X. Liu,
   {\em Analytical and exact critical phenomena of $d$-dimensional singly spinning Kerr-AdS black holes},
   Phys. Rev. D \textbf{93}, 084015 (2016), [arXiv:1510.00085 [gr-qc]].

\bibitem{Hennigar}
  R. A. Hennigar, R. B. Mann, and E. Tjoa,
   {\em Superfluid Black Holes},
   Phys. Rev. Lett. \textbf{118}, 021301 (2017), [arXiv:1609.02564 [hep-th]].

\bibitem{ZouYue}
  M. Zhang, D.-C. Zou, and R.-H. Yue,
   {\em Reentrant phase transitions of topological AdS black holes in four-dimensional Born-Infeld-massive gravity},
   Adv. High Energy Phys. \textbf{2017}, 3819246 (2017),
     [arXiv:1707.04101 [hep-th]].

\bibitem{Hendi}
  S. H. Hendi, B. E. Panah, and S. Panahiyan,
   {\em Einstein-Born-Infeld-Massive Gravity: AdS-black hole solutions and their thermodynamical properties},
   J. High Energy Phys. \textbf{1511}, 157 (2015),
     [arXiv:1508.01311 [hep-th]].

\bibitem{Hendi2}
 S. H. Hendi, B. E. Panah, and S. Panahiyan,
   {\em Thermodynamical structure of AdS black holes in massive gravity with stringy gauge-gravity corrections},
   Class. Quant. Grav. \textbf{33}, 235007 (2016),
     [arXiv:1510.00108 [hep-th]].

\bibitem{Hendi3}
 S. H. Hendi, G.-Q. Li, J.-X. Mo, S. Panahiyan, and B. E. Panah,
   {\em New perspective for black hole thermodynamics in Gauss-Bonnet-Born-Infeld massive gravity},
   Eur. Phys. J. C \textbf{76}, 571 (2016),
     [arXiv:1608.03148 [gr-qc]].

\bibitem{Hendi4}
 S. H. Hendi, R. B. Mann, S. Panahiyan, and B. EslamPanah,
   {\em van der Waals like behaviour of topological AdS black holes in massive gravity},
   Phys. Rev. D \textbf{95}, 021501(R) (2017),
     [arXiv:1702.00432 [gr-qc]].

\bibitem{Momeni}
 D. Momeni, M. Faizal, K. Myrzakulov, and R. Myrzakulov,
   {\em Fidelity susceptibility as holographic PV-criticality},
   Phys. Lett. B \textbf{765}, 154 (2017),
     [arXiv:1604.06909 [hep-th]].

\bibitem{Chakraborty}
 S. Chakraborty and T. Padmanabhan,
   {\em Thermodynamical interpretation of the geometrical variables associated with null surfaces},
   Phys. Rev. D \textbf{92}, 104011 (2015),
     [arXiv:1508.04060 [gr-qc]].

\bibitem{Weisw}
 S.-W. Wei and Y.-X. Liu,
  {\em Insight into the Microscopic Structure of an AdS Black Hole from thermodynamic Phase Transition},
      Phys. Rev. Lett. \textbf{115}, 111302 (2015), [arXiv:1502.00386 [gr-qc]].

\bibitem{Pan}
 J. Jing and Q. Pan,
   {\em Quasinormal modes and second order thermodynamic phase transition for Reissner-Nordstr\"{o}m black hole},
   Phys. Lett. B \textbf{660}, 13 (2008),
 [arXiv:0802.0043 [gr-qc]].

\bibitem{BertiBerti}
 E. Berti and V. Cardoso,
   {\em Quasinormal modes and thermodynamic phase transitions},
   Phys. Rev. D \textbf{77}, 087501 (2008),
 [arXiv:0802.1889 [hep-th]].

\bibitem{He}
 X. He, S. Chen, B. Wang, R.-G. Cai, and C.-Y. Lin,
   {\em Quasinormal modes in the background of charged Kaluza-Klein black hole with squashed horizons},
   Phys. Lett. B \textbf{665}, 392 (2008),
 [arXiv:0802.2449 [hep-th]].

\bibitem{He2}
 X. He, B. Wang, and S. Chen,
   {\em Quasinormal modes of charged squashed Kaluza-Klein black holes in the G?del Universe},
   Phys. Rev. D \textbf{79}, 084005 (2009),
 [arXiv:0811.2322 [gr-qc]].

\bibitem{LinLin}
 K. Lin, J. Li, and N. Yang,
   {\em Dynamical behavior and nonminimal derivative coupling scalar field of Reissner-Nordstroem black hole with a global monopole},
   Gen. Rel. Grav. \textbf{43} (2011).

\bibitem{Sup}
 Q.-Y. Pan and R.-K. Su,
   {\em Quasinormal Modes of Phantom Scalar Perturbation in Background of Reissner-Nordstrom Black Hole},
   Commun. Theor. Phys. \textbf{55}, 221 (2011).

%---------------------------------------------

\bibitem{Liu}
  Y. Liu, D.-C. Zou, and B. Wang,
   {\em Signature of the Van der Waals like small-large charged AdS black hole phase transition in quasinormal modes},
     J. High Energy Phys. \textbf{1409}, 179 (2014), [arXiv:1405.2644 [hep-th]].

\bibitem{Liuzz}
  D.-C. Zou, Y. Liu, C.-Y. Zhang, and B. Wang,
   {\em Dynamical probe of thermodynamical properties in three-dimensional hairy AdS black holes},
      Europhys. Lett. \textbf{116}, 40005 (2016), [arXiv:1411.6740 [hep-th]].

\bibitem{Mahapatra}
  S. Mahapatra,
   {\em Thermodynamics, phase transition and quasinormal modes with Weyl corrections},
     J. High Energy Phys. \textbf{1604}, 142 (2016),
     [arXiv:1602.03007 [hep-th]].

\bibitem{Chabab}
  M. Chabab, H. El Moumni, S. Iraoui, and K. Masmar,
   {\em Behavior of quasinormal modes and high dimension RNAdS black hole phase transition},
     Eur. Phys. J. C \textbf{76}, 676 (2016), [arXiv:1606.08524 [hep-th]].

\bibitem{Zou2}
  D.-C. Zou, Y. Liu, and R.-H. Yue,
   {\em Behavior of quasinormal modes and Van der Waals-like phase transition of charged AdS black holes in massive gravity},
     Eur. Phys. J. C \textbf{77}, 365 (2017), [arXiv:1702.08118 [hep-th]].



\bibitem{Prasia0}
  R. Tharanath, N. Varghese, and V. C. Kuriakose,
   {\em Phase transition, Quasinormal modes and Hawking radiation of Schwarzschild black hole in Quintessence field},
     Mod. Phys. Lett. A \textbf{29}, 1450057 (2014), [arXiv:1404.0791 [gr-qc]].

\bibitem{Prasia2}
  P. Prasia and V. C. Kuriakose,
   {\em Quasi Normal Modes and P-V Criticallity for scalar perturbations in a class of dRGT massive gravity around Black Holes},
      Gen. Rel. Grav. \textbf{48}, 89 (2016), [arXiv:1606.01132 [gr-qc]].

\bibitem{Prasia}
  P. Prasia and V. C. Kuriakose,
   {\em Quasinormal modes and thermodynamics of linearly charged btz black holes in massive gravity in (Anti) de Sitter space time},
     Eur. Phys. J. C \textbf{77}, 27 (2017), [arXiv:1608.05299 [gr-qc]].

\bibitem{www}
  B. Liang, S.-W. Wei, and Y.-X. Liu,
  {\em Quasinormal Modes and Van der Waals like phase transition of charged AdS black holes in Lorentz symmetry breaking massive gravity},
  [arXiv:1712.01545 [gr-qc]].

\bibitem{ZengZeng}
  A.-C. Li, H.-Q. Shi, and D.-F. Zeng,
  {\em Phase structure and QNMs of a charged AdS dilaton black hole},
  Phys. Rev. D \textbf{97}, 026014 (2018)
  [arXiv:1711.04613 [hep-th]].

\bibitem{YueYue}
  M. Zhang and R.-H. Yue,
  {\em Phase Transition and Quasinormal Modes for Spherical Black Holes in 5D Gauss¨CBonnet Gravity},
  Chin. Phys. Lett. \textbf{35}, 040401 (2018).




\bibitem{WeiLiuLiu}
  S.-W. Wei and Y.-X. Liu,
    {\em Photon orbits and thermodynamic phase transition of d-dimensional charged AdS black holes},
    Phys. Rev. D \textbf{97}, 104027 (2018),
      [arXiv:1711.01522 [gr-qc]].

\bibitem{WeiLiuLiu2}
  S.-W. Wei, Y.-X. Liu, and Y.-Q. Wang,
    {\em Probing the relationship between the null geodesics and thermodynamic phase transition for rotating Kerr-AdS black holes},
    Phys. Rev. D \textbf{99}, 044013 (2019),
      [arXiv:1807.03455 [gr-qc]].

\bibitem{Bhamidipati}
  C. Bhamidipati and S. Mohapatra,
    {\em A Note on Circular Geodesics and Phase Transitions of Black Holes},
    Phys. Lett. B \textbf{791}, 367 (2019),
      [arXiv:1805.05088 [hep-th]].

\bibitem{Han}
  S.-Z. Han, J. Jiang, M. Zhang, and W.-B. Liu,
    {\em Photon orbits and thermodynamic phase transition in Gauss-Bonnet AdS black holes},
      [arXiv:1812.11862 [gr-qc]].

\bibitem{Bhamidipatib}
  M. Chabab, H. El Moumni, S. Iraoui, and K. Masmar,
    {\em Probing correlation between photon orbits and phase structure of charged AdS black hole in massive gravity background},
      [arXiv:1902.00557 [hep-th]].

\bibitem{ZhangHan}
  M. Zhang, S.-Z. Han, J. Jiang, and W.-B. Liu,
    {\em Circular orbit of a test particle and phase transition of a black hole},
    Phys. Rev. D \textbf{99}, 065016 (2019)
      [arXiv:1903.08293 [hep-th]].

\bibitem{Cvetic2}
  M. Cvetic, G. W. Gibbons, and C. N. Pope,
   {\em Photon spheres and sonic horizons in black holes from supergravity and other theories},
     Phys. Rev. D \textbf{94}, 106005 (2016), [arXiv:1608.02202 [gr-qc]].

\bibitem{Cvetic3}
  G. W. Gibbons, C. M. Warnick, and M. C. Werner,
   {\em Light-bending in Schwarzschild-de-Sitter: projective geometry of the optical metric},
     Class. Quant. Grav. \textbf{25}, 245009 (2008), [arXiv:0808.3074 [gr-qc]].

\bibitem{Tang}
 Z.-Y. Tang, Y. C. Ong, and B. Wang,
   {\em Lux in obscuro II: photon orbits of extremal AdS black holes revisited},
     Class. Quant. Grav. \textbf{34}, 245006 (2017),
     [arXiv:1705.09633 [gr-qc]].

\end{thebibliography}
\end{document}